\documentclass[aps,prd,reprint,preprintnumbers,superscriptaddress,nofootinbib,longbibliography,floatfix]{revtex4-1}

\usepackage{url}
\usepackage{xspace}
\usepackage{dsfont}
\usepackage{amssymb}
\usepackage{amsmath}
\usepackage{graphicx}
\usepackage[caption=false]{subfig}
\usepackage{multirow}
 \usepackage{physics}
 \usepackage{hyperref}
\hypersetup{
  colorlinks=true,
  citecolor=red,
  linkcolor=red,
  urlcolor=red
}

\usepackage{placeins}

\usepackage{color}
\definecolor{mit-red}{rgb}{0.64,.12,0.2}
\definecolor{orange}{rgb}{1.0,0.5,0.0}
\definecolor{darkred}{rgb}{1.0,0.1,0.1}
\definecolor{darkgreen}{rgb}{0.1,0.7,0.1}
\definecolor{darkblue}{rgb}{0.1,0.1,1.0}

\DeclareMathSymbol{\Xdsum}{\mathop}{largesymbols}{88}
\DeclareMathSymbol{\Xtsum}{\mathop}{largesymbols}{80}

\DeclareRobustCommand{\Sec}[1]{Sec.~\ref{sec:#1}}

\DeclareRobustCommand{\App}[1]{App.~\ref{app:#1}}

\DeclareRobustCommand{\Tab}[1]{Table~\ref{tab:#1}}

\DeclareRobustCommand{\Fig}[1]{Fig.~\ref{fig:#1}}
\DeclareRobustCommand{\Figs}[2]{Figs.~\ref{fig:#1} and \ref{fig:#2}}
\DeclareRobustCommand{\Eq}[1]{Eq.~(\ref{eq:#1})}

\DeclareRobustCommand{\Reference}[1]{Ref.~\cite{#1}}
\DeclareRobustCommand{\Refs}[1]{Refs.~\cite{#1}}

\def\cL{\mathcal{L}}
\def\cO{\mathcal{O}}
\def\cP{\mathcal{P}}

\newcommand{\Pythia}{{\sc Pythia}\xspace}
\newcommand{\EnergyFlow}{{\sc EnergyFlow}\xspace}

\newcommand{\FastJet}{{\sc FastJet}\xspace}

\newcommand{\Keras}{{\sc Keras}\xspace}
\newcommand{\TensorFlow}{{\sc TensorFlow}\xspace}
\newcommand{\Adam}{{\sc Adam}\xspace}

\begin{document}

\title{Moments of Clarity: \\Streamlining Latent Spaces in Machine Learning using Moment Pooling}

\preprint{MIT-CTP 5689}

\author{Rikab Gambhir}
\email{rikab@mit.edu}
\affiliation{Center for Theoretical Physics, Massachusetts Institute of Technology, Cambridge, MA 02139, USA}
\affiliation{The NSF AI Institute for Artificial Intelligence and Fundamental Interactions}

\author{Athis Osathapan}
\email{aosathap@bowdoin.edu}
\affiliation{The NSF AI Institute for Artificial Intelligence and Fundamental Interactions}
\affiliation{Bowdoin College, Brunswick, ME 04011, USA}

\author{Jesse Thaler}
\email{jthaler@mit.edu}
\affiliation{Center for Theoretical Physics, Massachusetts Institute of Technology, Cambridge, MA 02139, USA}
\affiliation{The NSF AI Institute for Artificial Intelligence and Fundamental Interactions}

\begin{abstract}
Many machine learning applications involve learning a latent representation of data, which is often high-dimensional and difficult to directly interpret.
In this work, we propose ``Moment Pooling'', a natural extension of Deep Sets networks which drastically decreases the latent space dimensionality of these networks while maintaining or even improving performance. 
Moment Pooling generalizes the summation in Deep Sets to arbitrary multivariate moments, which enables the model to achieve a much higher effective latent dimensionality for a fixed learned latent space dimension.
We demonstrate Moment Pooling on the collider physics task of quark/gluon jet classification by extending Energy Flow Networks (EFNs) to Moment EFNs.
We find that Moment EFNs with latent dimensions as small as 1 perform similarly to ordinary EFNs with higher latent dimension. 
This small latent dimension allows for the internal representation to be directly visualized and interpreted, which in turn enables the learned internal jet representation to be extracted in closed form.
\end{abstract}

\maketitle

\tableofcontents


\section{Introduction}

%
As modern machine learning (ML) models and their applications continue to grow in size and scope, their internal representations of data become increasingly more complex and difficult to decipher.
While there are a variety of ways to interpret what is ``learned'' in an ML model~\cite{d2,d3, d1, Molnar_2020, rudin2021interpretable, Dillon:2021gag, Bogatskiy:2022hub, gao2023interpretability, Bogatskiy:2023nnw, wayland2024mapping}, it is often difficult to draw concrete, first-principles conclusions on how these models internally represent learned data, as the latent space tends to be high-dimensional and complex.
This, in turn, makes it more difficult not only to trust ML models when applied outside their original training sets, but also to understand what additional domain insights may be driving the improved performance of these models.

ML methods have been gaining interest in collider physics, and have shown to perform remarkably well in a variety of collider physics and jet substructure tasks \cite{Denby:1987rk,Guest:2018yhq,Butter:2017cot, Albertsson:2018maf, PhysRevD.101.056019, Bourilkov:2019yoi, Gong:2022lye, Shlomi:2020gdn, Chakraborty:2020yfc, Butter:2020tvl, Kagan:2020yrm, Lim:2020igi, Dreyer:2020brq, Karagiorgi:2021ngt,  Schwartz:2021ftp,  Qu:2022mxj, Baldi:2022okj,Plehn:2022ftl,RevModPhys.91.045002, Butter:2022rso, Bogatskiy:2022czk, Atkinson:2022uzb, Bhardwaj:2024djv}. 
Recently, the Energy Flow Network (EFN)~\cite{efn} has emerged as a promising model, performing relatively well on jet tagging~\cite{Kasieczka:2019dbj} while being more robust than other models with respect to training set simulation choice~\cite{ATLAS:2022qby}.
EFNs are a generalization of Deep Sets~\cite{deepsets},\footnote{EFNs generalize Deep Sets in the sense that EFNs reduce to Deep Sets when weights are removed, discussed more below.} which use a set-based representation of the event, $\cP$, to construct observables with the ansatz:
\begin{align}
    \mathcal{O}\left( \cP \right) = F\left(\expval{ \Phi^a}_\cP \right) \label{eq:EFN},
\end{align}
where $\expval{\Phi}$ is the expectation value of $\Phi$ over the event $\cP$, defined below.
The function $\Phi:\mathbb{R}^d \xrightarrow{}\mathbb{R}^L$, usually parametrized as a dense neural network, is a per-particle $L$-dimensional latent representation of $x$, with the latent dimension indexed by $a = 1, ..., L$.
The function $F$ (another dense neural network) is then a function of this representation, which converts the latent representation into the observable $\mathcal{O}$.
The Deep Sets theorem, as discussed in \Refs{deepsets, efn}, guarantees that any (infrared and collinear (IRC)-safe) observable can be approximated arbitrarily well for a sufficiently expressive $F$ and $\Phi$, and large enough $L$.
However, the theorem makes no guarantees on the complexity of $\Phi$ or $F$, and may require a very large $L$.

In this paper, we introduce Moment Pooling, a natural extension of Deep Sets architectures that significantly reduces the number of latent dimensions $L$ needed while maintaining or improving its performance.
The Moment Pooling operation generalizes the expectation value of $\Phi$ in \Eq{EFN} to higher order multivariate moments:
\begin{equation}
\cO_k(\cP) \equiv F \big(\expval{\Phi^a}_\cP, \expval{\Phi^{a_1}\Phi^{a_2}}_\cP, ... \expval{\Phi^{a_1}...\Phi^{a_k}}_\cP \big),
\label{eq:MPA}
\end{equation}
where $k$ is the highest order moment considered.
%
%
This procedure is inspired by histogram pooling ~\cite{histogrampooling}, in which the $\Phi$ are histograms binned in $x$.
We focus primarily on applying Moment Pooling to EFNs in the collider physics context, where \Eq{MPA} defines an order $k$ \emph{Moment EFN}, which reduces to the ordinary EFN when $k = 1$.
Alternative modifications of EFNs are discussed in \Refs{Shen:2023ofd, Bright-Thonney:2023gdl}.

We show that for $k > 1$, a Moment EFN enables the same or better performance on quark/gluon jet classification as an EFN, but with a much smaller latent dimension $L$, allowing the same machine-learned observables to be constructed using fewer base functions.
With fewer latent dimensions, it is much easier to directly visualize the model's internal representations and therefore easier to directly interpret and find closed-form expressions for the learned observable.
As a concrete example, an order $k = 4$ Moment EFN with a \emph{single} latent dimension achieves comparable performance on quark/gluon jet classification to an ordinary EFN with 4 latent dimensions.
We are able to directly plot this latent dimension and find that it takes a remarkably simple closed form, the ``log angularity'' observable, which bears many similarities to jet angularities~\cite{Berger:2003iw, Berger:2004xf}.

The rest of the paper is organized as follows: 
In \Sec{math}, we give an overview of moment pooling and the Moment EFN architecture, show how it naturally arises as a generalization of Deep Sets, and introduce the idea of effective latent dimensions.
In \Sec{qg_discrimination}, we demonstrate how the Moment EFN may be used for quark/gluon discrimination, and how Moment EFNs outperform ordinary EFNs as $L$ and $k$ are varied.
In \Sec{black_box}, we analyze the latent spaces of small-$L$ Moment EFNs and attempt to understand them in terms of simple closed-form fits, allowing for analytic observables to be extracted from the model.
Finally, in \Sec{conclusions}, we present our conclusions and outlook.
Implementation details of the architecture may be found in \App{model_specifications}. 
An additional study involving regression on jet angularities, rather than classification, using Moment EFNs may be found in \App{jet_angularities}.
Additional studies complementing \Sec{qg_discrimination}, involving top/QCD discrimination and Moment Particle Flow Networks (PFNs) rather than EFNs, may be found in \App{additional}.

\section{Moment Pooling}\label{sec:math}

We begin with the construction of the Moment Pooling operation. 
We first define Moment Pooling as an extension of Deep Sets and apply it to EFNs, a form of weighted Deep Sets, to produce Moment EFNs in \Sec{moment_efn}.
Then, in \Sec{effective_latent_dim}, we discuss how Moment Pooling is capable of reducing the latent dimension of EFNs through the concept of effective latent dimensions.


\subsection{The Moment Energy Flow Network}\label{sec:moment_efn}

The Moment Pooling operation, as given by \Eq{MPA}, is a generalization of Deep Sets-style architectures. 
The form of \Eq{MPA} is motivated by the observation that the summation step over the latent representation $\Phi^a$ in Deep Sets architectures, generalized to weighted sums in EFNs, can be regarded as taking an \emph{expectation} value of the $L$-dimensional random variable $\Phi^a(x)$ defined over a base space $\mathbb{R}^d$, taken over $\cP$:
\begin{align}
    \expval{\Phi^a}_\cP \equiv \sum_{i \in \cP} z_i \Phi^a(x_i), \label{eq:expectation_value}
\end{align}
where $z_i$ are weights and $x_i \in \mathbb{R}^d$.
In the collider physics context, $z_i$ are (normalized) particle energies and $x_i = (y_i, \phi_i)$ are particle's rapidity $y$ and azimuthal angle $\phi$ on the detector, and $\cP$ is a probability distribution of energy on detector space, or an energy flow~\cite{Tkachov:1995kk,Sveshnikov:1995vi, Tkachov:1999py, Hofman:2008ar, Ba:2023hix}, over which we can take expectation values.\footnote{To align with the notation of \Reference{Ba:2023hix}, we have $\expval{\Phi^a}_\cP = \expval{\cP, \Phi^a}$.}

Applying \Eq{expectation_value} to \Eq{EFN}, we find:
\begin{align}
    \cO(\cP) &= F\left(\sum_{i\in\cP} z_i \Phi^a(x_i)\right),
\end{align}
which is how an EFN is typically written~\cite{efn}.
Note that an ordinary Deep Sets network, as presented in \Reference{deepsets}, is simply a special case of the EFN where $z_i = 1$ for all $i$.

Given that EFNs are functions $F$ of the expectation value of $\Phi^a$, it is natural to extend them to also include higher-order moments of $\Phi^a$, arriving at the Moment Energy Flow Network.
More precisely, the Moment EFN of \Eq{MPA} simply extends $F$ from being a function of only the expectation value of $\Phi^a$ to a function of up to $k$ moments of $\Phi^a$, which reduces to the ordinary EFN for $k=1$.
As an explicit example, the $k = 2$ Moment EFN takes the form:
\begin{align}
    \cO_2(\cP) = F(\expval{\Phi^a}_\cP, \expval{\Phi^{a_1} \Phi^{a_2}}_\cP),
\end{align}
where $\expval{\Phi^{a_1}\Phi^{a_2}}$ is the second moment of $\Phi$, which is:
\begin{align}
    \expval{\Phi^{a_1}\Phi^{a_2}}_\cP &= \sum_{i\in\cP} z_i \Phi^{a_1}(x_i)\Phi^{a_2}(x_i).
\end{align}
This quantity is related to the covariance between the random variables $\Phi^{a_1}$ and $\Phi^{a_2}$:
\begin{align}
    \expval{\Phi^{a_1}\Phi^{a_2}}_\cP &= \left[\text{Cov}(\Phi, \Phi)\right]^{a_1 a_2}_\cP + \expval{\Phi^{a_1}}_\cP\expval{\Phi^{a_2}}_\cP.
\end{align}
Similarly, the $k=3$ and $k=4$ Moment EFNs contain the skew and kurtosis of $\Phi^a$, respectively.
In principle, it is also possible to instead define a ``Cumulant EFN'' with ``Cumulant Pooling'', where $F$ is a function of the first $k$ cumulants rather than the first $k$ moments, though we will not pursue this here.
In general, keeping only the first $k$ moments can be thought of as an unpixelated generalization of max- or mean-pooling procedure in convolutional neural networks, wherein one ``coarse grains'' the distribution $\Phi$, hence the term ``moment pooling''.
For example, by keeping only the first two moments of $\Phi$, we have effectively smoothed out the higher-moment information of our point cloud and kept only the Gaussian component.

It is important to emphasize that $\Phi^a$ remains a function of a single particle, and that the moments are taken over the set of particles, \emph{not} pairs or $m$-tuples of particles. 
In other words, $\Phi^a(x_i)$ only provides information about the $i$'th particle, and the moments $\expval{\prod \Phi^a}$ describe only how that information is distributed across an event, \emph{not} explicit inter-particle correlations.
This is in contrast to graph-based approaches, such as ParticleNet~\cite{PhysRevD.101.056019}, IRC-safe graph networks~\cite{Atkinson:2022uzb, Bhardwaj:2024djv}, or Energy Flow Polynomials~\cite{Komiske:2017aww}, which explicitly construct inter-particle correlations of the form $h(x_i, x_j, x_k, ...)$.
As a final point of contrast, for an event with $N$ particles, a graph-based approach with $m$ edges has to consider $\mathcal{O}(N^m)$ terms, while the Moment EFN still only has $N$ terms in its sum for each moment.

The above discussion has focused on extending EFNs to have multiple moments.
However, it is straightforward to drop IRC-safety and generalize to the Particle Flow Network (PFN)~\cite{efn}, or indeed any realization of Deep Sets architectures.\footnote{A note on nomenclature: A PFN is identical to an ordinary Deep Sets network. We use the term ``PFN'' to refer to this architecture in the particle physics context, and ``Deep Sets'' to refer generically to sets-based architectures. }
This can be accomplished by simply modifying the definition of the expectation value \Eq{expectation_value} to remove the energy weighting:
\begin{align}
    \expval{\Phi}_\cP^{\text{PFN}} = \sum_{i \in \cP} \Phi(p_i), \label{eq:PFN}
\end{align}
where $\Phi$ is a function of the per-particle information $p$, which can include the particle's energy, momentum, charge, flavor, and other information.
Here, $\cP$ can no longer be regarded as the distribution of energy over the detector space, but rather just as an abstract unnormalized distribution of particle information.
Example studies involving Moment PFNs rather than Moment EFNs may be found in \App{additional}.

Finally, some notes about our conventions and notation for the rest of the paper: 
First, we will use the terms ``energy'' and transverse momentum ``$p_T$'' interchangeably, as nothing we say here depends on this distinction -- our studies here focus on the Large Hadron Collider (LHC), where it is typical to speak of transverse momenta rather than energies.
Second, although detectors are often cylindrical or spherical and these models can be extended to accommodate this, we will only consider local rectangular patches $\sim\mathbb{R}^2$ in the rapidity-azimuth plane.
Third, we will always implicitly include the $k = 0$ moment, $\expval{1}_\cP$, which is the total energy of the event.
For normalized events, this contains no information, but we find it convenient to include.
When we speak of a $k = 1$ Moment EFN, or equivalently an ``ordinary'' EFN, we are still including the $k = 0$ moment, which differs from the conventions of \Reference{efn} slightly.
Practically speaking, this makes no numeric difference.
Finally, we will occasionally find it convenient to speak of $\Phi^a$ not as a single $L$-dimensional function of $x$, but as $L$ separate 1-dimensional functions $\Phi$, and suppress the $a$ indices.

\subsection{The Effective Latent Dimension}\label{sec:effective_latent_dim}

Given that, by the Deep Sets theorem~\cite{deepsets}, EFNs are already capable of approximating any IRC-safe observable arbitrarily well, why should we bother making them more complicated by adding moments?
A Moment EFN is able to approximate the same observable with a much smaller learned latent space dimension than an ordinary EFN, by taking advantage of its large \emph{effective latent dimension}.
The effective latent dimension of an order $k$ Moment EFN with $L$ latent dimensions is the total number of distinct inputs to the function $F$, and is given by:
\begin{align}
    L_{\mathrm{eff}} = {{L+k} \choose {k}},
    \label{eq:effective_latent_dim}
\end{align}
which asymptotically goes as $L_{\mathrm{eff}} \sim \frac{L^k}{k!}$ for large $L$.

An order $k$ Moment EFN with $L$ different $\Phi$ functions (indexed by $a$) acts like an ordinary EFN with $L_{\mathrm{eff}}$ different $\Phi'$ functions (indexed by $A$), in the sense that if we identify:
\begin{align}
    \Phi'^0(x) &= 1 \nonumber \\
    \Phi'^A(x) &= \Phi^a \text{ for } A = 1,...,L \nonumber\\
        \Phi'^A(x) &= \Phi^{a_1}\Phi^{a_2} \text{ for } A = L+1,...,\frac{L^2 + 3L}{2} \nonumber\\
        & ... \nonumber\\
    \Phi'^A(x) &= \prod_{i=1}^k\Phi^{a_i} \text{ for } A = {L + k -1 \choose k -1},...,{L + k \choose k}-1 ,\label{eq:moment_efn_identification}
\end{align}
and assign both models exactly the same $F$ network, then the two models are completely identical.
The Moment EFN only needs $L$ different learnable functions $\Phi$ to express this, while the ordinary EFN needs $L_{\mathrm{eff}}$ different learnable functions $\Phi'$, with $L \ll L_{\mathrm{eff}}$.
We can think of a Moment EFN as effectively ``compressing'' or ``encoding'' $L_{\mathrm{eff}}$ pieces of information into just $L$ functions, with \Eq{moment_efn_identification} being the ``decoder''.

\Eq{moment_efn_identification} highlights another distinguishing feature of the Moment EFN: explicit nonlinear products. 
Typically, $\Phi$ is parametrized as a dense neural network, which consists of affine transformations interleaved with nonlinear activation functions. 
It is difficult for these models to approximate highly nonlinear functions -- it is a lot easier for a dense neural network to learn $\Phi(x) = x$ than to learn $\Phi(x) = x^k$.
Moment EFNs involve an explicit product of $\Phi$ functions, which directly enable functions of the type $x^k$ to be easily represented. 
See \App{jet_angularities} for a concrete example of how this multiplication structure can aid in learning jet angularities, which are observables involving nonlinear powers of particle coordinates.

For a fixed $L$, a Moment EFN with greater $k$ is always at least as expressive as Moment EFN with a smaller $k$, and therefore should perform at least as well.
This is because $F$ can be chosen to simply ignore the extra effective latent dimensions.
Practically speaking, increasing $k$ with $L$ fixed makes the models slightly more complex to train, as there are more parameters to optimize over, so this monotonicity in performance may be imperfect in practice. 
To mitigate this, we employ a ``pre-training'' procedure on our models so that extra effective latent dimensions do not damage performance -- see \App{model_specifications} for details.

On the other hand, if we fix $L_{\mathrm{eff}}$ and the function $F$, there is a small loss of expressivity when using moments, and therefore not all observables can have their representations efficiently compressed this way.
A Moment EFN with $L$ latent dimensions is not quite as expressive as the equivalent ordinary EFN with $L_{\mathrm{eff}}$ latent dimensions, so the identification in \Eq{moment_efn_identification} is not always possible. 
This is because the latent dimensions of an ordinary EFN are completely uncorrelated, whereas the moments of a random variable $\Phi$ are correlated -- for example, it is always the case that $\expval{\Phi^2} > \expval{\Phi}^2$. 
As an explicit counter-example, suppose we were interested in constructing the energy-weighted average position of a jet in detector space: the observable $\cO(\cP) \equiv (y_{\text{avg}}, \phi_{\text{avg}})$.
This is easy to accomplish with an $L = 2$ ordinary EFN, with $\Phi'(x_i) = (y_i, \phi_i)$ and trivial $F$.
However, this is difficult to accomplish with the equivalent order $k=2$ Moment EFN with $L = 1$, even if we allowed $F$ to vary, because $y_{\text{avg}}$ and $\phi_{\text{avg}}$ are independent, whereas any possible $\Phi(x)$ we construct would have $\expval{\Phi}$ and $\expval{\Phi^2}$ correlated.\footnote{This is technically possible if $\Phi$ is a space-filling curve, but not only this this discontinuous and therefore not IRC-safe, it would be incredibly difficult to learn.}  

To summarize: While one should always expect a higher order Moment EFN with $L$ latent dimensions to more accurately approximate an observable than a lower order Moment EFN with $L$ latent dimensions, it is not guaranteed that a higher order Moment EFN with $L_{\mathrm{eff}}$ effective latent dimensions will outperform a lower order EFN with $L_{\mathrm{eff}}$ effective latent dimensions.
When this does happen, this is a statement that the observable being estimated has some simpler structure, which we will see is the case for IRC-safe quark/gluon jet discrimination in \Sec{qg_discrimination}.
Interestingly, this is \emph{not} the case for top/QCD jet discrimination -- see \App{additional} for a concrete example.

As a brief aside, the explicit product structure of Moment EFNs is reminiscent of the self-attention mechanism~\cite{bahdanau2016neural, cheng2016long} in transformer models~\cite{vaswani2023attention, Qu:2022mxj}.
Schematically, in the $k = 2$ product $\Phi^{a_1}(x)\Phi^{a_2}(x)$ we can think of $\Phi^{a_1}(x)$ as telling the network how much to ``pay attention'' to $\Phi^{a_2}(x)$ (and vice-versa). 
Similarly, the self-attention mechanism in transformers is of the schematic form:%
\begin{align}
    \text{softmax}(Q(x)K(x)) \cdot V(x),
\end{align}
which, ignoring the softmax (primarily used to interpret the result as a weight), is a cubic product of the form $\Phi^{a_1}(x)\Phi^{a_2}(x)\Phi^{a_3}(x)$.


\section{Case Study: Quark/Gluon Discrimination}\label{sec:qg_discrimination}

We now apply the Moment EFN to the task of quark/gluon jet tagging~\cite{Gallicchio:2011xq,Gras:2017jty}, to show how its performance varies with different choices of $L$ and $k$ compared to the ordinary EFN.
Additional details about the model specifications and training procedures can be found in \App{model_specifications}, and similar studies using a Moment PFN instead of a Moment EFN and for discriminating top-initiated jets from QCD jets can be found in \App{additional}.

\subsection{Dataset}\label{sec:qg_dataset}

\begin{figure*}[tbhp]
    \centering
    \subfloat[]{
        \includegraphics[width = 0.45\textwidth]{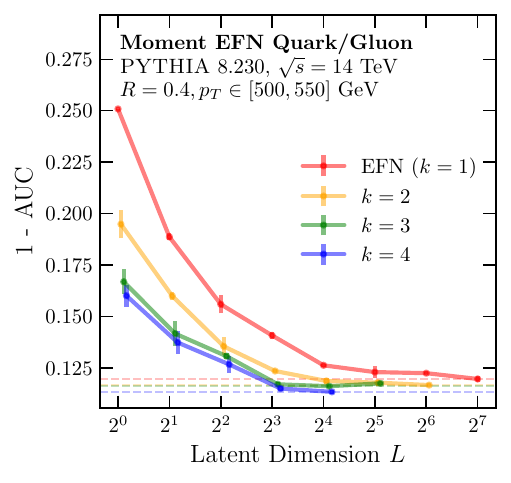}    \label{fig:latent_dim}
    }
    \subfloat[]{
    \includegraphics[width = 0.45\textwidth]{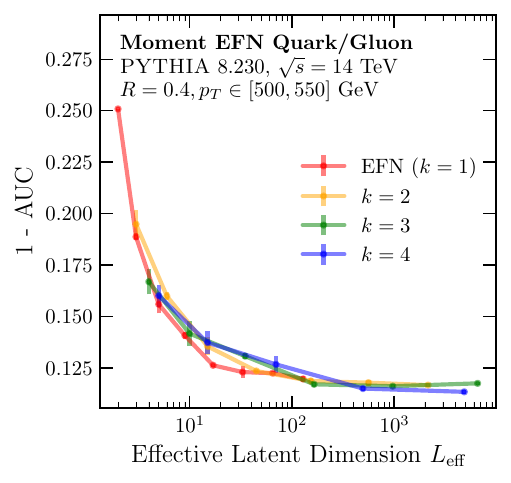}    \label{fig:effective_latent_dim}
    }
    \caption{The performance (AUC) on quark/gluon discrimination of the $k$ Moment EFN as a function of the (a) latent dimension $L$ and (b) effective latent dimension $L_{\mathrm{eff}}$ for different values of $k$. The thin horizontal dashed lines indicate the best value of the AUC achieved. For each model, the standard deviation and mean of the AUC across 3 trainings is shown.}
\end{figure*}

We use the same quark and gluon jet dataset as described in \Reference{efn}.
This dataset consists of $Z$ plus jet events at $\sqrt{s} = 14$ TeV generated using \Pythia 8.226~\cite{Sjostrand:2006za, Sjostrand:2014zea} with multiple parton interactions turned on.
The $Z$ is forced to decay invisibly to neutrinos, and the remaining particles at then clustered into $R = 0.4$ anti-$k_T$~\cite{Cacciari:2008gp} (AK4) jets using \FastJet 3.3.0~\cite{Cacciari:2011ma}.
Only jets with transverse momentum $p_T \in [500, 550]$ GeV and rapidity $|y| < 2$ are kept. 
Each jet is then labeled as a quark or gluon depending on the underlying hard process that generated it, with quarks generated using the \texttt{WeakBosonAndParton:qg2gmZq} process and gluons generated from the \texttt{WeakBosonAndParton:qqbar2gmZg} process.\footnote{These quark and gluon labels are technically unphysical, and there exist more physical operational definitions of the quark and gluon content of a jet~\cite{Metodiev:2018ftz, Komiske:2018vkc}, but this is largely unimportant for our study here.}
No detector simulation is applied.
Each jet is then prepossessed, such that the sum of the particle $p_T$'s is normalized to 1 and the $p_T$-weighted average position of the jet is $(0, 0)$ in the rapidity-azimuth plane.
In the following studies, we use 1M total jets to train, and 50k jets each for validation and testing.

\subsection{Performance}

\begin{table*}[t]
    \centering
    \begin{tabular}{l|c|c|c|c}
         Model & AUC & $1/\epsilon_g$ at $\epsilon_q = 0.3$ & $1/\epsilon_g$ at $\epsilon_q = 0.5$ & Trainable Parameters \\
         \hline
         \hline
$k = 1, L = 1$  EFN& 0.743 $\pm$ 0.001 & 54.2 $\pm$ 1.7 & 12.0 $\pm$ 0.1 & 31106 \\
$k = 2, L = 1$ Moment EFN& 0.802 $\pm$ 0.002 & 53.9 $\pm$ 3.5 & 16.7 $\pm$ 0.2 & 31206 \\
$k = 3, L = 1$ Moment EFN& 0.831 $\pm$ 0.000 & 52.2 $\pm$ 2.6 & 18.5 $\pm$ 0.2 & 31306 \\
$k = 4, L = 1$ Moment EFN& \textbf{0.841 $\pm$ 0.004} & \textbf{61.6 $\pm$ 2.8} & \textbf{21.3 $\pm$ 0.1} & 31406 \\
\hline
$k = 1, L = 4$  EFN& 0.843 $\pm$ 0.004 & 64.0 $\pm$ 2.1 & 24.5 $\pm$ 1.0 & 31745 \\
\hline
\hline
$k = 1, L = 128$  EFN& 0.879 $\pm$ 0.001 & 69.2 $\pm$ 1.8 & 31.4 $\pm$ 0.8 & 89653 \\
$k = 2, L = 64$ Moment EFN& 0.886 $\pm$ 0.001 & \textbf{83.0 $\pm$ 1.3} & 30.7 $\pm$ 0.6 & 260085 \\
$k = 3, L = 32$ Moment EFN& 0.886 $\pm$ 0.001 & 72.6 $\pm$ 2.1 & \textbf{33.6 $\pm$ 1.0} & 690645 \\
$k = 4, L = 16$ Moment EFN& \textbf{0.887 $\pm$ 0.001} & 81.6 $\pm$ 3.4 & 32.7 $\pm$ 0.4 & 517461 \\
\hline
\hline
    \end{tabular}
    \caption{The AUC and gluon rejection factor at quark efficiencies of 0.3 and 0.5 for the $k = 1, 2, 3$, and $4$ Moment EFNs trained in \Sec{qg_discrimination}. Here, we show results for the $L = 1$ networks, alongside the $k = 1, L = 4$ ordinary EFN, which achieves comparable results to the $k = 4, L = 1$ Moment EFN. We also show the highest latent dimension $L$ considered at each order ($L = 128, 64, 32$, and $16$ respectively). For each metric, the model with the best performance is bolded.}
    \label{tab:auc}
\end{table*}

For orders $k = 1$ through $4$, we train Moment EFNs for a wide range of latent dimensions $L$ from $1$ to $128$ in powers of 2.
Due to memory and training time considerations, we consider only up to $L = 2^{8-k}$, since otherwise $L_{\mathrm{eff}}$ becomes prohibitively large.
For each model, we report its performance using the ``area under curve'' (AUC)\footnote{More precisely, if $P_i(x)$ is the cumulative distribution function for the model output $x$ assuming the distribution $i$ (either $q$ or $g$ in this case), then the AUC is defined to be $1 - \int_0^1 d\lambda \, P_g(P_q^{-1}(\lambda))$. An AUC of 0.5 indicates random guessing, and an AUC of 1.0 is a perfect classifier.} metric across three retrainings.
We also report the gluon rejection factor at quark efficiencies of $0.3$ and $0.5$ for the largest $L$ and smallest $L$ models in \Tab{auc} for ease of comparison with other quark/gluon discrimination studies~\cite{Komiske:2016rsd,ATL-PHYS-PUB-2017-017,Cheng:2017rdo,Luo:2017ncs,Kasieczka:2018lwf,Komiske:2018cqr,Lee:2019cad,Lee:2019ssx,Moreno:2019bmu,Qu:2019gqs,Mikuni:2020wpr,Dreyer:2020brq,Bogatskiy:2022czk,Qu:2022mxj,He:2023cfc, Dolan:2023abg}.

Each Moment EFN model is trained as an ordinary classifier to minimize the binary cross entropy between the quark and gluon classes. 
Both classes appear in the dataset with $50\%$ probability. 
The specific details of the models and training procedure may be found in \App{model_specifications}.

The resulting model performances on the quark/gluon discrimination task, as a function of $L$ and $L_{\mathrm{eff}}$, are shown in \Figs{latent_dim}{effective_latent_dim} respectively.
From these plots we can make four key observations:
\begin{enumerate}
    \item \textbf{At fixed $L$, AUC improves with $k$:} As expected, increasing the order of the Moment EFN improves its performance for fixed $L$, since the ansatz is more expressive. This effect is particularly pronounced near $L = 1$, with the AUC improving from 0.75 to 0.84 from $k = 1$ to $k = 4$.
    \item \textbf{Higher $k$ saturates faster:} As $k$ increases, the value of $L$ required to saturate performance drops. The ordinary EFN saturates around $L = 128$, whereas the order $k = 4$ saturates around $L = 16$. To achieve peak performance, you don't need as high an $L$ with a Moment EFN.
    \item \textbf{Peak AUC improves with $k$:} The highest AUC achieved by these models improves slightly with $k$, as indicated by the dashed lines in \Fig{latent_dim}. In \Fig{effective_latent_dim}, we can see that this is primarily driven by the extremely high effective latent dimensions reached by higher $k$ Moment EFNs.
    \item \textbf{$L_{\mathrm{eff}}$ drives performance:} The AUC correlates very strongly with $L_{\mathrm{eff}}$, regardless of the order $k$. This suggests that ``encoding'' and ``decoding'', as per \Eq{moment_efn_identification}, is occurring, and that the quark/gluon discriminant is ``compressible'' into fewer elementary functions. In particular, if an ordinary EFN requires $L_{\mathrm{eff}}$ latent dimensions to achieve a desired performance in quark/gluon discrimination, an order $k$ Moment EFN would only require $L \sim k! \,L_{\mathrm{eff}}^{1/k}$ latent dimensions to achieve the same performance.
    
\end{enumerate}

All of these observations point to Moment EFNs being able to achieve the same (or better) performance as ordinary EFNs but with a significantly smaller learned latent space dimension.
The quark/gluon discriminator can be efficiently compressed, with the peak classifier going from being composed of $\sim 128$ functions to only $\sim16$ functions while gaining a slight performance bump in the process.
It is also especially remarkable that an order $k = 4$ Moment EFN is able to achieve an AUC of 0.84 with just a \emph{single} latent dimension, equivalent to an ordinary EFN with 4 latent dimensions, and only a few points away from the best possible EFN score of 0.88.
We have checked for all studies shown here that going to $k = 5$ and beyond does not offer any significant improvement over $k = 4$.

Note that these 4 observations are \emph{not} generically true across different classification tasks: As shown in \App{additional}, the improvement in Observation 1 is not always perfectly monotonic in $k$, especially if the models fail to converge, and Observation 4 especially is not true for top/QCD jet discrimination -- as noted in \Sec{effective_latent_dim}, a higher $k$ Moment EFN may be less expressive than a lower $k$ Moment EFN with the same $L_{\mathrm{eff}}$, causing performance to potentially worsen with $k$ for fixed $L_{\mathrm{eff}}$.

\section{Opening the Black Box}\label{sec:black_box}

One practical advantage of the smaller latent dimension afforded by Moment EFNs is that lower dimensional spaces are easier to visualize and interpret.
An $L = 4$ ordinary EFN involves the complex interplay of 4 independent functions on detector space, whereas the equivalent order $k = 4$, $L = 1$ Moment EFN achieving the same performance only has a single function to look at (and moreover, we will see that this single function is radially symmetric!). 
For small $L$, we can even obtain closed form expressions for the latent spaces of Moment EFNs, and due to the effective latent space, we can use this to extend our understanding of ordinary EFNs for larger values of $L$ than we could have otherwise.

The rest of this section proceeds as follows: In \Sec{log_angularities}, we take the order $k = 4$, $L = 1$ models trained in \Sec{qg_discrimination}, visualize their internal representations, and find a closed-form expression for their latent spaces, resulting in observables we call ``log angularities''.
In \Sec{F_networks}, we show to what extent the $F$ network can also be cast into closed form.
Finally, in \Sec{larger_L}, we briefly discuss $L \geq 2$ models.

\subsection{$L = 1$ and Log Angularities}\label{sec:log_angularities}

When $L = 1$, it is feasible to study exactly what the model learned and extract a single closed-form, one-dimensional observable representing the entire latent space~\cite{Datta:2017lxt, efn}.
The order $k = 4$, $L = 1$ Moment EFN is able to achieve an AUC of 0.84 using a single learned representation, and our goal is to understand and extract this representation.
Because of the effective latent space, interpreting the latent space of the order $k = 4$, $L = 1$ Moment EFN is equivalent to interpreting \emph{all four} of the latent dimensions of an $L = 4$ ordinary EFN.
This study is modeled after the study performed in \Reference{efn}, which constructs \emph{two} independent observable using an $L = 2$ ordinary EFN and achieves an AUC $\sim 0.80$.

\begin{figure*}[p]
    \centering
    \subfloat[]{
         \includegraphics[width=0.45\textwidth]{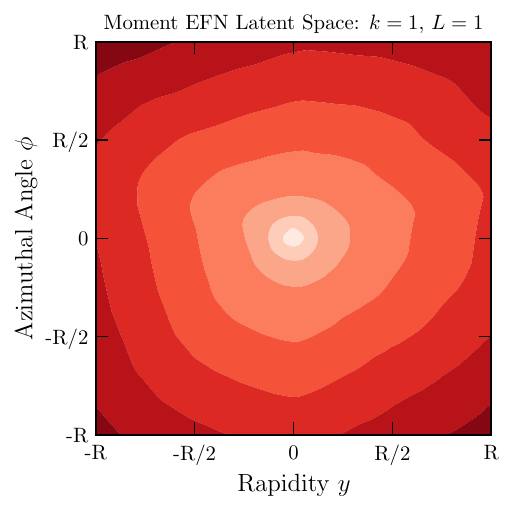}
        \label{fig:k1_latent_space_L1}
    }
    \subfloat[]{
        \includegraphics[width=0.45\textwidth]{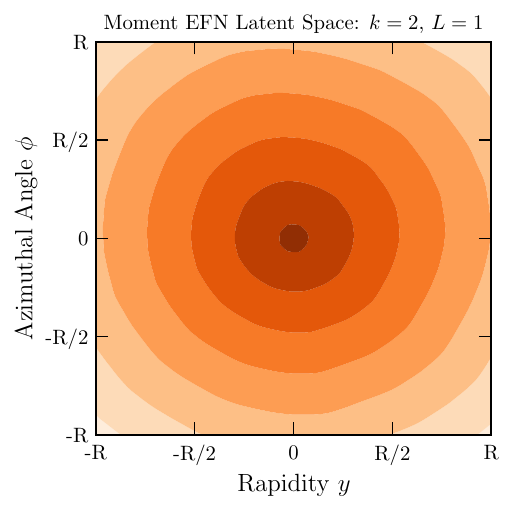}
        \label{fig:k2_latent_space_L1}
    }
    \vspace{0pt}
    \subfloat[]{
         \includegraphics[width=0.45\textwidth]{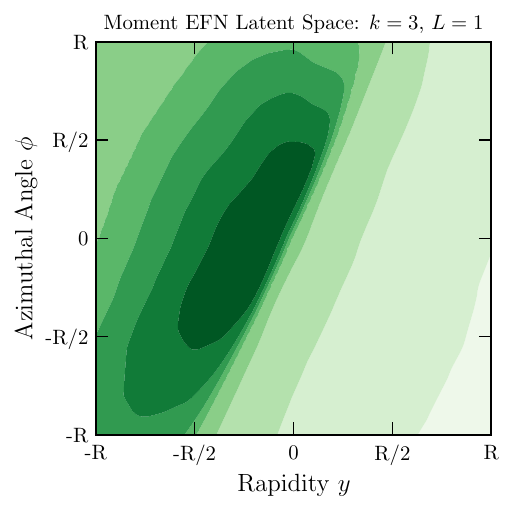}
        \label{fig:k3_latent_space_L1}
    }
    \subfloat[]{
        \includegraphics[width=0.45\textwidth]{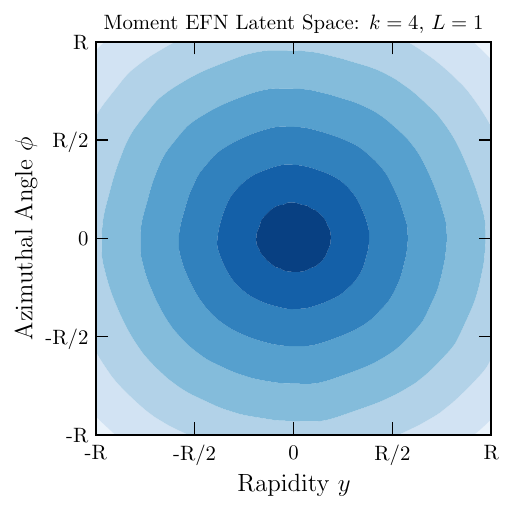}
        \label{fig:k4_latent_space_L1}
    }
    \caption{
         Examples of learned Moment EFN latent space embeddings $\Phi(x)$, for (a) $k = 1$, (b) $k = 2$, (c) $k = 3$, and (d) $k = 4$. 
         Each figure represents the best model of the $L = 1$ trainings from \Sec{qg_discrimination}. 
         The overall normalization is arbitrary.
         The $k=3$ example features mirror rather than radial symmetry, which can generically occur for $k =2$, $3$, and $4$.
        }
    \label{fig:latent_spaces_L1}
\end{figure*}

Since $\Phi(x)$ is a function of the rapidity-azimuth plane, we can directly plot the latent spaces of the best $L = 1$ networks in 2D following the procedure outlined in \Reference{efn}, where it is possible to visualize the entire latent space at once. 
We show examples of this in \Fig{latent_spaces_L1} for $k = 1$ through $4$.
We first note that while the $k = 1, 2$ and $4$ networks learn a radially symmetric latent space, the $k = 3$ network does not -- it instead exhibits an approximate mirror symmetry.
This is not a feature unique to $k = 3$, as this mirror symmetric latent space occasionally occurs for $k = 2$ and $k = 4$ networks as well in some retrainings, though seemingly without loss in performance. 
Since QCD jets are approximately radially symmetric~\cite{Ellis:1992qq, CDF:1992cus}, the fact that this ``symmetry breaking'' doesn't affect performance is not surprising, since only radial information is necessary for classification. 
The precise mechanism that causes this to occur likely is sensitive to the training dynamics of the models.

\begin{figure}[tbp!]
    \centering
    \includegraphics[width = 0.95\linewidth]{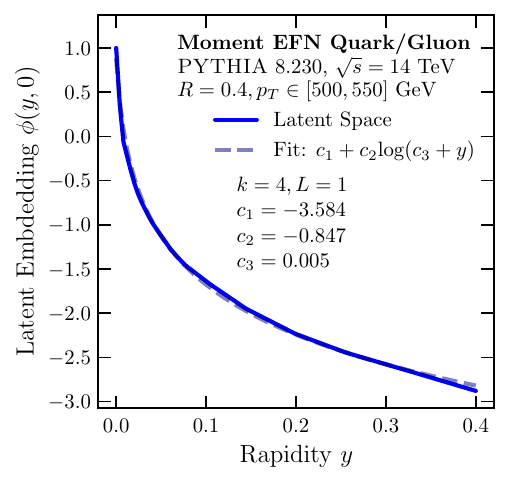}
    \caption{A radial slice of the latent space for the best performing $k = 4, L = 1$ Moment EFN, as a function of rapidity ($y$). The radial slice is taken at an azimuthal angle of zero. The latent space is shown as a dark blue line, and the logarithmic fit from \Eq{fit} is shown as a blue dashed line. }
    \label{fig:latent_fit}
\end{figure}

We now focus our attention to the highest performing models: the order $k = 4$ Moment EFNs.
In \Fig{latent_fit}, we show a radial slice of the $k = 4, L = 1$ latent space (shown fully in \Fig{k4_latent_space_L1}).
The radial slice is taken at an azimuthal angle of 0 as a function of the rapidity $y$, though this choice is arbitrary.
Even in cases where ``symmetry breaking'' occurs in the latent space, we find that the radial profile is largely the same up to normalization on any projection not along the mirror symmetry axis.
Motivated by the form of the radial profile in \Fig{latent_fit}, we fit the function:
\begin{align}
    \Phi_{\cL}(r) = c_1 + c_2 \log(c_3 + r), \label{eq:fit}
\end{align}
where $r$ is the radial distance in the rapdity-azimuth plane from $(0, 0)$, as defined by the energy-weighted average position of the jet.

This function provides an excellent fit to the latent function with $c_1 = -3.584, c_2 = -0.847$, and $c_3 = 0.005$.
The values of $c_1$ and $c_2$ are largely unimportant, since they will be subject to affine transformations within the first layer of the $F$ network, and these parameters vary significantly across retrainings.
On the other hand, $c_3$ is consistently a small number in the range of 0.002 to 0.01, and is embedded in a logarithm which is more nontrivial for $F$ to unravel.

The function $\Phi_{\cL}$ has a divergence as $r \to 0$ (i.e. as particles become collinear with the jet center), but this divergence is regulated by the $c_3$ parameter. 
Interestingly, $c_3$ is within an $\mathcal{O}(1)$ factor of $\frac{\Lambda_{QCD}}{p_T R} \sim 0.001$, suggesting that the nonzero value of $c_3$ is due to genuine nonperturbative physics near the jet core learned by the Moment EFN.

The moments of the function $\Phi_\mathcal{L}$ can be used to construct jet shape observables of the form:
\begin{align}
    \cL^{(n)}(\cP) &= \expval{\Phi_{\cL}^n}_\cP, \nonumber\\
    &= \sum_{i \in \cP}z_i\left(c_1 + c_2\log(c_3 + r_i)\right)^n.\label{eq:analytic_observables}
\end{align}
We call the observables $\cL^{(n)}$ \emph{log angularities}, since they resemble ordinary jet angularities $\lambda^{\beta}(\cP) = \sum_{i\in \cP} z_i r_i^\beta$ for $c_1 = c_2 = 0$.
It is possible to generically set $c_1 = c_2 = 0$ by taking linear combinations of $\cL^{(n)}$ for different $n$, but we elect to keep these parameters as it reduces the amount of total linear transformations our 3 hidden-layer dense networks have to do.
These log angularities are interesting observables in their own right, especially in the $c_3 \to 0$ limit and are closely related to the $\beta \to 0$ limit of ordinary angularities~\cite{Larkoski:2013eya, Bright-Thonney:2023gdl}, though we save a more in-depth theoretical discussion of log angularities for future work and here focus on their use as quark/gluon taggers.

We can use these analytic observables as inputs to a simple dense neural network classifier of the form:
\begin{align}
    F^{(k)}(\cL^{(1)}, ..., \cL^{(k)}). \label{eq:dense_neural_net}
\end{align}
If the $k = 4$ dense neural net classifier has the same performance as the full order $k = 4$ Moment EFN, then we can claim not only to have found a fully analytic form of the $k = 4$ latent space, but equivalently to have found a fully analytic form of the four different $L = 4$ ordinary EFN latent space dimensions.

\begin{table*}[t]
    \centering
    \begin{tabular}{l|c|c|c|c}
         Model & AUC & $1/\epsilon_g$ at $\epsilon_q = 0.3$ & $1/\epsilon_g$ at $\epsilon_q = 0.5$ & Trainable Parameters \\
         \hline
         \hline
$k = 1$ Log Angularity DNN& 0.730 $\pm$ 0.001 & 50.5 $\pm$ 1.2 & 8.5 $\pm$ 0.9 & $3 + 20702$ \\
$k = 2$ Log Angularity DNN& 0.784 $\pm$ 0.001 & \textbf{72.0 $\pm$ 1.7} & 13.5 $\pm$ 0.8 & $3 + 20802$ \\
$k = 3$ Log Angularity DNN& 0.816 $\pm$ 0.001 & 59.9 $\pm$ 1.7 & \textbf{19.4 $\pm$ 1.1} & $3 + 20902$ \\
$k = 4$, Log Angularity DNN& \textbf{0.821 $\pm$ 0.002} & 55.6 $\pm$ 2.0 & 18.5 $\pm$ 1.2 & $3 + 21002$ \\
\hline
$k = 4$, $c_3\to0$ DNN& 0.799 $\pm$ 0.001 & 60.7 $\pm$ 1.8 & 15.5 $\pm$ 1.0 & $2 + 21002$ \\
\hline
\hline
    \end{tabular}
    \caption{The same as \Tab{auc}, but with dense neural networks on log angularities as defined in \Eq{dense_neural_net}. The ``$3 +$'' and ``$2+$'' in the trainable parameters column refer to the parameters in the log angularity fit.}
    \label{tab:auc_dense}
\end{table*}

\begin{figure}[t]
    \centering
    \includegraphics[width = 0.95\linewidth]{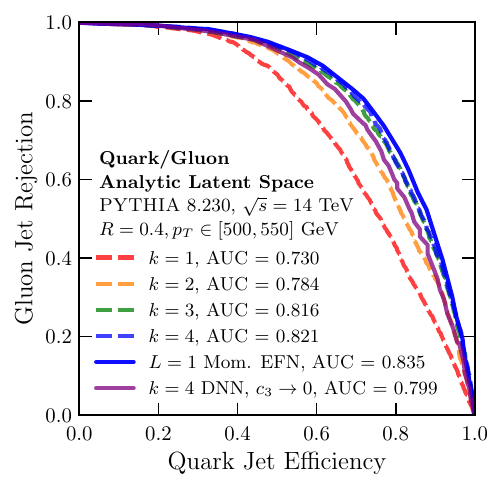}
    \caption{ROC curves showing the performance of the analytic jet shape observables, as defined in \Eq{analytic_observables}, as a quark/gluon classifier. The jet shapes are passed into a dense neural net $F$ as in \Eq{dense_neural_net}. The ROC of the original $k = 4, L = 1$ Moment EFN, from which the fits were derived, is shown in dark blue. Also shown in purple is a variant of the $k = 4$ DNN, except where the $c_3$ parameter is taken to zero.}
    \label{fig:analytic_auc}
\end{figure}

In \Fig{analytic_auc}, we show ROC curves\footnote{The Receiver Operating Characteristic (ROC) curve of a classifier quantifies the background rejection rate as a function of the signal acceptance rate, and is given by $\text{ROC}(\lambda) = 1-P_g(P_q^{-1}(\lambda))$.} of the classifier defined by \Eq{dense_neural_net} for $k = 1$ through $4$.
The $F$ networks used here have precisely the same architecture and training procedure as those used for the Moment EFNs in \Sec{qg_discrimination}, described in \App{model_specifications}.
These results are also summarized in \Tab{auc_dense}.
We also show, in purple, the $k = 4$ DNN classifier taking $c_3$ to 0.
From this plot, we can make several observations:
\begin{enumerate}
    \item \textbf{The $k = 4$ dense model is as good as the $k = 4$ Moment EFN:} We can replace the neural network latent dimension $\Phi(x)$ with the much simpler $\Phi_{\cL}(r)$ when $L = 1$. Moreover, since the $k = 4$, $L = 1$ Moment EFN is just as good as the $L = 4$ ordinary EFN, the single function $\Phi_{\cL}(r)$ and its powers captures the same information as 4 dimensions worth of latent space in an ordinary EFN.
    \item \textbf{The $k = 1, 2$, and $3$ dense models are not as good as their corresponding order $k$ Moment EFNs:} The AUCs of the dense models are slightly lower than the corresponding $L = 1$ Moment EFNs in \Tab{auc} for $k < 4$. This suggests that while the \emph{combination} of $\cL^{(1)}, \cL^{(2)}, \cL^{(3)}, \cL^{(4)}$ are optimal, individually they are not, and $\cL^{(1)}$ by itself is not the most optimal single-variable observable for IRC-safe quark/gluon discrimination.
    \item \textbf{The $c_3$ parameter matters:} Taking the parameter $c_3$ to zero reduces the AUC of the $k = 4$ models significantly. Interpreting $c_3$ as a nonperturbative parameter regulating a collinear divergence in the logarithm, this loss in performance can be viewed as the learned effect of nonperturbative physics in quark/gluon discrimination.
\end{enumerate}

Thus, at least for $k = 4$, we have successfully cast the latent space of not only the Moment EFN, but the equivalent $L = 4$ ordinary EFN, into closed form.

\subsection{$L = 1$ $F$ Networks}
\label{sec:F_networks}

Next, we attempt to go further by attempting to also find closed-form expressions for the dense neural network classifiers $F$, building off of the analysis performed in \Sec{log_angularities} where we found closed-form expressions for the latent space network $\Phi$.
This would result in a fully closed-form quark/gluon jet classifier.
However, analyzing $F$ is inherently more difficult than the individual $\Phi$ functions, as for the $k = 4$, $L = 1$ model of interest, $F$ is a function of 4 inputs.
Moreover, we know $F$ must be nontrivial in all 4 inputs, since otherwise there would be no difference between the $L = 1$ networks at different values of $k$.

\begin{figure*}[p]
    \centering
    \subfloat[]{
        \includegraphics[width = \columnwidth]{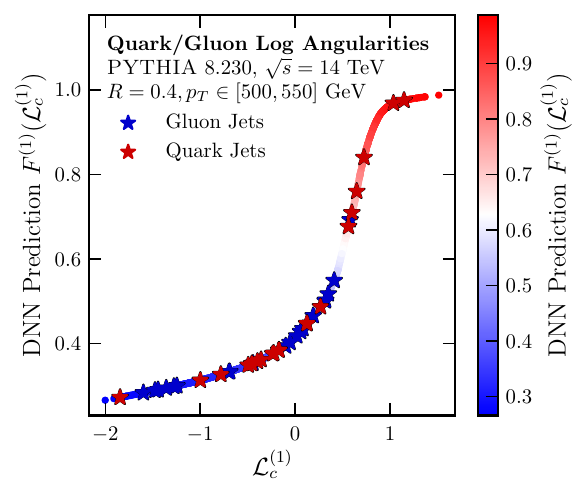}
        \label{fig:log_angularities_f1}
    }
    \subfloat[]{
    \includegraphics[width = \columnwidth]{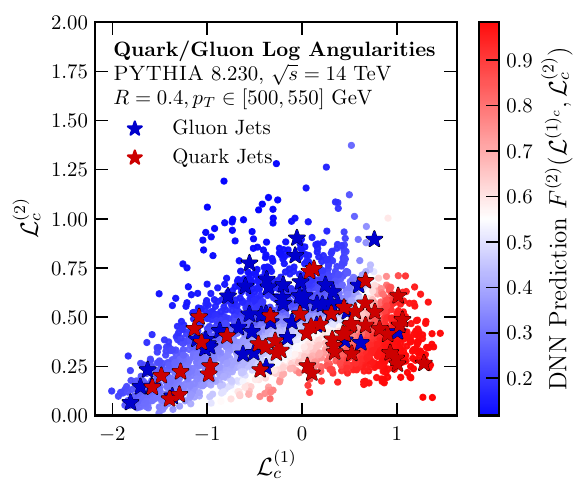}
        \label{fig:log_angularities_f2}
    }
    \vspace{0pt}
    \subfloat[]{
         \includegraphics[width = 1.25\columnwidth]{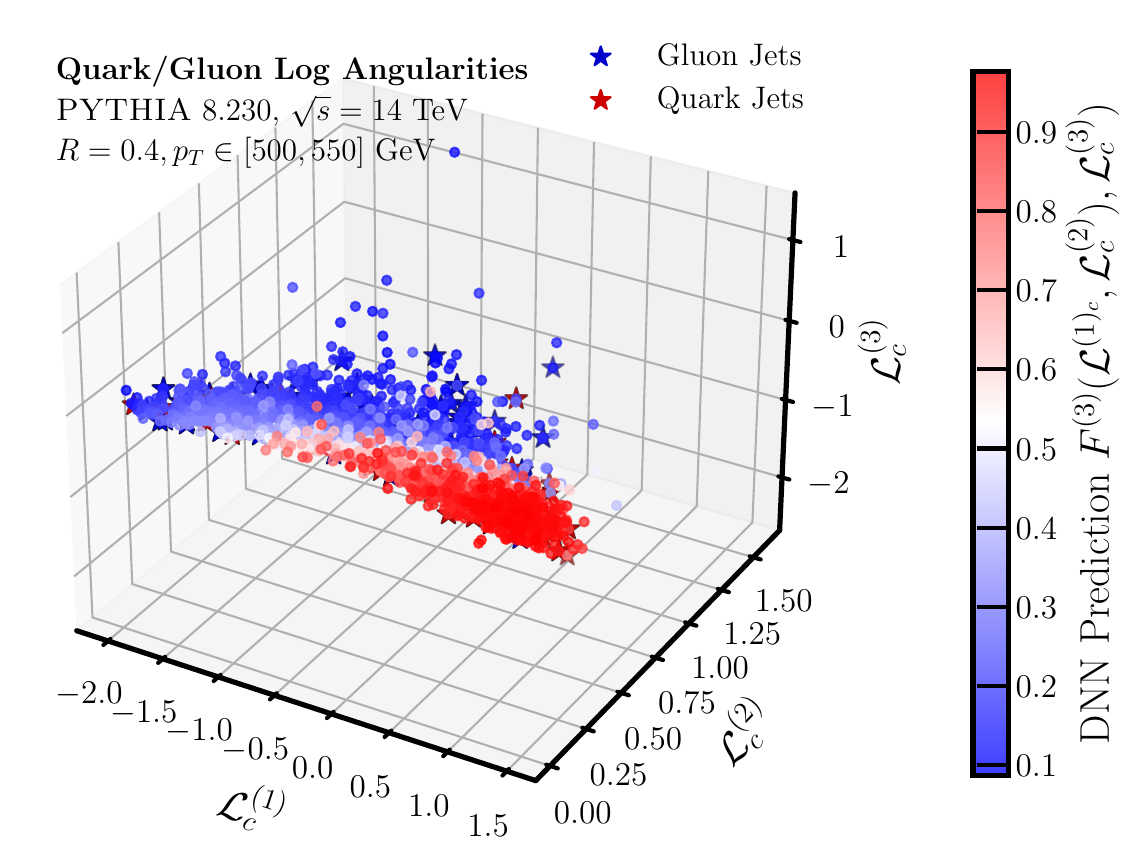}
        \label{fig:log_angularities_f3}
    }
    \caption{
        The distribution of the dense neural network output (a) $F^{(1)}$, (b) $F^{(2)}$, and (c) $F^{(3)}$ as a function of the first, first two, and first three (cumulant) log angularities respectively. The true quark/gluon label for several random jets are indicated with colored stars.
            }
    \label{fig:log_angularities_f}
\end{figure*}

To aid in determining a functional form of $F$, we begin by plotting the output of the dense neural networks (\Eq{dense_neural_net}) considered in \Sec{log_angularities} as a function of $\cL^{(k)}$, for $k = 1,\ 2$, and 3.
To be explicit, we plot $F^{(1)}(\cL^{(1)})$ in \Fig{log_angularities_f1}, plot $F^{(2)}(\cL^{(1)}, \cL^{(2)})$ in \Fig{log_angularities_f2}, and plot $F^{(3)}(\cL^{(1)}, \cL^{(2)}, \cL^{(3)})$ in \Fig{log_angularities_f3}.\footnote{Unfortunately, we are unable to display the full plot of $F^{(4)}(\cL^{(1)}, \cL^{(2)}, \cL^{(3)}, \cL^{(4)})$, since the PDF file format is not yet available in more than two dimensions.}
We will find it convenient to work with ``cumulant'' log angularities $\cL_c$, rather than ordinary log angularities, as this makes the distributions in \Fig{log_angularities_f} and the resulting fits simpler. 
The cumulant log angularities are defined as:
\begin{align}
    \cL_c^{(1)} &= \cL^{(1)}, \\
    \cL_c^{(2)} &= \cL^{(2)} - [\cL^{(1)}]^2, \\
    \cL_c^{(3)} &= \cL^{(3)} - 3\cL^{(1)}\cL^{(2)} + 2[\cL^{(1)}]^3, \\
    \cL_c^{(4)} &= \cL^{(4)} - 4\cL^{(1)}\cL^{(3)} - 3[\cL^{(2)}]^2  \nonumber\\
    &\quad+ 12\cL^{(2)}[\cL^{(1)}]^2 -6[\cL^{(1)}]^4 .
\end{align}

In all three plots, we see that the DNN output is, for the most part, cleanly divided into distinct regions in $\cL_c^{(i)}$ space.
This motivates using a weighted distance from a learned reference point in $\cL_c^{(i)}$ space as a classifier, with the ansatz:
\begin{align}
    F^{(k)}(\cL^{(1)},...,\cL^{(k)}) = \sigma\left(w_0 + \sum_{i=1}^k w_i \left(\cL_c^{(i)} - b_i\right)^2\right), \label{eq:analytic_F}
\end{align}
where $w_i$ and $b_i$ are parameters to be minimized, and $\sigma$ is the sigmoid function. 
This classifier arises naturally as the (sigmoid of) the log-likelihood if we assume that the $\cL_c$ are Gaussian distributed with means $b_i$ and variances $1/(2w_i)$, motivated by the observation that in \Fig{log_angularities_f} the distributions form rough contiguous blobs.
The number of parameters in this classifier is naively $4 + 2k$, with 3 from the log angularity fit, 1 from $w_0$, and $2k$ from $w_i$ and $b_i$.
However, because any monotonic function of a classifier is an equally good classifier, this can be reduced to $2 + 2k$ by removing $w_0$ and an overall scale from the $w_i$'s.

\begin{table*}[p]
    \centering
    \begin{tabular}{l|c|c|c|c}
         Model & AUC & $1/\epsilon_g$ at $\epsilon_q = 0.3$ & $1/\epsilon_g$ at $\epsilon_q = 0.5$ & Trainable Parameters \\
         \hline
         \hline
        $k = 1$ Log Angularity Closed Form& 0.725 $\pm$ 0.001 & 36.2 $\pm$ 0.2 & 7.0 $\pm$ 0.0 & $3+1$ \\
        $k = 2$ Log Angularity Closed Form& 0.780 $\pm$ 0.002 & 57.4 $\pm$ 0.2 & 10.6 $\pm$ 0.1 & $3+3$ \\
        $k = 3$ Log Angularity Closed Form& 0.781 $\pm$ 0.002 & \textbf{57.8 $\pm$ 2.8} & 12.1 $\pm$ 0.1 & $3+5$ \\
        $k = 4$ Log Angularity Closed Form&\textbf{ 0.793 $\pm$ 0.002 }& 54.6 $\pm$ 1.7 &\textbf{ 12.7 $\pm$ 0.4 }& \textbf{$3+7$} \\
        \hline
    \end{tabular}
    \caption{The same as \Tab{auc}, but with the closed-form expressions defined in \Eq{analytic_F}. The ``$3 +$'' in the trainable parameters column refers to the parameters in the log angularity fit.}
    \label{tab:auc_closed}
\end{table*}

\begin{figure}[t]
    \centering
    \includegraphics[width = 0.90\linewidth]{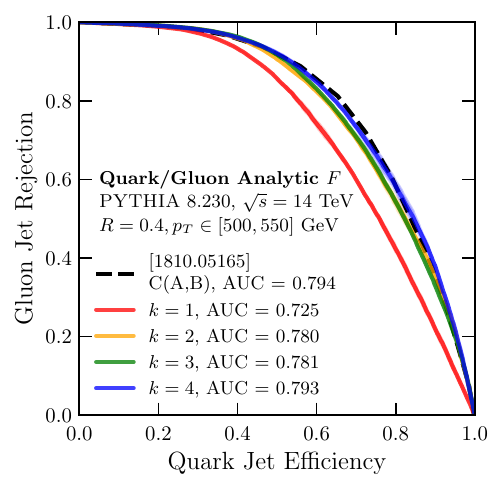}
    \caption{ROC curves showing the performance of the closed form observables, as defined in \Eq{analytic_F}, as a quark/gluon classifier. For comparison, we also show the ROC curve of the observable $C(A,B)$, a closed-form observable based on fits to an $L = 2$ EFN as defined in Ref.\ \cite{efn}.}
    \label{fig:analytic_F}
\end{figure}

In \Fig{analytic_F}, we show ROC curves corresponding to the observables defined in \Eq{analytic_F}.
We also summarize these results in \Tab{auc_closed}.
For $k > 1$, the performance saturates around an AUC of $0.78-0.79$, which is roughly the performance of an $L = 2$ EFN.
Unlike the case where $F$ was a dense network, however, the performance does not improve significantly beyond $k = 2$ -- moreover, these results are largely uncharged for various modifications to the functional form of \Eq{analytic_F}, including extending to up to degree $4$ polynomials in $\cL$ and $\cL_c$ and even up to degree $4$ rational polynomials.\footnote{There are a variety of ways one can combine cumulants one could try instead -- for example, the Gram-Charlier A series or the Edgeworth Series to approximate probability distributions given cumulants~\cite{10.1214/aoms/1177706528}.}
This suggests that the contributions of higher-order moments beyond the first two are more complex and unable to be easily parametrized -- that is, while it is easy to \emph{encode} the latent space information into a smaller $L$ via \Eq{moment_efn_identification}, the \emph{decoding} of the effective latent space in $F$ is nontrivial.

For comparison, we also show the ROC curve of the observable $C(A,B)$, an observable with a similar functional form defined in Ref.\ \cite{efn} based off of fits to an $L = 2$ ordinary EFN.\footnote{The EFNs in Ref.\ \cite{efn} use the ReLU rather than LeakyReLU~\cite{Maas2013RectifierNI} activation, which slightly changes the small $L$ behavior relative to the studies here.}
This observable also saturates at roughly the same AUC.
Thus, while we have succeeded in developing a fully closed-form solution suitable for up to $k=2$, equivalent to 2 latent dimensions of an ordinary EFN, 
finding a closed form expression for $F$ beyond 2 effective latent dimensions is nontrivial.

\subsection{Beyond $L =1$}\label{sec:larger_L}

Finally, we briefly look at Moment EFNs with $L > 1$ to see if we can attempt to gain some insight into them, just like the $L = 1$ Moment EFNs.
Unlike the case with $L = 1$ Moment EFNs, the analysis of $ L > 1$ models is much more difficult, both due to the higher dimension and because the product structure allows radial symmetry to be more easily broken, even for $L = 2$.

\begin{figure*}[p]
    \centering
    \subfloat[]{
         \includegraphics[width=0.45\textwidth]{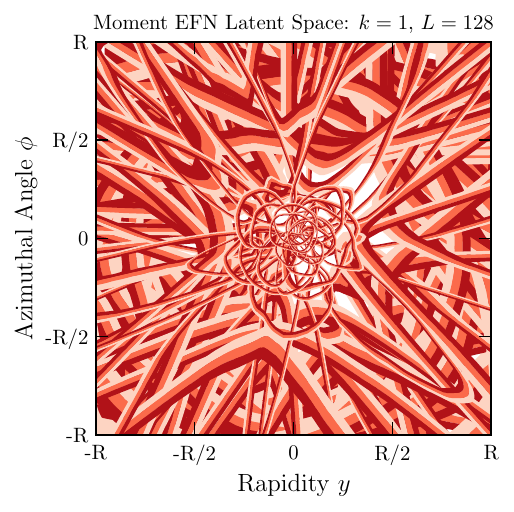}
        \label{fig:k1_latent_space}
    }
    \subfloat[]{
        \includegraphics[width=0.45\textwidth]{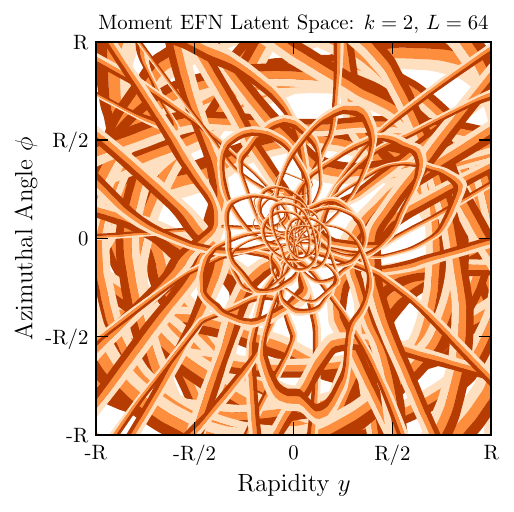}
        \label{fig:k2_latent_space}
    }
    \vspace{0pt}
    \subfloat[]{
         \includegraphics[width=0.45\textwidth]{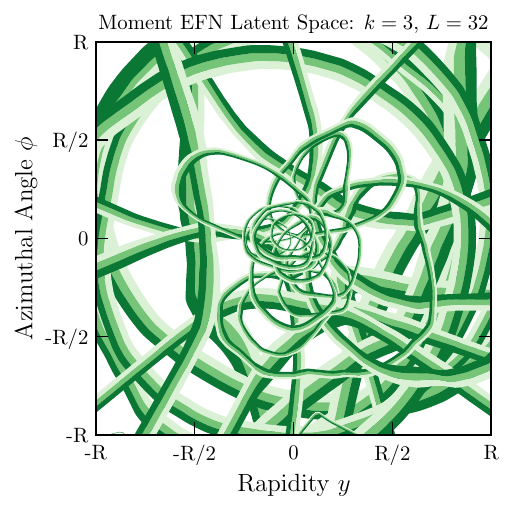}
        \label{fig:k3_latent_space}
    }
    \subfloat[]{
        \includegraphics[width=0.45\textwidth]{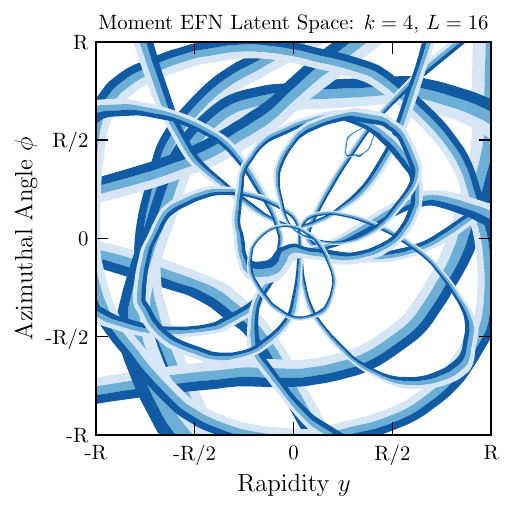}
        \label{fig:k4_latent_space}
    }
    \caption{
         The contours of the learned Moment EFN latent space embeddings $\Phi$, for (a) $k = 1$, (b) $k = 2$, (c) $k = 3$, and (d) $k = 4$. Each figure represents the best model for the highest value of $L$ considered for each $k$: 128, 64, 32, and 16 respectively. Each curve represents the $45\%-55\%$ contours of each of the $L$ different $\Phi$ functions. The overall normalization is arbitrary.
        }
    \label{fig:latent_spaces}
\end{figure*}

In order to visualize the latent spaces for $L > 1$, we follow the procedure outlined in \Reference{efn}.
We can overlay all $L$ learned $\Phi$ functions of each model at once by drawing the $45\%-55\%$ contours of each $\Phi$ function.
In these plots, the overall normalization and sign of the $\Phi$ functions is unimportant, as the $F$ network easily learn to rescale its inputs via simple linear transformations.
As an example, in \Fig{latent_spaces}, we show the $L = 128, 64, 32,$ and $16$ (the highest $L$ considered for each $k$) dimensional latent spaces of the highest performing models for the order $k = 1, 2, 3, $ and $4$ Moment EFNs, respectively. 
Like the EFN, the Moment EFN is able to pick up on the collinear singularity of QCD, as filters closer to the center are more closely resolved.
Note that while on the whole, the entire ensemble of contours appears to be radially symmetric, the individual contours are not -- this in contrast with the $L = 1$ Moment EFNs, where the $\Phi$ functions were genuinely radially symmetric (or at the very least, broken to mirror symmetric).

\begin{figure*}[t]
    \centering
    \subfloat[]{
         \includegraphics[width=0.45\textwidth]{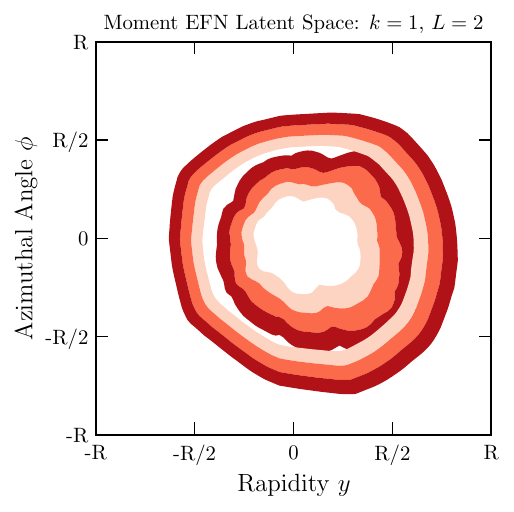}
        \label{fig:k1_latent_space2}
    }
    \subfloat[]{
        \includegraphics[width=0.45\textwidth]{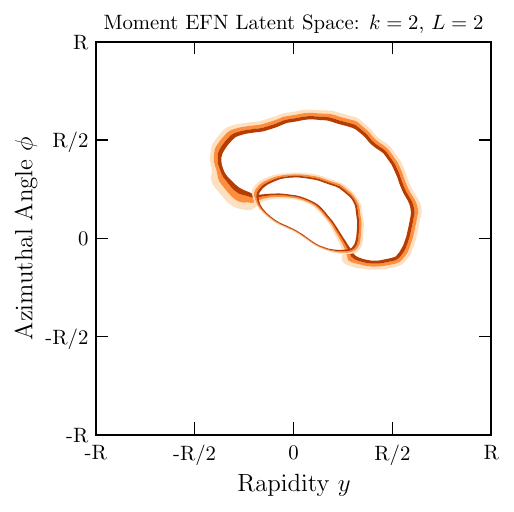}
        \label{fig:k2_latent_space2}
    }
    \vspace{0pt}
    \subfloat[]{
         \includegraphics[width=0.45\textwidth]{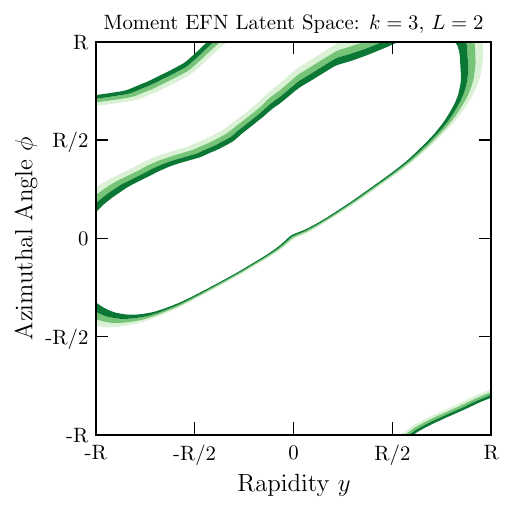}
        \label{fig:k3_latent_space2}
    }
    \subfloat[]{
        \includegraphics[width=0.45\textwidth]{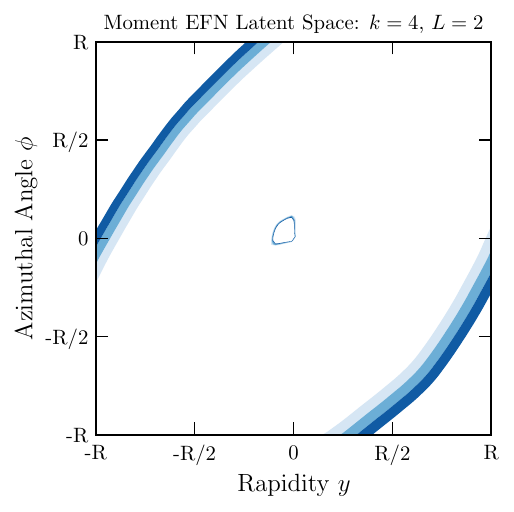}
        \label{fig:k4_latent_space2}
    }
    \caption{
         The contours of the learned $L = 2$ Moment EFN latent space embeddings $\Phi$, for (a) $k = 1$, (b) $k = 2$, (c) $k = 3$, and (d) $k = 4$. Each curve represents the $45\%-55\%$ contours of each of the 2 different $\Phi$ functions. The overall normalization is arbitrary.
        }
    \label{fig:latent_spaces2}
\end{figure*}

After $L = 1$ from \Fig{latent_spaces_L1}, the next most natural thing to study are the $L = 2$ latent spaces, which we show in \Fig{latent_spaces2}.
Not only is $L = 2$ less radially symmetric than $L = 1$ in general, we also notice the $k > 1, L = 2$ Moment EFNs exhibit less radial symmetry than the $L = 2$ ordinary EFN. 
With product structures, it is easier to form nontrivial representations of the rotation group that later combine to form the trivial representation. 
As an example, the functions $\Phi^1(x) = x^{(1)}$ and $\Phi^2(x) = x^{(2)}$ are themselves not radially symmetric, but the $k = 2$ moment combination $F(\Phi) = (\Phi^1)^2 + (\Phi^2)^2$ is.
Nevertheless, some models occasionally still seem to retain some approximate radial symmetry, as is the case for the order $k = 4$ model in \Fig{k4_latent_space2}.

%
%

\begin{figure}[t]
    \centering
    \includegraphics[width = 0.95\linewidth]{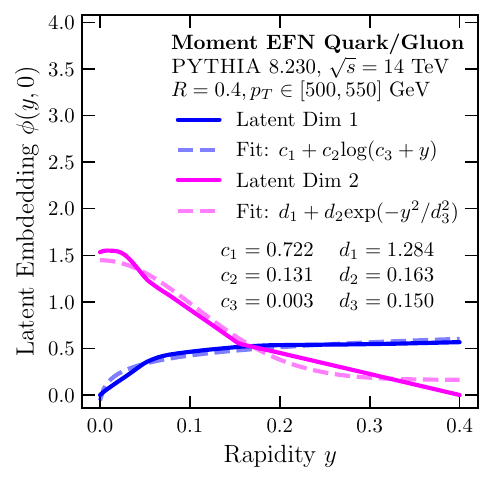}
    \caption{The same as \Fig{latent_fit}, but for the best performing $k = 4, L = 2$ Moment EFN. The two latent dimensions are shown in blue and pink.}
    \label{fig:latent_fit2}
\end{figure}

We now move to analyze and fit the radial projections of the order $k = 4, L = 2$ Moment EFN in hopes of finding closed-form expressions for the (now two) latent dimensions, exactly as was done in \Sec{log_angularities}.
This model has $L_{\mathrm{eff}} = 15$, which is roughly equivalent to an ordinary EFN with 14 latent dimensions -- in principle, fitting these two functions should therefore enable us to extract the same information as 14 latent dimensions worth of information in an ordinary EFN by taking moments.
These radial projections are shown in \Fig{latent_fit2}. 
One of the two latent dimensions takes the form of a log angularity, as defined in \Eq{analytic_observables}, and the $c_3$ parameter is also in the expected range.
We will call the associated jet shape observable $\cL'$, with the prime indicative of the slightly different values of the fit parameters $c_i$.
For the second latent dimension, we fit the form:
\begin{align}
    \Phi_{\mathcal{E}}(r) =  d_1 + d_2\exp(-r^2/d_3^2),
\end{align}
motivated both by the form of the plot and the use of a similar form in the analysis of the $L = 2$ EFN in \Reference{efn}.
Just as with log angularities, we can define moment-based jet shape observables based off this fit, which we will denote $\mathcal{E}^{(n)}$, and associated cumulants $\mathcal{E}^{(n)}_c$.
We can also define mixed moments $\mathcal{M}^{(mn)}$ of the form $\mathcal{M}^{(mn)} \equiv \expval{\Phi_{\cL'}^m\Phi_{\mathcal{E}}^n}$.

\begin{figure*}[tpbh]
    \centering
    \subfloat[]{
         \includegraphics[width=0.45\textwidth]{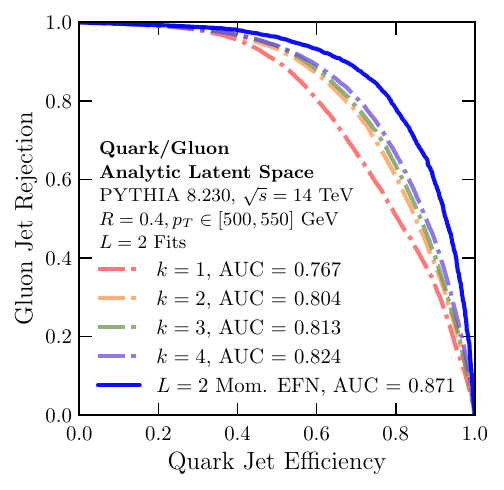}
        \label{fig:analytic_auc_2}
    }
    \subfloat[]{
        \includegraphics[width=0.45\textwidth]{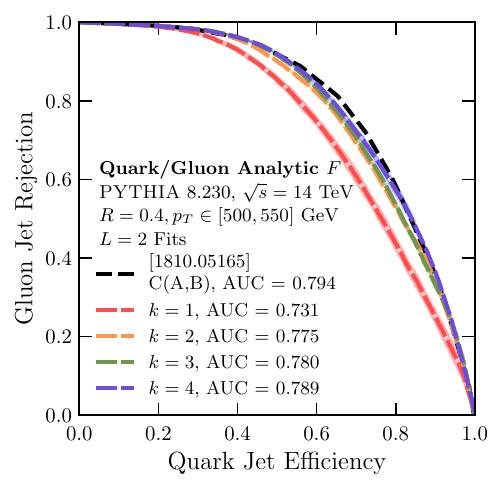}
        \label{fig:analytic_F2}
    }
    \caption{
         (a) The same as \Fig{analytic_auc}, but with the $L = 2$ fits and dense networks. (b) The same as \Fig{analytic_F}, but for the closed-form $F$ defined using $L = 2$ in \Eq{analytic_F2}.
        }
    \label{fig:analytic_2}
\end{figure*}

We can use the observables $\cL'^{(n)}$ and $\mathcal{E}^{(n)}$ as inputs to dense neural networks $F$ as a measure of how well the latent spaces are approximated by these fits.
Here, the dense network $F^{(k)}$ takes in \emph{all} $L_{\mathrm{eff}}(k)$ multivariate moments of the two observables -- for instance, the $k = 2$ dense network has the form $F^{(2)}({\cL'}^{(1)}, \mathcal{E}^{(1)}, {\cL'}^{(2)}, \mathcal{E}^{(2)},\mathcal{M}^{1,1})$.
We show the performance of these networks in \Fig{analytic_auc_2}.
We see here that unlike the $L = 1$ analysis performed in \Sec{log_angularities}, the two observables here are not enough to reproduce the full $L = 2$, $k = 4$ Moment EFN -- in fact, they are only about as good as the $L = 1$, $k = 4$ result at best.
This means that the fits are not enough to capture the full latent space, suggesting that non radially symmetric information is important.

Finally, as was done in \Sec{F_networks}, we may attempt to build closed-form taggers from these fits. 
To accommodate the second observable, we extend the form of our fit to:
\begin{align}
    F^{(k)}(\cL^{(i)},\mathcal{E}^{(i)}) = \sigma\bigg(&w_0 + \sum_{i=1}^k w^{\cL'}_i \left(\cL_c^{(i)} - b_i^{\cL'}\right)^2\nonumber\\
    &+ \sum_{i=1}^k w^{\mathcal{E}}_i \left(\mathcal{E}_c^{(i)} - b_i^{\mathcal{E}}\right)^2\bigg). \label{eq:analytic_F2}
\end{align}
For simplicity, we have not considered mixed moments between the two observables. 
We show the performance of these taggers in \Fig{analytic_F2}.
The observables here are no better than the $L = 1$ analytic observables shown in \Fig{analytic_F} -- in fact, they are slightly worse, due to the extra numerical cost of optimizing additional parameters.\footnote{This holds for many modifications of \Eq{analytic_F2}, including rational polynomials.}
That is, unlike the EFN observable $C(A,B)$, the additional parameterization power granted by an additional latent dimension in a Moment EFN is more complex than can be captured by a simple functional dependence.

\section{Conclusions and Outlook}
\label{sec:conclusions}

In this paper, we presented an extension of the Deep Sets framework in the form of Moment Pooling. 
Moment Pooling arises when the summation operation in Deep Sets is generalized to arbitrary multivariate moments.
We have shown an implementation of Moment Pooling in Moment EFNs, a particular Deep Sets framework useful in particle physics.
For a fixed number of latent dimensions $L$, an order $k$ Moment EFN is able to reach a much higher \emph{effective} latent dimension by recycling the same $L$ functions in product structures, and conversely, a Moment EFN is capable of reducing the $L$ required to achieve the same performance as an ordinary EFN.

We find that Moment EFNs are able to achieve the same (or better) performance as ordinary EFNs for quark/gluon discrimination, but with significantly fewer latent dimensions.
In particular, an order $k = 4$ Moment EFN with only 16 latent dimensions is able to achieve slightly better performance than an EFN with 128 latent dimensions, indicative of ``data compression'' and structure in the quark/gluon discrimination task.
Similarly, an order $k = 4$ Moment EFN with only a single latent dimension is able to achieve an AUC of 0.84, equivalent to an ordinary EFN with 4 latent dimensions.
By analyzing the latent space of this model, we are able to find a simple, closed-form expression in the form of the log angularity shape observable, whose moments contain the same information as 4 latent dimensions of an ordinary EFN.
However, we find that analyzing the $F$ network in closed-form is difficult, and that simple parameterizations cannot go beyond the performance of up to an $L = 2$ model, which suggests some complexity in the way the information in log angularities is decoded.

We conclude by discussing possible avenues of further study.
One can ask if the latent space structure granted by the moment architecture helps in learning richer jet representations -- that is, if the latent representation is a genuine representation of a jet from which multiple attributes and observables can be estimated, not just a quark/gluon discriminant.
The latent spaces of EFNs are highly degenerate, and as we have seen, the 4 latent dimensions of an $L = 4$ EFN may be replaced by 4 moments of log angularities without any loss of performance on quark/gluon tagging.
Second, we have \emph{only} considered moments, in the sense that we have restricted our architecture to only have terms of the form $\sum_{i\in \cP} z_i (\Phi^{1}(x_i))^{w_1}(\Phi^{2}(x_i))^{w_2}...(\Phi^{L}(x_i))^{w_L}$, for pre-determined positive integers $w_l$ such that $\sum_l w_l \leq k$.
One could imagine generalizing this even further, for example by letting the powers $w_l$ be real and negative, or even learned.
Third, our study here primarily focused on IRC-safe observables and EFNs, but the Moment Pooling concept is applicable to any Deep Sets-style architecture. 
In particular, non-IRC safe observables such as multiplicity are known to aid in quark/gluon tagging, and indeed the studies in \App{additional} show there is potential for gain when applying Moment Pooling to non-IRC safe PFNs. 
It may be possible, therefore, to use Moment Pooling or related concepts to help compress the latent spaces of more sophisticated point-cloud based approaches, such as ParticleNet~\cite{PhysRevD.101.056019}, Particle Transformer~\cite{Qu:2022mxj}, or PELICAN~\cite{Bogatskiy:2022czk}.

The Moment Pooling operation and Moment EFN are a step towards generalizing existing models while adding to their interpretability. 
We look forward to further developments in the direction of flexible models with interpretable internal representations.

\section*{Code and Data}

The general Moment EFN, along with several variations (including cumulant-based models and Moment PFNs), is available at \url{https://github.com/athiso/moment}.
The code used to perform all the analyses and make all the figures featured in this paper is available at \url{https://github.com/rikab/MomentAnalysis} in the form of \textsc{Jupyter} notebooks~\cite{Kluyver:2016aa}.

\section*{Acknowledgments}

We would like to thank Samuel Alipour-fard, Sean Benevedes, Miles Cranmer, Andrew Larkoski, Benjamin Nachman, and Sokratis Trifinopoulos for useful and interesting discussions and comments. 

R.G. and J.T. are supported by the National Science Foundation under Cooperative Agreement PHY-2019786 (The NSF AI Institute for Artificial Intelligence and Fundamental Interactions, \url{http://iaifi.org/}), and by the U.S. DOE Office of High Energy Physics under grant number DE-SC0012567. 
J.T. is additionally supported by the Simons Foundation through
Investigator grant 929241.
A.O. was supported by the Bowdoin College Summer Internship Program.

\appendix

\section{Model and Training Specifications}\label{app:model_specifications}

In this appendix, we provide details for the models and training procedures used in \Sec{qg_discrimination} and \App{additional}. 
All models are implemented as modified versions of the EFN/PFN models in the \EnergyFlow Python package~\cite{efn}, built with \Keras~\cite{keras} using the \TensorFlow~\cite{tensorflow} backend.
Each training is performed using an NVIDIA A100.

The key difference between the Moment EFN and the ordinary EFN is the addition of the \texttt{MomentPooling} layer between the $\Phi$ and $F$ functions. 
The \texttt{MomentPooling} layer is a deterministic function $\mathbb{R}^L \xrightarrow[]{} \mathbb{R}^{L_{\mathrm{eff}}}$, where $L_{\mathrm{eff}}$ depends on both $L$ and the order $k$, that maps the $L$-component $\Phi$ to the list of all multivariate moments up to order $k$, taken over the event $\cP$:
\begin{align}
    \Phi^a \mapsto \left(\expval{\Phi^a}_\cP, \expval{\Phi^{a_1}\Phi^{a_2}}_\cP, ... \expval{\Phi^{a_1}...\Phi^{a_k}}_\cP \right).
\end{align}

Implementation-wise, the \texttt{MomentPooling} layer takes in a \TensorFlow tensor of shape $(N_{\text{batch}}, N_{\text{particles}}, L)$, where $N_{\text{batch}}$ is the number of events to be computed in parallel, $N_{\text{particles}}$ is the maximum number of particles per event,\footnote{While Deep Sets, in theory, allows for an unbounded number of particles, it is more practical for speed and memory to have a fixed cap.} and $L$ is the latent dimension -- this tensor is obtained as the output of the $\Phi$ network. 
The moments are then computed recursively: We first define the $k = 0$ moment tensor, a tensor of shape $(N_{\text{batch}}, N_{\text{particles}}, 1)$ with every entry equal to 1. 
Given the $k$'th moment tensor, which is of shape $(N_{\text{batch}}, N_{\text{particles}}, L_{\mathrm{eff}}(k))$, the $k +1$'th moment tensor is obtained by performing the outer product of the original input tensor, to obtain a new tensor of shape $(N_{\text{batch}}, N_{\text{particles}}, L_{\mathrm{eff}}(k), L)$. 
Note that this outer product will contain redundant moments, since the order of indices does not matter -- thus, we only perform the outer product on the indices corresponding to the upper triangular part of the $k+1$-dimensional hypercube with $L$ indices per dimension.
The tensor is then flattened and concatenated to the $k$'th moment tensor to form the $k+1$'th moment tensor of shape $(N_{\text{batch}}, N_{\text{particles}}, L_{\mathrm{eff}}(k+1))$.
At the end of the recursive procedure, we perform the $z$-weighted sum over the $N_{\text{particles}}$ dimension, so that the final output has shape $(N_{\text{batch}}, L_{\mathrm{eff}}(k))$.
This recursive procedure allows for some computations to be reused when computing higher-order moments, simplifying the \TensorFlow computational graphs and saving time on backpropagation versus recomputing all moments from scratch.

Following \Reference{efn}, all of our models consist of a $\Phi$ network with three layers of sizes $100$, $100,$ and $L$ respectively (with $L$ being the latent dimension), and an $F$ network with four layers of size $100, 100, 100,$ and $1$ respectively.
For both networks, the final layer is the output layer. 
In between $\Phi$ and $F$ is the \texttt{MomentPooling} layer.
We use LeakyReLU~\cite{Maas2013RectifierNI} with $\alpha = 0.3$ for all activation functions,\footnote{This is to avoid the Dying ReLU problem~\cite{dyingrelu}, especially for smaller $L$.} except for the final layer of $F$, where we use a sigmoid function for the classifier output.

To aid our models in learning efficient representations, especially for $k > 1$, we use a  ``pre-training'' procedure. 
This procedure helps to ensure that each additional moment added is used to learn ``new'' information that helps the model and to mitigate the effect of more nontrivial training for higher order $k$ Moment EFNs.
To train an order $k$ Moment EFN, first, we first train order $k-1$ Moment EFN with the exact same value of $L$ and the same hyperparameters for the $\Phi$ and $F$ networks, using the training procedure defined below.  
Then, we initialize an order $k$ network whose $\Phi$ and $F$ weights are identical to the order $k-1$ model's weights, \emph{except} for the weights attached to the \texttt{MomentPooling} layer, as the size of this layer has changed.
The weights connecting the first $L_{\mathrm{eff}}(k-1)$ outputs of the \texttt{MomentPooling} layer to the first $F$ layer are the same as the $k-1$ network weights.
Finally, the order $k$ model can be trained.
This pretraining procedure is recursive: to train the order $k -1$ model, we initialize its weights from an order $k - 2$ model, and so on. 
The weights of the $k = 1$ models, plus all other undetermined weights (namely the rest of the weights connecting the \texttt{MomentPooling} layer to the first $F$ layer)  are initialized using the default He-uniform~\cite{he2015delving} distribution.
To ensure that any improvements are not the result of having more epochs to train, in our comparative studies lower $k$ models have the training procedure described below applied to them multiple times so that all models train for the same total number of epochs.
That is, since we study up to $k = 4$ Moment EFNs, all of our models are trained effectively for $4 \times$ the number of stated epochs.

Each model is trained for 50 epochs (times 4, with the recursive pre-training) with a batch size of 512 -- we allow for early stopping with a patience parameter of 8, though early stopping never occurred in any of our trainings.
We have checked that allowing for a maximum of 150 epochs, rather than 50, per training round makes no significant difference in our results.
We use 1M total jets for training, 50k for validation, and 50k for testing.
We train to minimize the binary crossentropy loss using the \Adam optimizer~\cite{kingma2017adam} with a learning rate of $0.001$.
The training times of each network per epoch on an NVIDIA A100 are shown in \Tab{training_time}.
Each model is re-initialized and re-trained 3 times, the ROC curves and AUC saved for each of the trainings.
Note that we do not train models with both larger $k$ and larger $L$, as the training time and memory requirements become excessive.

\begin{table}[tph]
    \centering
    \begin{tabular}{c|c|c|c|c}
         Latent Dim.\ $L$ & EFN & $k = 2$ & $k = 3$ & $k = 4$ \\
         \hline
         \hline
         $2^0$ & 2s & 3s & 3s & 4s \\
         $2^1$ & 3s & 3s & 3s & 4s \\
         $2^2$ & 3s & 3s & 4s & 5s \\
         $2^3$ & 3s & 4s & 6s & 11s \\
         $2^4$ & 3s & 6s & 16s & 66s \\
         $2^5$ & 4s & 13s & 91s & \\
         $2^6$ & 4s & 38s & & \\
         $2^7$ & 5s & & & \\
         \hline
    \end{tabular}
    \caption{The time per epoch, in seconds, for the $k = 1$ through $4$ Moment EFNs to train on the quark/gluon discrimination dataset.}
    \label{tab:training_time}
\end{table}

\section{Regression with Jet Angularities}
\label{app:jet_angularities}

In this appendix, we demonstrate the improved performance and reduced complexity of Moment EFNs in regression tasks by exploring their relationship to jet angularities~\cite{Berger:2003iw, Berger:2004xf}.

Jet angularities are well-studied QCD observables that quantify the radial distribution of energy within a jet. 
The product structure of the Moment EFN is especially suited for angularities, since a jet angularity can be thought of as an energy-weighted radial moment.
For a jet $\cP$, the $\beta$-angularity of the jet, $\lambda^{(\beta)}$, is defined as:
\begin{align}
    \lambda^{(\beta)}(\cP) \equiv \sum_{i\in \cP} z_i |x_i|^\beta,
    \label{eq:angularity}
\end{align}
where $x_i = (y_i, \phi_i)$ are the particle coordinates defined from an appropriately defined jet center $x_0$, which we will take to be the energy-weighted average position of the jet.

In the language of moments, jet angularities take a very natural form:
\begin{align}
        \lambda^{(\beta)}(\cP) = \expval{|x_i|^\beta}.
\end{align}
If $\beta$ is an even integer, then $\lambda^{(\beta)}$ can be expressed using a completely \emph{linear} Moment EFN with $k=\beta$ and $L=2$:
\begin{align}
    \Phi^1(x_i) &= y_i, \quad  \Phi^2(x_i) = \phi_i, \\
    F(\expval{\Phi}_\cP) &= \sum_{a_1 = 1}^2 ... \sum_{a_{\beta/2} = 1}^2 \expval{\left(\Phi^{a_1}\right)^2...\left(\Phi^{a_{\beta/2}}\right)^2}_\cP, \label{eq:linear_angularity}
\end{align}
where $a = 1, 2$ corresponds to the two dimensions of $x$. 
In particular, for $\beta = 2$, this becomes:
\begin{align}
    F(\expval{\Phi}_\cP) &= \expval{y^2} + \expval{\phi^2}. 
\end{align}
The $\beta = 2$ angularity is especially nice, as it relates to the jet mass:
\begin{align}
    \lambda^{(2)} \approx \frac{m_J^2}{(\sum E_i)^2}.
\end{align}

\begin{figure}[tbp]
    \centering
    \includegraphics[width = 0.95\linewidth]{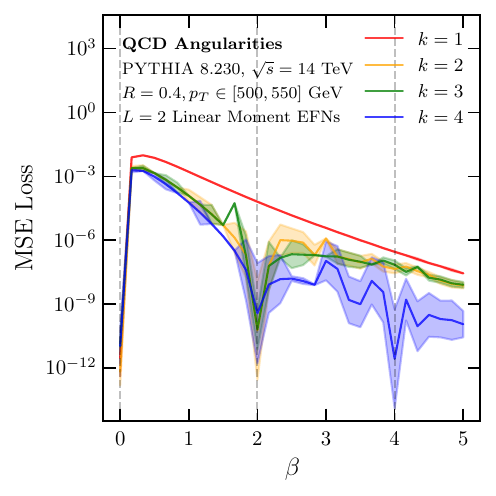}
    \caption{The (average) MSE loss of the $k$ Moment EFN trained to regress the jet angularity $\lambda^{(\beta)}$ as a function of $\beta$. Each line is the average of the MSE across three retrainings (in log space), and the bands are the standard deviation (in log space). }
    \label{fig:beta}
\end{figure}

The fact that angularities can be represented as a linear function of moments is significant, as dense neural networks, especially those using variants of ReLU, are at their core are piecewise-linear approximators.\footnote{This is still qualitatively true for other activation functions.}
The nonlinear part of the angularities, namely the $|x|^\beta$ function, is encoded in the pre-specified moment functions, leaving only a purely linear function to learn.
Thus, one would expect a strictly linear Moment EFN to learn even integer $\beta$ angularities \emph{exactly} (that is, with a mean squared error loss of zero) for $k \geq \beta$, and in general that $k > 1$ Moment EFNs outperform $k = 1$ EFNs on generic regression tasks.

To illustrate the improved performance of the Moment EFN on regression tasks, we train $L = 2$ Moment EFNs from $k = 1$ through $4$ to learn the angularity $\lambda^{(\beta)}$, for values of $\beta \in [0,5]$. 
The Moment EFNs are strictly linear: The functions $\Phi$ and $F$ are \emph{strictly linear} functions from $\mathbb{R}^2 \xrightarrow{} \mathbb{R}^L$ and $\mathbb{R}^{L_{\mathrm{eff}}} \xrightarrow{} \mathbb{R}$, respectively, with no hidden layers, activation functions, or even layer biases.
We use the exact same quark/gluon dataset described above in \Sec{qg_dataset} for this study, though we ignore the quark/gluon label. On each jet, we compute the $\beta$-angularity as defined in \Eq{angularity}, for $\beta$ from $0$ to $5$ in increments of $\Delta\beta = \frac{1}{6}$.
The models are trained to minimize the mean-squared error (MSE) over the training set.

In \Fig{beta}, we plot the average MSE of each model, taken over the three retrainings, on the test set as a function of $\beta$.
First, we note that the $k > 1$ models are consistently better than the $k = 1$ model for $\beta > 0$. 
Second, there are large downward spikes in the loss: at $\beta = 0$ for all models; at $\beta = 2$ for $k = 2$, $3$, and $4$; and at $\beta = 4$ for $k = 4$.
The first two spikes are especially close to zero, with an MSE loss of $\sim10^{-12}$ approaching floating point precision.
This is precisely the behavior expected by \Eq{linear_angularity}, as the linear model is able to achieve an exact (up to machine precision) fit for even integer $\beta$ with $k \geq \beta$.


\section{Additional Collider Classification Studies} \label{app:additional}

In this appendix, we present additional studies to supplement \Sec{qg_discrimination}, both by replacing the EFN in the moment architecture with a PFN, and by replacing the quark/gluon discrimination task with a top/QCD jet discrimination task.

The Moment PFN models are identical to those described in \App{model_specifications}, except $\Phi$ is now a function of the particle $(z_i, x_i)$ rather than just $x_i$, and $\Phi$ is no longer weighted by $z$.
Importantly, this changes the definition of the moment pooling operation, since, as shown in \Eq{PFN} moments are \emph{not} energy-weighted.

For our top/QCD dataset, we use the same top-tagging set as in \Refs{Butter:2017cot,Kasieczka:2019dbj}, commonly used as a benchmark in tagger studies. 
This set consists of a top quark jet signal and a mixed light quark and gluon jet background, generated in \Pythia 8.2.15~\cite{Sjostrand:2006za, Sjostrand:2014zea} at 14 TeV, and passed through the \textsc{Delphes} 3.3.2~\cite{deFavereau:2013fsa} \textsc{ATLAS} detector simulation. 
The jets are clustered using the anti-$k_T$ algorithm~\cite{Cacciari:2008gp} with $R = 0.8$, and satisfy $p_{T} \in [550, 650]$ GeV and $|y| \leq 2$.  
We do not perform any additional preprocessing here -- in particular, we do not rotate the jets in the rapidity-azimuth plane for these studies.

\begin{figure*}[tbhp]
    \centering
    \subfloat[]{
         \includegraphics[width=0.45\textwidth]{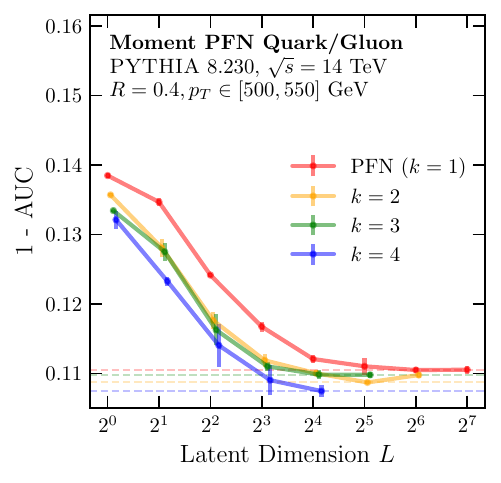}
        \label{fig:qg_PFN_latent_dim}
    }
    \subfloat[]{
        \includegraphics[width=0.45\textwidth]{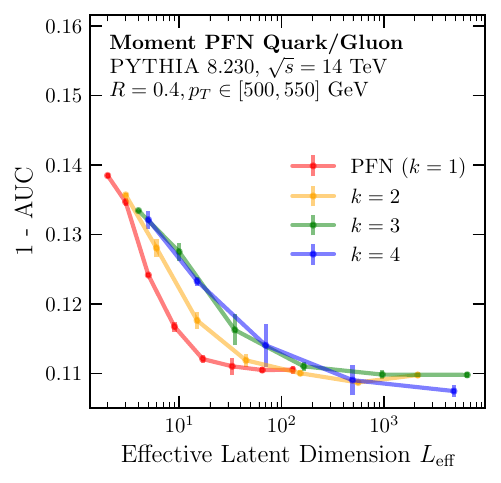}
        \label{fig:qg_PFN_effective_latent_dim}
    }
    \caption{
         The same as \Figs{latent_dim}{effective_latent_dim}, but with a Moment PFN rather than a Moment EFN.
        }
    \label{fig:qg_PFN}
\end{figure*}

\begin{figure*}[tbhp]
    \centering
    \subfloat[]{
         \includegraphics[width=0.45\textwidth]{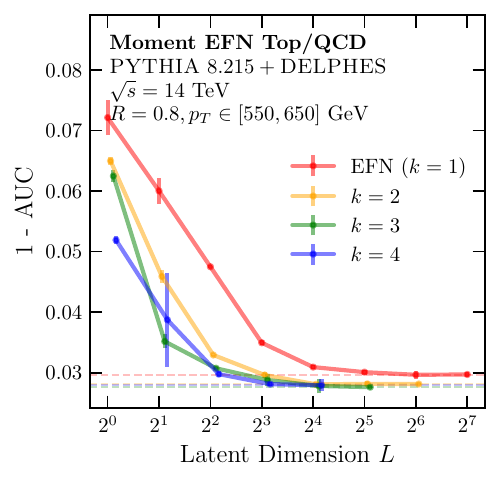}
        \label{fig:top_EFN_latent_dim}
    }
    \subfloat[]{
        \includegraphics[width=0.45\textwidth]{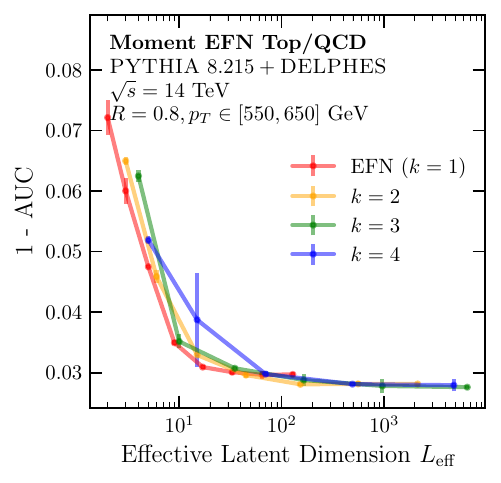}
        \label{fig:top_EFN_effective_latent_dim}
    }
    \caption{
         The same as \Figs{latent_dim}{effective_latent_dim}, but with the top/QCD dataset rather than the quark/gluon dataset. Note the $k = 4, L = 2$ outlier, potentially due to a failed convergence.
        }
    \label{fig:top_EFN}
\end{figure*}

\begin{figure*}[tbhp]
    \centering
    \subfloat[]{
         \includegraphics[width=0.45\textwidth]{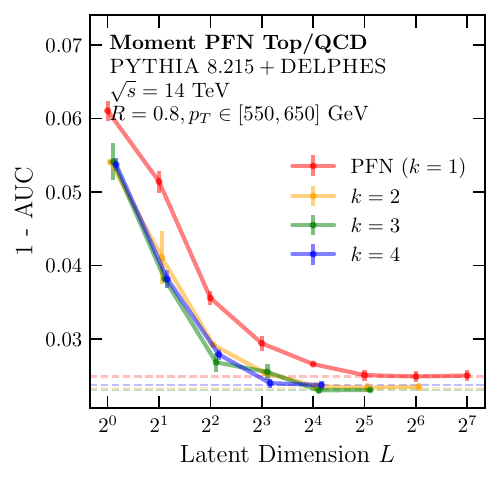}
        \label{fig:top_PFN_latent_dim}
    }
    \subfloat[]{
        \includegraphics[width=0.45\textwidth]{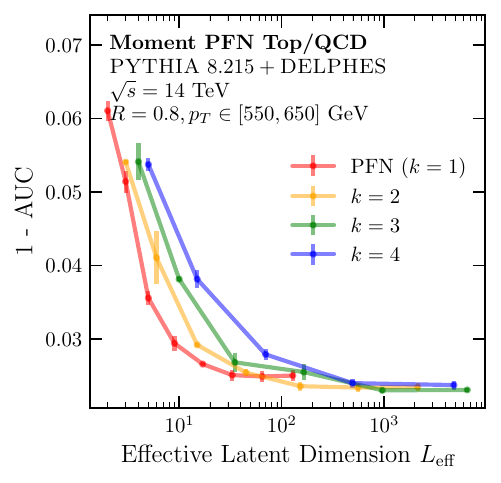}
        \label{fig:top_PFN_effective_latent_dim}
    }
    \caption{
         The same as \Figs{latent_dim}{effective_latent_dim}, but with a Moment PFN rather than a Moment EFN and the top/QCD dataset rather than the quark/gluon dataset.
        }
    \label{fig:top_PFN}
\end{figure*}

In Figs.\ \ref{fig:qg_PFN}--\ref{fig:top_PFN}, we show the AUC as a function of latent dimension and effective latent dimension for quark/gluon tagging with Moment PFNs, top tagging with Moment EFNs, and top tagging with Moment PFNs respectively.
These figures are meant to complement \Figs{latent_dim}{effective_latent_dim} for quark/gluon tagging with Moment EFNs.
We also summarize the results of these classifiers in Tables \ref{tab:qg_PFN}--\ref{tab:top_PFN} for ease of comparison with other studies using the same datasets.

We first observe that $k \geq 2$ architectures perform the same or better as their $k = 1$ counterparts for a given value of $L$, though this improvement is not always monotonic. 
In particular, in \Fig{top_EFN}, we can see that the $k = 4, L = 2$ Moment EFNS are indeed slightly worse than the $k = 3, L = 2$ Moment EFNs, with a high variance. 
This is likely driven by a failed network convergence, reflective of the fact that $k = 4$ models are difficult to train. 
Second, unlike the quark/gluon Moment EFNs, performance per effective latent dimension is \emph{not} independent of $k$.
For the quark/gluon Moment PFNs, the performance per effective latent dimension tends to \emph{increase} with $k$, but for the top/QCD Moment EFNs and Moment PFNs, the performance per effective latent dimension tends to \emph{decrease} with $k$.
This would seem to imply that the top/QCD discriminant is not ``compressible'', in that its latent space cannot be easily factorized into products of just a few functions, and that many independent functions are genuinely required.

\begin{table*}[tpbh]
    \centering
    \begin{tabular}{l|c|c|c|c}
         Model & AUC & $1/\epsilon_g$ at $\epsilon_q = 0.3$ & $1/\epsilon_g$ at $\epsilon_q = 0.5$ & Trainable Parameters \\

 \hline
\hline
$k = 1, L = 1$  PFN& 0.862 $\pm$ 0.000 & 50.8 $\pm$ 1.0 & 19.4 $\pm$ 0.2 & 31206 \\
$k = 2, L = 1$ Moment PFN& 0.864 $\pm$ 0.001 & 55.7 $\pm$ 0.9 & 20.3 $\pm$ 0.1 & 31306 \\
$k = 3, L = 1$ Moment PFN& 0.866 $\pm$ 0.001 & 60.9 $\pm$ 2.3 & 21.3 $\pm$ 0.5 & 31406 \\
$k = 4, L = 1$ Moment PFN& \textbf{0.867 $\pm$ 0.002} & \textbf{62.8 $\pm$ 1.2} & \textbf{22.0 $\pm$ 0.2} & 31506 \\
 \hline
 \hline
$k = 1, L = 128$  PFN& 0.889 $\pm$ 0.002 & 76.9 $\pm$ 0.6 & 31.6 $\pm$ 0.5 & 89753 \\
$k = 2, L = 64$ Moment PFN& 0.890 $\pm$ 0.001 & \textbf{83.0 $\pm$ 1.3} & 32.2 $\pm$ 1.4 & 260185 \\
$k = 3, L = 32$ Moment PFN& 0.890 $\pm$ 0.000 & 76.4 $\pm$ 2.7 & 33.2 $\pm$ 0.4 & 690745 \\
$k = 4, L = 16$ Moment PFN& \textbf{0.893 $\pm$ 0.001 }& 78.4 $\pm$ 4.0 & \textbf{34.1 $\pm$ 0.8} & 517561 \\
\hline
\hline
    \end{tabular}
    \caption{The same as \Tab{auc}, but with a Moment PFN rather than a Moment EFN.}
    \label{tab:qg_PFN}
\end{table*}

\begin{table*}[tph]
    \centering
    \begin{tabular}{l|c|c|c|c}
         Model & AUC & $1/\epsilon_{QCD}$ at $\epsilon_{top} = 0.3$ & $1/\epsilon_{QCD}$ at $\epsilon_{top} = 0.5$ & Trainable Parameters \\
         \hline
         \hline
 $k = 1, L = 1$  EFN& 0.922 $\pm$ 0.002 & 29.5 $\pm$ 0.7 & 16.8 $\pm$ 0.3 & 31106 \\
$k = 2, L = 1$ Moment EFN& 0.930 $\pm$ 0.002 & 37.6 $\pm$ 0.6 & 20.2 $\pm$ 0.4 & 31206 \\
$k = 3, L = 1$ Moment EFN& 0.933 $\pm$ 0.001 & 41.2 $\pm$ 2.3 & 20.6 $\pm$ 0.6 & 31306 \\
$k = 4, L = 1$ Moment EFN& \textbf{0.946 $\pm$ 0.001} & \textbf{70.3 $\pm$ 5.8} & \textbf{29.1 $\pm$ 1.1} & 31406 \\
\hline
\hline
$k = 1, L = 128$  EFN& 0.970 $\pm$ 0.000 & 465.0 $\pm$ 40.0 & 108.8 $\pm$ 2.4 & 89653 \\
$k = 2, L = 64$ Moment EFN& 0.972 $\pm$ 0.001 & 407.9 $\pm$ 16.7 & 131.0 $\pm$ 3.5 & 260085 \\
$k = 3, L = 32$ Moment EFN& \textbf{0.974 $\pm$ 0.000} & 416.1 $\pm$ 40.0 & \textbf{138.4 $\pm$ 11.7} & 690645 \\
$k = 4, L = 16$ Moment EFN& 0.972 $\pm$ 0.001 & \textbf{579.3 $\pm$ 118.4} & 132.2 $\pm$ 8.2 & 517461 \\
\hline
\hline
    \end{tabular}
    \caption{The same as \Tab{auc}, but with the top/QCD dataset rather than the quark/gluon dataset.}
    \label{tab:top_EFN}
\end{table*}

\begin{table*}[tph]
    \centering
    \begin{tabular}{l|c|c|c|c}
         Model & AUC & $1/\epsilon_{QCD}$ at $\epsilon_{top} = 0.3$ & $1/\epsilon_{QCD}$ at $\epsilon_{top} = 0.5$ & Trainable Parameters \\
         \hline
         \hline
 $k = 1, L = 1$  PFN& 0.937 $\pm$ 0.001 & 43.3 $\pm$ 0.5 & 25.1 $\pm$ 0.1 & 31206 \\
$k = 2, L = 1$ Moment PFN& 0.944 $\pm$ 0.000 & 59.8 $\pm$ 0.7 & 29.0 $\pm$ 0.2 & 31306 \\
$k = 3, L = 1$ Moment PFN& 0.944 $\pm$ 0.001 & \textbf{62.8 $\pm$ 1.3} & 29.1 $\pm$ 0.9 & 31406 \\
$k = 4, L = 1$ Moment PFN& \textbf{0.944 $\pm$ 0.000} & 62.2 $\pm$ 0.4 & \textbf{30.4 $\pm$ 0.9} & 31506 \\
\hline
\hline
$k = 1, L = 128$  PFN& 0.975 $\pm$ 0.001 & 351.8 $\pm$ 12.5 & 138.8 $\pm$ 7.0 & 89753 \\
$k = 2, L = 64$ Moment PFN& 0.976 $\pm$ 0.001 & 303.2 $\pm$ 26.9 & \textbf{156.4 $\pm$ 11.1} & 260185 \\
$k = 3, L = 32$ Moment PFN& 0.975 $\pm$ 0.000 & 321.2 $\pm$ 0.0 & 136.1 $\pm$ 6.2 & 690745 \\
$k = 4, L = 16$ Moment PFN& \textbf{0.976 $\pm$ 0.001} & \textbf{351.8 $\pm$ 12.5} & 146.0 $\pm$ 17.0 & 517561 \\
\hline
\hline
    \end{tabular}
    \caption{The same as \Tab{auc}, but with a Moment PFN rather than a Moment EFN and the top/QCD dataset rather
than the quark/gluon dataset.}
    \label{tab:top_PFN}
\end{table*}

\bibliography{refs, HEPML}

\begin{thebibliography}{85}%
\makeatletter
\providecommand \@ifxundefined [1]{%
 \@ifx{#1\undefined}
}%
\providecommand \@ifnum [1]{%
 \ifnum #1\expandafter \@firstoftwo
 \else \expandafter \@secondoftwo
 \fi
}%
\providecommand \@ifx [1]{%
 \ifx #1\expandafter \@firstoftwo
 \else \expandafter \@secondoftwo
 \fi
}%
\providecommand \natexlab [1]{#1}%
\providecommand \enquote  [1]{``#1''}%
\providecommand \bibnamefont  [1]{#1}%
\providecommand \bibfnamefont [1]{#1}%
\providecommand \citenamefont [1]{#1}%
\providecommand \href@noop [0]{\@secondoftwo}%
\providecommand \href [0]{\begingroup \@sanitize@url \@href}%
\providecommand \@href[1]{\@@startlink{#1}\@@href}%
\providecommand \@@href[1]{\endgroup#1\@@endlink}%
\providecommand \@sanitize@url [0]{\catcode `\\12\catcode `\$12\catcode
  `\&12\catcode `\#12\catcode `\^12\catcode `\_12\catcode `\%12\relax}%
\providecommand \@@startlink[1]{}%
\providecommand \@@endlink[0]{}%
\providecommand \url  [0]{\begingroup\@sanitize@url \@url }%
\providecommand \@url [1]{\endgroup\@href {#1}{\urlprefix }}%
\providecommand \urlprefix  [0]{URL }%
\providecommand \Eprint [0]{\href }%
\providecommand \doibase [0]{http://dx.doi.org/}%
\providecommand \selectlanguage [0]{\@gobble}%
\providecommand \bibinfo  [0]{\@secondoftwo}%
\providecommand \bibfield  [0]{\@secondoftwo}%
\providecommand \translation [1]{[#1]}%
\providecommand \BibitemOpen [0]{}%
\providecommand \bibitemStop [0]{}%
\providecommand \bibitemNoStop [0]{.\EOS\space}%
\providecommand \EOS [0]{\spacefactor3000\relax}%
\providecommand \BibitemShut  [1]{\csname bibitem#1\endcsname}%
\let\auto@bib@innerbib\@empty
\bibitem [{\citenamefont {Chakraborty}\ \emph {et~al.}(2017)\citenamefont
  {Chakraborty}, \citenamefont {Tomsett}, \citenamefont {Raghavendra},
  \citenamefont {Harborne}, \citenamefont {Alzantot}, \citenamefont {Cerutti},
  \citenamefont {Srivastava}, \citenamefont {Preece}, \citenamefont {Julier},
  \citenamefont {Rao}, \citenamefont {Kelley}, \citenamefont {Braines},
  \citenamefont {Sensoy}, \citenamefont {Willis},\ and\ \citenamefont
  {Gurram}}]{d2}%
  \BibitemOpen
  \bibfield  {author} {\bibinfo {author} {\bibfnamefont {Supriyo}\ \bibnamefont
  {Chakraborty}}, \bibinfo {author} {\bibfnamefont {Richard}\ \bibnamefont
  {Tomsett}}, \bibinfo {author} {\bibfnamefont {Ramya}\ \bibnamefont
  {Raghavendra}}, \bibinfo {author} {\bibfnamefont {Daniel}\ \bibnamefont
  {Harborne}}, \bibinfo {author} {\bibfnamefont {Moustafa}\ \bibnamefont
  {Alzantot}}, \bibinfo {author} {\bibfnamefont {Federico}\ \bibnamefont
  {Cerutti}}, \bibinfo {author} {\bibfnamefont {Mani}\ \bibnamefont
  {Srivastava}}, \bibinfo {author} {\bibfnamefont {Alun}\ \bibnamefont
  {Preece}}, \bibinfo {author} {\bibfnamefont {Simon}\ \bibnamefont {Julier}},
  \bibinfo {author} {\bibfnamefont {Raghuveer~M.}\ \bibnamefont {Rao}},
  \bibinfo {author} {\bibfnamefont {Troy~D.}\ \bibnamefont {Kelley}}, \bibinfo
  {author} {\bibfnamefont {Dave}\ \bibnamefont {Braines}}, \bibinfo {author}
  {\bibfnamefont {Murat}\ \bibnamefont {Sensoy}}, \bibinfo {author}
  {\bibfnamefont {Christopher~J.}\ \bibnamefont {Willis}}, \ and\ \bibinfo
  {author} {\bibfnamefont {Prudhvi}\ \bibnamefont {Gurram}},\ }\bibfield
  {title} {\enquote {\bibinfo {title} {Interpretability of deep learning
  models: A survey of results},}\ }in\ \href {\doibase
  10.1109/UIC-ATC.2017.8397411} {\emph {\bibinfo {booktitle} {2017 IEEE
  SmartWorld, Ubiquitous Intelligence \& Computing, Advanced \& Trusted
  Computed, Scalable Computing \& Communications, Cloud \& Big Data Computing,
  Internet of People and Smart City Innovation
  (SmartWorld/SCALCOM/UIC/ATC/CBDCom/IOP/SCI)}}}\ (\bibinfo {year} {2017})\
  pp.\ \bibinfo {pages} {1--6}\BibitemShut {NoStop}%
\bibitem [{\citenamefont {Gilpin}\ \emph {et~al.}(2018)\citenamefont {Gilpin},
  \citenamefont {Bau}, \citenamefont {Yuan}, \citenamefont {Bajwa},
  \citenamefont {Specter},\ and\ \citenamefont {Kagal}}]{d3}%
  \BibitemOpen
  \bibfield  {author} {\bibinfo {author} {\bibfnamefont {Leilani~H.}\
  \bibnamefont {Gilpin}}, \bibinfo {author} {\bibfnamefont {David}\
  \bibnamefont {Bau}}, \bibinfo {author} {\bibfnamefont {Ben~Z.}\ \bibnamefont
  {Yuan}}, \bibinfo {author} {\bibfnamefont {Ayesha}\ \bibnamefont {Bajwa}},
  \bibinfo {author} {\bibfnamefont {Michael}\ \bibnamefont {Specter}}, \ and\
  \bibinfo {author} {\bibfnamefont {Lalana}\ \bibnamefont {Kagal}},\ }\href
  {\doibase 10.48550/ARXIV.1806.00069} {\enquote {\bibinfo {title} {Explaining
  explanations: An overview of interpretability of machine learning},}\ }
  (\bibinfo {year} {2018})\BibitemShut {NoStop}%
\bibitem [{\citenamefont {Zhang}\ \emph {et~al.}(2021)\citenamefont {Zhang},
  \citenamefont {Tino}, \citenamefont {Leonardis},\ and\ \citenamefont
  {Tang}}]{d1}%
  \BibitemOpen
  \bibfield  {author} {\bibinfo {author} {\bibfnamefont {Yu}~\bibnamefont
  {Zhang}}, \bibinfo {author} {\bibfnamefont {Peter}\ \bibnamefont {Tino}},
  \bibinfo {author} {\bibfnamefont {Ales}\ \bibnamefont {Leonardis}}, \ and\
  \bibinfo {author} {\bibfnamefont {Ke}~\bibnamefont {Tang}},\ }\bibfield
  {title} {\enquote {\bibinfo {title} {A survey on neural network
  interpretability},}\ }\href {\doibase 10.1109/tetci.2021.3100641} {\bibfield
  {journal} {\bibinfo  {journal} {{IEEE} Transactions on Emerging Topics in
  Computational Intelligence}\ }\textbf {\bibinfo {volume} {5}},\ \bibinfo
  {pages} {726--742} (\bibinfo {year} {2021})}\BibitemShut {NoStop}%
\bibitem [{\citenamefont {Molnar}\ \emph {et~al.}(2020)\citenamefont {Molnar},
  \citenamefont {Casalicchio},\ and\ \citenamefont {Bischl}}]{Molnar_2020}%
  \BibitemOpen
  \bibfield  {author} {\bibinfo {author} {\bibfnamefont {Christoph}\
  \bibnamefont {Molnar}}, \bibinfo {author} {\bibfnamefont {Giuseppe}\
  \bibnamefont {Casalicchio}}, \ and\ \bibinfo {author} {\bibfnamefont {Bernd}\
  \bibnamefont {Bischl}},\ }\enquote {\bibinfo {title} {Interpretable machine
  learning – a brief history, state-of-the-art and challenges},}\ in\ \href
  {\doibase 10.1007/978-3-030-65965-3_28} {\emph {\bibinfo {booktitle}
  {Communications in Computer and Information Science}}}\ (\bibinfo
  {publisher} {Springer International Publishing},\ \bibinfo {year} {2020})\
  p.\ \bibinfo {pages} {417–431}\BibitemShut {NoStop}%
\bibitem [{\citenamefont {Rudin}\ \emph {et~al.}(2021)\citenamefont {Rudin},
  \citenamefont {Chen}, \citenamefont {Chen}, \citenamefont {Huang},
  \citenamefont {Semenova},\ and\ \citenamefont
  {Zhong}}]{rudin2021interpretable}%
  \BibitemOpen
  \bibfield  {author} {\bibinfo {author} {\bibfnamefont {Cynthia}\ \bibnamefont
  {Rudin}}, \bibinfo {author} {\bibfnamefont {Chaofan}\ \bibnamefont {Chen}},
  \bibinfo {author} {\bibfnamefont {Zhi}\ \bibnamefont {Chen}}, \bibinfo
  {author} {\bibfnamefont {Haiyang}\ \bibnamefont {Huang}}, \bibinfo {author}
  {\bibfnamefont {Lesia}\ \bibnamefont {Semenova}}, \ and\ \bibinfo {author}
  {\bibfnamefont {Chudi}\ \bibnamefont {Zhong}},\ }\href@noop {} {\enquote
  {\bibinfo {title} {Interpretable machine learning: Fundamental principles and
  10 grand challenges},}\ } (\bibinfo {year} {2021}),\ \Eprint
  {http://arxiv.org/abs/2103.11251} {arXiv:2103.11251 [cs.LG]} \BibitemShut
  {NoStop}%
\bibitem [{\citenamefont {Dillon}\ \emph {et~al.}(2021)\citenamefont {Dillon},
  \citenamefont {Kasieczka}, \citenamefont {Olischlager}, \citenamefont
  {Plehn}, \citenamefont {Sorrenson},\ and\ \citenamefont
  {Vogel}}]{Dillon:2021gag}%
  \BibitemOpen
  \bibfield  {author} {\bibinfo {author} {\bibfnamefont {Barry~M.}\
  \bibnamefont {Dillon}}, \bibinfo {author} {\bibfnamefont {Gregor}\
  \bibnamefont {Kasieczka}}, \bibinfo {author} {\bibfnamefont {Hans}\
  \bibnamefont {Olischlager}}, \bibinfo {author} {\bibfnamefont {Tilman}\
  \bibnamefont {Plehn}}, \bibinfo {author} {\bibfnamefont {Peter}\ \bibnamefont
  {Sorrenson}}, \ and\ \bibinfo {author} {\bibfnamefont {Lorenz}\ \bibnamefont
  {Vogel}},\ }\bibfield  {title} {\enquote {\bibinfo {title} {{Symmetries,
  Safety, and Self-Supervision}},}\ }\href@noop {} {\  (\bibinfo {year}
  {2021})},\ \Eprint {http://arxiv.org/abs/2108.04253} {arXiv:2108.04253
  [hep-ph]} \BibitemShut {NoStop}%
\bibitem [{\citenamefont {Bogatskiy}\ \emph
  {et~al.}(2022{\natexlab{a}})\citenamefont {Bogatskiy} \emph
  {et~al.}}]{Bogatskiy:2022hub}%
  \BibitemOpen
  \bibfield  {author} {\bibinfo {author} {\bibfnamefont {Alexander}\
  \bibnamefont {Bogatskiy}} \emph {et~al.},\ }\bibfield  {title} {\enquote
  {\bibinfo {title} {{Symmetry Group Equivariant Architectures for Physics}},}\
  }in\ \href@noop {} {\emph {\bibinfo {booktitle} {{2022 Snowmass Summer
  Study}}}}\ (\bibinfo {year} {2022})\ \Eprint
  {http://arxiv.org/abs/2203.06153} {arXiv:2203.06153 [cs.LG]} \BibitemShut
  {NoStop}%
\bibitem [{\citenamefont {Gao}\ and\ \citenamefont
  {Guan}(2023)}]{gao2023interpretability}%
  \BibitemOpen
  \bibfield  {author} {\bibinfo {author} {\bibfnamefont {Lei}\ \bibnamefont
  {Gao}}\ and\ \bibinfo {author} {\bibfnamefont {Ling}\ \bibnamefont {Guan}},\
  }\href@noop {} {\enquote {\bibinfo {title} {Interpretability of machine
  learning: Recent advances and future prospects},}\ } (\bibinfo {year}
  {2023}),\ \Eprint {http://arxiv.org/abs/2305.00537} {arXiv:2305.00537
  [cs.MM]} \BibitemShut {NoStop}%
\bibitem [{\citenamefont {Bogatskiy}\ \emph {et~al.}(2023)\citenamefont
  {Bogatskiy}, \citenamefont {Hoffman}, \citenamefont {Miller}, \citenamefont
  {Offermann},\ and\ \citenamefont {Liu}}]{Bogatskiy:2023nnw}%
  \BibitemOpen
  \bibfield  {author} {\bibinfo {author} {\bibfnamefont {Alexander}\
  \bibnamefont {Bogatskiy}}, \bibinfo {author} {\bibfnamefont {Timothy}\
  \bibnamefont {Hoffman}}, \bibinfo {author} {\bibfnamefont {David~W.}\
  \bibnamefont {Miller}}, \bibinfo {author} {\bibfnamefont {Jan~T.}\
  \bibnamefont {Offermann}}, \ and\ \bibinfo {author} {\bibfnamefont
  {Xiaoyang}\ \bibnamefont {Liu}},\ }\bibfield  {title} {\enquote {\bibinfo
  {title} {{Explainable Equivariant Neural Networks for Particle Physics:
  PELICAN}},}\ }\href@noop {} {\  (\bibinfo {year} {2023})},\ \Eprint
  {http://arxiv.org/abs/2307.16506} {arXiv:2307.16506 [hep-ph]} \BibitemShut
  {NoStop}%
\bibitem [{\citenamefont {Wayland}\ \emph {et~al.}(2024)\citenamefont
  {Wayland}, \citenamefont {Coupette},\ and\ \citenamefont
  {Rieck}}]{wayland2024mapping}%
  \BibitemOpen
  \bibfield  {author} {\bibinfo {author} {\bibfnamefont {Jeremy}\ \bibnamefont
  {Wayland}}, \bibinfo {author} {\bibfnamefont {Corinna}\ \bibnamefont
  {Coupette}}, \ and\ \bibinfo {author} {\bibfnamefont {Bastian}\ \bibnamefont
  {Rieck}},\ }\href@noop {} {\enquote {\bibinfo {title} {Mapping the multiverse
  of latent representations},}\ } (\bibinfo {year} {2024}),\ \Eprint
  {http://arxiv.org/abs/2402.01514} {arXiv:2402.01514 [cs.LG]} \BibitemShut
  {NoStop}%
\bibitem [{\citenamefont {Denby}(1988)}]{Denby:1987rk}%
  \BibitemOpen
  \bibfield  {author} {\bibinfo {author} {\bibfnamefont {Bruce~H.}\
  \bibnamefont {Denby}},\ }\bibfield  {title} {\enquote {\bibinfo {title}
  {{Neural Networks and Cellular Automata in Experimental High-energy
  Physics}},}\ }\href {\doibase 10.1016/0010-4655(88)90004-5} {\bibfield
  {journal} {\bibinfo  {journal} {Comput. Phys. Commun.}\ }\textbf {\bibinfo
  {volume} {49}},\ \bibinfo {pages} {429--448} (\bibinfo {year}
  {1988})}\BibitemShut {NoStop}%
\bibitem [{\citenamefont {Guest}\ \emph {et~al.}(2018)\citenamefont {Guest},
  \citenamefont {Cranmer},\ and\ \citenamefont {Whiteson}}]{Guest:2018yhq}%
  \BibitemOpen
  \bibfield  {author} {\bibinfo {author} {\bibfnamefont {Dan}\ \bibnamefont
  {Guest}}, \bibinfo {author} {\bibfnamefont {Kyle}\ \bibnamefont {Cranmer}}, \
  and\ \bibinfo {author} {\bibfnamefont {Daniel}\ \bibnamefont {Whiteson}},\
  }\bibfield  {title} {\enquote {\bibinfo {title} {{Deep Learning and its
  Application to LHC Physics}},}\ }\href {\doibase
  10.1146/annurev-nucl-101917-021019} {\  (\bibinfo {year} {2018}),\
  10.1146/annurev-nucl-101917-021019},\ \Eprint
  {http://arxiv.org/abs/1806.11484} {arXiv:1806.11484 [hep-ex]} \BibitemShut
  {NoStop}%
\bibitem [{\citenamefont {Butter}\ \emph {et~al.}(2018)\citenamefont {Butter},
  \citenamefont {Kasieczka}, \citenamefont {Plehn},\ and\ \citenamefont
  {Russell}}]{Butter:2017cot}%
  \BibitemOpen
  \bibfield  {author} {\bibinfo {author} {\bibfnamefont {Anja}\ \bibnamefont
  {Butter}}, \bibinfo {author} {\bibfnamefont {Gregor}\ \bibnamefont
  {Kasieczka}}, \bibinfo {author} {\bibfnamefont {Tilman}\ \bibnamefont
  {Plehn}}, \ and\ \bibinfo {author} {\bibfnamefont {Michael}\ \bibnamefont
  {Russell}},\ }\bibfield  {title} {\enquote {\bibinfo {title} {{Deep-learned
  Top Tagging with a Lorentz Layer}},}\ }\href {\doibase
  10.21468/SciPostPhys.5.3.028} {\bibfield  {journal} {\bibinfo  {journal}
  {SciPost Phys.}\ }\textbf {\bibinfo {volume} {5}},\ \bibinfo {pages} {028}
  (\bibinfo {year} {2018})},\ \Eprint {http://arxiv.org/abs/1707.08966}
  {arXiv:1707.08966 [hep-ph]} \BibitemShut {NoStop}%
\bibitem [{\citenamefont {Albertsson}\ \emph {et~al.}(2018)\citenamefont
  {Albertsson} \emph {et~al.}}]{Albertsson:2018maf}%
  \BibitemOpen
  \bibfield  {author} {\bibinfo {author} {\bibfnamefont {Kim}\ \bibnamefont
  {Albertsson}} \emph {et~al.},\ }\bibfield  {title} {\enquote {\bibinfo
  {title} {{Machine Learning in High Energy Physics Community White Paper}},}\
  }\href {\doibase 10.1088/1742-6596/1085/2/022008} {\  (\bibinfo {year}
  {2018}),\ 10.1088/1742-6596/1085/2/022008},\ \Eprint
  {http://arxiv.org/abs/1807.02876} {arXiv:1807.02876 [physics.comp-ph]}
  \BibitemShut {NoStop}%
\bibitem [{\citenamefont {Qu}\ and\ \citenamefont
  {Gouskos}(2020{\natexlab{a}})}]{PhysRevD.101.056019}%
  \BibitemOpen
  \bibfield  {author} {\bibinfo {author} {\bibfnamefont {Huilin}\ \bibnamefont
  {Qu}}\ and\ \bibinfo {author} {\bibfnamefont {Loukas}\ \bibnamefont
  {Gouskos}},\ }\bibfield  {title} {\enquote {\bibinfo {title} {Jet tagging via
  particle clouds},}\ }\href {\doibase 10.1103/PhysRevD.101.056019} {\bibfield
  {journal} {\bibinfo  {journal} {Phys. Rev. D}\ }\textbf {\bibinfo {volume}
  {101}},\ \bibinfo {pages} {056019} (\bibinfo {year}
  {2020}{\natexlab{a}})}\BibitemShut {NoStop}%
\bibitem [{\citenamefont {Bourilkov}(2020)}]{Bourilkov:2019yoi}%
  \BibitemOpen
  \bibfield  {author} {\bibinfo {author} {\bibfnamefont {Dimitri}\ \bibnamefont
  {Bourilkov}},\ }\bibfield  {title} {\enquote {\bibinfo {title} {{Machine and
  Deep Learning Applications in Particle Physics}},}\ }\href {\doibase
  10.1142/S0217751X19300199} {\bibfield  {journal} {\bibinfo  {journal} {Int.
  J. Mod. Phys. A}\ }\textbf {\bibinfo {volume} {34}},\ \bibinfo {pages}
  {1930019} (\bibinfo {year} {2020})},\ \Eprint
  {http://arxiv.org/abs/1912.08245} {arXiv:1912.08245 [physics.data-an]}
  \BibitemShut {NoStop}%
\bibitem [{\citenamefont {Gong}\ \emph {et~al.}(2022)\citenamefont {Gong},
  \citenamefont {Meng}, \citenamefont {Zhang}, \citenamefont {Qu},
  \citenamefont {Li}, \citenamefont {Qian}, \citenamefont {Du}, \citenamefont
  {Ma},\ and\ \citenamefont {Liu}}]{Gong:2022lye}%
  \BibitemOpen
  \bibfield  {author} {\bibinfo {author} {\bibfnamefont {Shiqi}\ \bibnamefont
  {Gong}}, \bibinfo {author} {\bibfnamefont {Qi}~\bibnamefont {Meng}}, \bibinfo
  {author} {\bibfnamefont {Jue}\ \bibnamefont {Zhang}}, \bibinfo {author}
  {\bibfnamefont {Huilin}\ \bibnamefont {Qu}}, \bibinfo {author} {\bibfnamefont
  {Congqiao}\ \bibnamefont {Li}}, \bibinfo {author} {\bibfnamefont {Sitian}\
  \bibnamefont {Qian}}, \bibinfo {author} {\bibfnamefont {Weitao}\ \bibnamefont
  {Du}}, \bibinfo {author} {\bibfnamefont {Zhi-Ming}\ \bibnamefont {Ma}}, \
  and\ \bibinfo {author} {\bibfnamefont {Tie-Yan}\ \bibnamefont {Liu}},\
  }\bibfield  {title} {\enquote {\bibinfo {title} {{An Efficient Lorentz
  Equivariant Graph Neural Network for Jet Tagging}},}\ }\href@noop {} {\
  (\bibinfo {year} {2022})},\ \Eprint {http://arxiv.org/abs/2201.08187}
  {arXiv:2201.08187 [hep-ph]} \BibitemShut {NoStop}%
\bibitem [{\citenamefont {Shlomi}\ \emph {et~al.}(2020)\citenamefont {Shlomi},
  \citenamefont {Battaglia},\ and\ \citenamefont {Vlimant}}]{Shlomi:2020gdn}%
  \BibitemOpen
  \bibfield  {author} {\bibinfo {author} {\bibfnamefont {Jonathan}\
  \bibnamefont {Shlomi}}, \bibinfo {author} {\bibfnamefont {Peter}\
  \bibnamefont {Battaglia}}, \ and\ \bibinfo {author} {\bibfnamefont
  {Jean-Roch}\ \bibnamefont {Vlimant}},\ }\bibfield  {title} {\enquote
  {\bibinfo {title} {{Graph Neural Networks in Particle Physics}},}\ }\href
  {\doibase 10.1088/2632-2153/abbf9a} {\  (\bibinfo {year} {2020}),\
  10.1088/2632-2153/abbf9a},\ \Eprint {http://arxiv.org/abs/2007.13681}
  {arXiv:2007.13681 [hep-ex]} \BibitemShut {NoStop}%
\bibitem [{\citenamefont {Chakraborty}\ \emph {et~al.}(2020)\citenamefont
  {Chakraborty}, \citenamefont {Lim}, \citenamefont {Nojiri},\ and\
  \citenamefont {Takeuchi}}]{Chakraborty:2020yfc}%
  \BibitemOpen
  \bibfield  {author} {\bibinfo {author} {\bibfnamefont {Amit}\ \bibnamefont
  {Chakraborty}}, \bibinfo {author} {\bibfnamefont {Sung~Hak}\ \bibnamefont
  {Lim}}, \bibinfo {author} {\bibfnamefont {Mihoko~M.}\ \bibnamefont {Nojiri}},
  \ and\ \bibinfo {author} {\bibfnamefont {Michihisa}\ \bibnamefont
  {Takeuchi}},\ }\bibfield  {title} {\enquote {\bibinfo {title} {{Neural
  Network-based Top Tagger with Two-Point Energy Correlations and Geometry of
  Soft Emissions}},}\ }\href {\doibase 10.1007/JHEP07(2020)111} {\  (\bibinfo
  {year} {2020}),\ 10.1007/JHEP07(2020)111},\ \Eprint
  {http://arxiv.org/abs/2003.11787} {arXiv:2003.11787 [hep-ph]} \BibitemShut
  {NoStop}%
\bibitem [{\citenamefont {Butter}\ and\ \citenamefont
  {Plehn}(2020)}]{Butter:2020tvl}%
  \BibitemOpen
  \bibfield  {author} {\bibinfo {author} {\bibfnamefont {Anja}\ \bibnamefont
  {Butter}}\ and\ \bibinfo {author} {\bibfnamefont {Tilman}\ \bibnamefont
  {Plehn}},\ }\bibfield  {title} {\enquote {\bibinfo {title} {{Generative
  Networks for LHC events}},}\ }\href@noop {} {\  (\bibinfo {year} {2020})},\
  \Eprint {http://arxiv.org/abs/2008.08558} {arXiv:2008.08558 [hep-ph]}
  \BibitemShut {NoStop}%
\bibitem [{\citenamefont {Kagan}(2020)}]{Kagan:2020yrm}%
  \BibitemOpen
  \bibfield  {author} {\bibinfo {author} {\bibfnamefont {Michael}\ \bibnamefont
  {Kagan}},\ }\bibfield  {title} {\enquote {\bibinfo {title} {{Image-Based Jet
  Analysis}},}\ }\href@noop {} {\  (\bibinfo {year} {2020})},\ \Eprint
  {http://arxiv.org/abs/2012.09719} {arXiv:2012.09719 [physics.data-an]}
  \BibitemShut {NoStop}%
\bibitem [{\citenamefont {Lim}\ and\ \citenamefont
  {Nojiri}(2020)}]{Lim:2020igi}%
  \BibitemOpen
  \bibfield  {author} {\bibinfo {author} {\bibfnamefont {Sung~Hak}\
  \bibnamefont {Lim}}\ and\ \bibinfo {author} {\bibfnamefont {Mihoko~M.}\
  \bibnamefont {Nojiri}},\ }\bibfield  {title} {\enquote {\bibinfo {title}
  {{Morphology for Jet Classification}},}\ }\href@noop {} {\  (\bibinfo {year}
  {2020})},\ \Eprint {http://arxiv.org/abs/2010.13469} {arXiv:2010.13469
  [hep-ph]} \BibitemShut {NoStop}%
\bibitem [{\citenamefont {Dreyer}\ and\ \citenamefont
  {Qu}(2021)}]{Dreyer:2020brq}%
  \BibitemOpen
  \bibfield  {author} {\bibinfo {author} {\bibfnamefont {Fr\'ed\'eric~A.}\
  \bibnamefont {Dreyer}}\ and\ \bibinfo {author} {\bibfnamefont {Huilin}\
  \bibnamefont {Qu}},\ }\bibfield  {title} {\enquote {\bibinfo {title} {{Jet
  tagging in the Lund plane with graph networks}},}\ }\href {\doibase
  10.1007/JHEP03(2021)052} {\bibfield  {journal} {\bibinfo  {journal} {JHEP}\
  }\textbf {\bibinfo {volume} {03}},\ \bibinfo {pages} {052} (\bibinfo {year}
  {2021})},\ \Eprint {http://arxiv.org/abs/2012.08526} {arXiv:2012.08526
  [hep-ph]} \BibitemShut {NoStop}%
\bibitem [{\citenamefont {Karagiorgi}\ \emph {et~al.}(2021)\citenamefont
  {Karagiorgi}, \citenamefont {Kasieczka}, \citenamefont {Kravitz},
  \citenamefont {Nachman},\ and\ \citenamefont {Shih}}]{Karagiorgi:2021ngt}%
  \BibitemOpen
  \bibfield  {author} {\bibinfo {author} {\bibfnamefont {Georgia}\ \bibnamefont
  {Karagiorgi}}, \bibinfo {author} {\bibfnamefont {Gregor}\ \bibnamefont
  {Kasieczka}}, \bibinfo {author} {\bibfnamefont {Scott}\ \bibnamefont
  {Kravitz}}, \bibinfo {author} {\bibfnamefont {Benjamin}\ \bibnamefont
  {Nachman}}, \ and\ \bibinfo {author} {\bibfnamefont {David}\ \bibnamefont
  {Shih}},\ }\bibfield  {title} {\enquote {\bibinfo {title} {{Machine Learning
  in the Search for New Fundamental Physics}},}\ }\href@noop {} {\  (\bibinfo
  {year} {2021})},\ \Eprint {http://arxiv.org/abs/2112.03769} {arXiv:2112.03769
  [hep-ph]} \BibitemShut {NoStop}%
\bibitem [{\citenamefont {Schwartz}(2021)}]{Schwartz:2021ftp}%
  \BibitemOpen
  \bibfield  {author} {\bibinfo {author} {\bibfnamefont {Matthew~D.}\
  \bibnamefont {Schwartz}},\ }\bibfield  {title} {\enquote {\bibinfo {title}
  {{Modern Machine Learning and Particle Physics}},}\ }\href@noop {} {\
  (\bibinfo {year} {2021})},\ \Eprint {http://arxiv.org/abs/2103.12226}
  {arXiv:2103.12226 [hep-ph]} \BibitemShut {NoStop}%
\bibitem [{\citenamefont {Qu}\ \emph {et~al.}(2022)\citenamefont {Qu},
  \citenamefont {Li},\ and\ \citenamefont {Qian}}]{Qu:2022mxj}%
  \BibitemOpen
  \bibfield  {author} {\bibinfo {author} {\bibfnamefont {Huilin}\ \bibnamefont
  {Qu}}, \bibinfo {author} {\bibfnamefont {Congqiao}\ \bibnamefont {Li}}, \
  and\ \bibinfo {author} {\bibfnamefont {Sitian}\ \bibnamefont {Qian}},\
  }\bibfield  {title} {\enquote {\bibinfo {title} {{Particle Transformer for
  Jet Tagging}},}\ }\href@noop {} {\  (\bibinfo {year} {2022})},\ \Eprint
  {http://arxiv.org/abs/2202.03772} {arXiv:2202.03772 [hep-ph]} \BibitemShut
  {NoStop}%
\bibitem [{\citenamefont {Baldi}\ \emph {et~al.}(2022)\citenamefont {Baldi},
  \citenamefont {Sadowski},\ and\ \citenamefont {Whiteson}}]{Baldi:2022okj}%
  \BibitemOpen
  \bibfield  {author} {\bibinfo {author} {\bibfnamefont {Pierre}\ \bibnamefont
  {Baldi}}, \bibinfo {author} {\bibfnamefont {Peter}\ \bibnamefont {Sadowski}},
  \ and\ \bibinfo {author} {\bibfnamefont {Daniel}\ \bibnamefont {Whiteson}},\
  }\bibfield  {title} {\enquote {\bibinfo {title} {{Deep Learning From Four
  Vectors}},}\ }\href@noop {} {\  (\bibinfo {year} {2022})},\ \Eprint
  {http://arxiv.org/abs/2203.03067} {arXiv:2203.03067 [hep-ex]} \BibitemShut
  {NoStop}%
\bibitem [{\citenamefont {Plehn}\ \emph {et~al.}(2022)\citenamefont {Plehn},
  \citenamefont {Butter}, \citenamefont {Dillon},\ and\ \citenamefont
  {Krause}}]{Plehn:2022ftl}%
  \BibitemOpen
  \bibfield  {author} {\bibinfo {author} {\bibfnamefont {Tilman}\ \bibnamefont
  {Plehn}}, \bibinfo {author} {\bibfnamefont {Anja}\ \bibnamefont {Butter}},
  \bibinfo {author} {\bibfnamefont {Barry}\ \bibnamefont {Dillon}}, \ and\
  \bibinfo {author} {\bibfnamefont {Claudius}\ \bibnamefont {Krause}},\
  }\bibfield  {title} {\enquote {\bibinfo {title} {{Modern Machine Learning for
  LHC Physicists}},}\ }\href@noop {} {\  (\bibinfo {year} {2022})},\ \Eprint
  {http://arxiv.org/abs/2211.01421} {arXiv:2211.01421 [hep-ph]} \BibitemShut
  {NoStop}%
\bibitem [{\citenamefont {Carleo}\ \emph {et~al.}(2019)\citenamefont {Carleo},
  \citenamefont {Cirac}, \citenamefont {Cranmer}, \citenamefont {Daudet},
  \citenamefont {Schuld}, \citenamefont {Tishby}, \citenamefont
  {Vogt-Maranto},\ and\ \citenamefont {Zdeborov\'a}}]{RevModPhys.91.045002}%
  \BibitemOpen
  \bibfield  {author} {\bibinfo {author} {\bibfnamefont {Giuseppe}\
  \bibnamefont {Carleo}}, \bibinfo {author} {\bibfnamefont {Ignacio}\
  \bibnamefont {Cirac}}, \bibinfo {author} {\bibfnamefont {Kyle}\ \bibnamefont
  {Cranmer}}, \bibinfo {author} {\bibfnamefont {Laurent}\ \bibnamefont
  {Daudet}}, \bibinfo {author} {\bibfnamefont {Maria}\ \bibnamefont {Schuld}},
  \bibinfo {author} {\bibfnamefont {Naftali}\ \bibnamefont {Tishby}}, \bibinfo
  {author} {\bibfnamefont {Leslie}\ \bibnamefont {Vogt-Maranto}}, \ and\
  \bibinfo {author} {\bibfnamefont {Lenka}\ \bibnamefont {Zdeborov\'a}},\
  }\bibfield  {title} {\enquote {\bibinfo {title} {Machine learning and the
  physical sciences},}\ }\href {\doibase 10.1103/RevModPhys.91.045002}
  {\bibfield  {journal} {\bibinfo  {journal} {Rev. Mod. Phys.}\ }\textbf
  {\bibinfo {volume} {91}},\ \bibinfo {pages} {045002} (\bibinfo {year}
  {2019})}\BibitemShut {NoStop}%
\bibitem [{\citenamefont {Badger}\ \emph {et~al.}(2022)\citenamefont {Badger}
  \emph {et~al.}}]{Butter:2022rso}%
  \BibitemOpen
  \bibfield  {author} {\bibinfo {author} {\bibfnamefont {Simon}\ \bibnamefont
  {Badger}} \emph {et~al.},\ }\bibfield  {title} {\enquote {\bibinfo {title}
  {{Machine Learning and LHC Event Generation}},}\ }\href@noop {} {\  (\bibinfo
  {year} {2022})},\ \Eprint {http://arxiv.org/abs/2203.07460} {arXiv:2203.07460
  [hep-ph]} \BibitemShut {NoStop}%
\bibitem [{\citenamefont {Bogatskiy}\ \emph
  {et~al.}(2022{\natexlab{b}})\citenamefont {Bogatskiy}, \citenamefont
  {Hoffman}, \citenamefont {Miller},\ and\ \citenamefont
  {Offermann}}]{Bogatskiy:2022czk}%
  \BibitemOpen
  \bibfield  {author} {\bibinfo {author} {\bibfnamefont {Alexander}\
  \bibnamefont {Bogatskiy}}, \bibinfo {author} {\bibfnamefont {Timothy}\
  \bibnamefont {Hoffman}}, \bibinfo {author} {\bibfnamefont {David~W.}\
  \bibnamefont {Miller}}, \ and\ \bibinfo {author} {\bibfnamefont {Jan~T.}\
  \bibnamefont {Offermann}},\ }\bibfield  {title} {\enquote {\bibinfo {title}
  {{PELICAN: Permutation Equivariant and Lorentz Invariant or Covariant
  Aggregator Network for Particle Physics}},}\ }\href@noop {} {\  (\bibinfo
  {year} {2022}{\natexlab{b}})},\ \Eprint {http://arxiv.org/abs/2211.00454}
  {arXiv:2211.00454 [hep-ph]} \BibitemShut {NoStop}%
\bibitem [{\citenamefont {Atkinson}\ \emph {et~al.}(2022)\citenamefont
  {Atkinson}, \citenamefont {Bhardwaj}, \citenamefont {Englert}, \citenamefont
  {Konar}, \citenamefont {Ngairangbam},\ and\ \citenamefont
  {Spannowsky}}]{Atkinson:2022uzb}%
  \BibitemOpen
  \bibfield  {author} {\bibinfo {author} {\bibfnamefont {Oliver}\ \bibnamefont
  {Atkinson}}, \bibinfo {author} {\bibfnamefont {Akanksha}\ \bibnamefont
  {Bhardwaj}}, \bibinfo {author} {\bibfnamefont {Christoph}\ \bibnamefont
  {Englert}}, \bibinfo {author} {\bibfnamefont {Partha}\ \bibnamefont {Konar}},
  \bibinfo {author} {\bibfnamefont {Vishal~S.}\ \bibnamefont {Ngairangbam}}, \
  and\ \bibinfo {author} {\bibfnamefont {Michael}\ \bibnamefont {Spannowsky}},\
  }\bibfield  {title} {\enquote {\bibinfo {title} {{IRC-Safe Graph Autoencoder
  for Unsupervised Anomaly Detection}},}\ }\href {\doibase
  10.3389/frai.2022.943135} {\bibfield  {journal} {\bibinfo  {journal} {Front.
  Artif. Intell.}\ }\textbf {\bibinfo {volume} {5}},\ \bibinfo {pages} {943135}
  (\bibinfo {year} {2022})},\ \Eprint {http://arxiv.org/abs/2204.12231}
  {arXiv:2204.12231 [hep-ph]} \BibitemShut {NoStop}%
\bibitem [{\citenamefont {Bhardwaj}\ \emph {et~al.}(2024)\citenamefont
  {Bhardwaj}, \citenamefont {Englert}, \citenamefont {Naskar}, \citenamefont
  {Ngairangbam},\ and\ \citenamefont {Spannowsky}}]{Bhardwaj:2024djv}%
  \BibitemOpen
  \bibfield  {author} {\bibinfo {author} {\bibfnamefont {Akanksha}\
  \bibnamefont {Bhardwaj}}, \bibinfo {author} {\bibfnamefont {Christoph}\
  \bibnamefont {Englert}}, \bibinfo {author} {\bibfnamefont {Wrishik}\
  \bibnamefont {Naskar}}, \bibinfo {author} {\bibfnamefont {Vishal~S.}\
  \bibnamefont {Ngairangbam}}, \ and\ \bibinfo {author} {\bibfnamefont
  {Michael}\ \bibnamefont {Spannowsky}},\ }\bibfield  {title} {\enquote
  {\bibinfo {title} {{Equivariant, Safe and Sensitive -- Graph Networks for New
  Physics}},}\ }\href@noop {} {\  (\bibinfo {year} {2024})},\ \Eprint
  {http://arxiv.org/abs/2402.12449} {arXiv:2402.12449 [hep-ph]} \BibitemShut
  {NoStop}%
\bibitem [{\citenamefont {Komiske}\ \emph
  {et~al.}(2019{\natexlab{a}})\citenamefont {Komiske}, \citenamefont
  {Metodiev},\ and\ \citenamefont {Thaler}}]{efn}%
  \BibitemOpen
  \bibfield  {author} {\bibinfo {author} {\bibfnamefont {Patrick~T.}\
  \bibnamefont {Komiske}}, \bibinfo {author} {\bibfnamefont {Eric~M.}\
  \bibnamefont {Metodiev}}, \ and\ \bibinfo {author} {\bibfnamefont {Jesse}\
  \bibnamefont {Thaler}},\ }\bibfield  {title} {\enquote {\bibinfo {title}
  {{Energy Flow Networks: Deep Sets for Particle Jets}},}\ }\href {\doibase
  10.1007/JHEP01(2019)121} {\bibfield  {journal} {\bibinfo  {journal} {JHEP}\
  }\textbf {\bibinfo {volume} {01}},\ \bibinfo {pages} {121} (\bibinfo {year}
  {2019}{\natexlab{a}})},\ \Eprint {http://arxiv.org/abs/1810.05165}
  {arXiv:1810.05165 [hep-ph]} \BibitemShut {NoStop}%
\bibitem [{\citenamefont {Butter}\ \emph {et~al.}(2019)\citenamefont {Butter}
  \emph {et~al.}}]{Kasieczka:2019dbj}%
  \BibitemOpen
  \bibfield  {author} {\bibinfo {author} {\bibfnamefont {Anja}\ \bibnamefont
  {Butter}} \emph {et~al.},\ }\bibfield  {title} {\enquote {\bibinfo {title}
  {{The Machine Learning Landscape of Top Taggers}},}\ }\href {\doibase
  10.21468/SciPostPhys.7.1.014} {\bibfield  {journal} {\bibinfo  {journal}
  {SciPost Phys.}\ }\textbf {\bibinfo {volume} {7}},\ \bibinfo {pages} {014}
  (\bibinfo {year} {2019})},\ \Eprint {http://arxiv.org/abs/1902.09914}
  {arXiv:1902.09914 [hep-ph]} \BibitemShut {NoStop}%
\bibitem [{ATL(2022)}]{ATLAS:2022qby}%
  \BibitemOpen
  \bibfield  {title} {\enquote {\bibinfo {title} {{Constituent-Based Top-Quark
  Tagging with the ATLAS Detector}},}\ }\href@noop {} {\  (\bibinfo {year}
  {2022})}\BibitemShut {NoStop}%
\bibitem [{\citenamefont {Zaheer}\ \emph {et~al.}(2017)\citenamefont {Zaheer},
  \citenamefont {Kottur}, \citenamefont {Ravanbakhsh}, \citenamefont {Poczos},
  \citenamefont {Salakhutdinov},\ and\ \citenamefont {Smola}}]{deepsets}%
  \BibitemOpen
  \bibfield  {author} {\bibinfo {author} {\bibfnamefont {Manzil}\ \bibnamefont
  {Zaheer}}, \bibinfo {author} {\bibfnamefont {Satwik}\ \bibnamefont {Kottur}},
  \bibinfo {author} {\bibfnamefont {Siamak}\ \bibnamefont {Ravanbakhsh}},
  \bibinfo {author} {\bibfnamefont {Barnabas}\ \bibnamefont {Poczos}}, \bibinfo
  {author} {\bibfnamefont {Ruslan}\ \bibnamefont {Salakhutdinov}}, \ and\
  \bibinfo {author} {\bibfnamefont {Alexander}\ \bibnamefont {Smola}},\ }\href
  {\doibase 10.48550/ARXIV.1703.06114} {\enquote {\bibinfo {title} {Deep
  sets},}\ } (\bibinfo {year} {2017})\BibitemShut {NoStop}%
\bibitem [{\citenamefont {Cranmer}\ \emph {et~al.}(2021)\citenamefont
  {Cranmer}, , \citenamefont {Kreisch}, \citenamefont {Pisani}, \citenamefont
  {Villaescusa-Navarro}, \citenamefont {Spergel},\ and\ \citenamefont
  {Ho}}]{histogrampooling}%
  \BibitemOpen
  \bibfield  {author} {\bibinfo {author} {\bibfnamefont {Miles}\ \bibnamefont
  {Cranmer}}, , \bibinfo {author} {\bibfnamefont {Christina}\ \bibnamefont
  {Kreisch}}, \bibinfo {author} {\bibfnamefont {Alice}\ \bibnamefont {Pisani}},
  \bibinfo {author} {\bibfnamefont {Francisco}\ \bibnamefont
  {Villaescusa-Navarro}}, \bibinfo {author} {\bibfnamefont {David~N.}\
  \bibnamefont {Spergel}}, \ and\ \bibinfo {author} {\bibfnamefont {Shirley}\
  \bibnamefont {Ho}},\ }\bibfield  {title} {\enquote {\bibinfo {title}
  {Histogram pooling operators: An alternative for deep sets},}\ \ }(\bibinfo
  {year} {2021})\BibitemShut {NoStop}%
\bibitem [{\citenamefont {Shen}\ \emph {et~al.}(2023)\citenamefont {Shen},
  \citenamefont {Wang},\ and\ \citenamefont {Yang}}]{Shen:2023ofd}%
  \BibitemOpen
  \bibfield  {author} {\bibinfo {author} {\bibfnamefont {Wei}\ \bibnamefont
  {Shen}}, \bibinfo {author} {\bibfnamefont {Daohan}\ \bibnamefont {Wang}}, \
  and\ \bibinfo {author} {\bibfnamefont {Jin~Min}\ \bibnamefont {Yang}},\
  }\bibfield  {title} {\enquote {\bibinfo {title} {{Hierarchical high-point
  Energy Flow Network for jet tagging}},}\ }\href {\doibase
  10.1007/JHEP09(2023)135} {\bibfield  {journal} {\bibinfo  {journal} {JHEP}\
  }\textbf {\bibinfo {volume} {09}},\ \bibinfo {pages} {135} (\bibinfo {year}
  {2023})},\ \Eprint {http://arxiv.org/abs/2308.08300} {arXiv:2308.08300
  [hep-ph]} \BibitemShut {NoStop}%
\bibitem [{\citenamefont {Bright-Thonney}\ \emph {et~al.}(2023)\citenamefont
  {Bright-Thonney}, \citenamefont {Nachman},\ and\ \citenamefont
  {Thaler}}]{Bright-Thonney:2023gdl}%
  \BibitemOpen
  \bibfield  {author} {\bibinfo {author} {\bibfnamefont {Samuel}\ \bibnamefont
  {Bright-Thonney}}, \bibinfo {author} {\bibfnamefont {Benjamin}\ \bibnamefont
  {Nachman}}, \ and\ \bibinfo {author} {\bibfnamefont {Jesse}\ \bibnamefont
  {Thaler}},\ }\bibfield  {title} {\enquote {\bibinfo {title} {{Safe but
  Incalculable: Energy-weighting is not all you need}},}\ }\href@noop {} {\
  (\bibinfo {year} {2023})},\ \Eprint {http://arxiv.org/abs/2311.07652}
  {arXiv:2311.07652 [hep-ph]} \BibitemShut {NoStop}%
\bibitem [{\citenamefont {Berger}\ \emph {et~al.}(2003)\citenamefont {Berger},
  \citenamefont {Kucs},\ and\ \citenamefont {Sterman}}]{Berger:2003iw}%
  \BibitemOpen
  \bibfield  {author} {\bibinfo {author} {\bibfnamefont {Carola~F.}\
  \bibnamefont {Berger}}, \bibinfo {author} {\bibfnamefont {Tibor}\
  \bibnamefont {Kucs}}, \ and\ \bibinfo {author} {\bibfnamefont {George~F.}\
  \bibnamefont {Sterman}},\ }\bibfield  {title} {\enquote {\bibinfo {title}
  {{Event shape / energy flow correlations}},}\ }\href {\doibase
  10.1103/PhysRevD.68.014012} {\bibfield  {journal} {\bibinfo  {journal} {Phys.
  Rev. D}\ }\textbf {\bibinfo {volume} {68}},\ \bibinfo {pages} {014012}
  (\bibinfo {year} {2003})},\ \Eprint {http://arxiv.org/abs/hep-ph/0303051}
  {arXiv:hep-ph/0303051} \BibitemShut {NoStop}%
\bibitem [{\citenamefont {Berger}\ and\ \citenamefont
  {Magnea}(2004)}]{Berger:2004xf}%
  \BibitemOpen
  \bibfield  {author} {\bibinfo {author} {\bibfnamefont {Carola~F.}\
  \bibnamefont {Berger}}\ and\ \bibinfo {author} {\bibfnamefont {Lorenzo}\
  \bibnamefont {Magnea}},\ }\bibfield  {title} {\enquote {\bibinfo {title}
  {{Scaling of power corrections for angularities from dressed gluon
  exponentiation}},}\ }\href {\doibase 10.1103/PhysRevD.70.094010} {\bibfield
  {journal} {\bibinfo  {journal} {Phys. Rev. D}\ }\textbf {\bibinfo {volume}
  {70}},\ \bibinfo {pages} {094010} (\bibinfo {year} {2004})},\ \Eprint
  {http://arxiv.org/abs/hep-ph/0407024} {arXiv:hep-ph/0407024} \BibitemShut
  {NoStop}%
\bibitem [{\citenamefont {Tkachov}(1997)}]{Tkachov:1995kk}%
  \BibitemOpen
  \bibfield  {author} {\bibinfo {author} {\bibfnamefont {Fyodor~V.}\
  \bibnamefont {Tkachov}},\ }\bibfield  {title} {\enquote {\bibinfo {title}
  {{Measuring multi - jet structure of hadronic energy flow or What is a
  jet?}}}\ }\href {\doibase 10.1142/S0217751X97002899} {\bibfield  {journal}
  {\bibinfo  {journal} {Int. J. Mod. Phys. A}\ }\textbf {\bibinfo {volume}
  {12}},\ \bibinfo {pages} {5411--5529} (\bibinfo {year} {1997})},\ \Eprint
  {http://arxiv.org/abs/hep-ph/9601308} {arXiv:hep-ph/9601308} \BibitemShut
  {NoStop}%
\bibitem [{\citenamefont {Sveshnikov}\ and\ \citenamefont
  {Tkachov}(1996)}]{Sveshnikov:1995vi}%
  \BibitemOpen
  \bibfield  {author} {\bibinfo {author} {\bibfnamefont {N.~A.}\ \bibnamefont
  {Sveshnikov}}\ and\ \bibinfo {author} {\bibfnamefont {F.~V.}\ \bibnamefont
  {Tkachov}},\ }\bibfield  {title} {\enquote {\bibinfo {title} {{Jets and
  quantum field theory}},}\ }\href {\doibase 10.1016/0370-2693(96)00558-8}
  {\bibfield  {journal} {\bibinfo  {journal} {Phys. Lett. B}\ }\textbf
  {\bibinfo {volume} {382}},\ \bibinfo {pages} {403--408} (\bibinfo {year}
  {1996})},\ \Eprint {http://arxiv.org/abs/hep-ph/9512370}
  {arXiv:hep-ph/9512370} \BibitemShut {NoStop}%
\bibitem [{\citenamefont {Tkachov}(2002)}]{Tkachov:1999py}%
  \BibitemOpen
  \bibfield  {author} {\bibinfo {author} {\bibfnamefont {Fyodor~V.}\
  \bibnamefont {Tkachov}},\ }\bibfield  {title} {\enquote {\bibinfo {title} {{A
  Theory of jet definition}},}\ }\href {\doibase 10.1142/S0217751X02009977}
  {\bibfield  {journal} {\bibinfo  {journal} {Int. J. Mod. Phys. A}\ }\textbf
  {\bibinfo {volume} {17}},\ \bibinfo {pages} {2783--2884} (\bibinfo {year}
  {2002})},\ \Eprint {http://arxiv.org/abs/hep-ph/9901444}
  {arXiv:hep-ph/9901444} \BibitemShut {NoStop}%
\bibitem [{\citenamefont {Hofman}\ and\ \citenamefont
  {Maldacena}(2008)}]{Hofman:2008ar}%
  \BibitemOpen
  \bibfield  {author} {\bibinfo {author} {\bibfnamefont {Diego~M.}\
  \bibnamefont {Hofman}}\ and\ \bibinfo {author} {\bibfnamefont {Juan}\
  \bibnamefont {Maldacena}},\ }\bibfield  {title} {\enquote {\bibinfo {title}
  {{Conformal collider physics: Energy and charge correlations}},}\ }\href
  {\doibase 10.1088/1126-6708/2008/05/012} {\bibfield  {journal} {\bibinfo
  {journal} {JHEP}\ }\textbf {\bibinfo {volume} {05}},\ \bibinfo {pages} {012}
  (\bibinfo {year} {2008})},\ \Eprint {http://arxiv.org/abs/0803.1467}
  {arXiv:0803.1467 [hep-th]} \BibitemShut {NoStop}%
\bibitem [{\citenamefont {Ba}\ \emph {et~al.}(2023)\citenamefont {Ba},
  \citenamefont {Dogra}, \citenamefont {Gambhir}, \citenamefont {Tasissa},\
  and\ \citenamefont {Thaler}}]{Ba:2023hix}%
  \BibitemOpen
  \bibfield  {author} {\bibinfo {author} {\bibfnamefont {Demba}\ \bibnamefont
  {Ba}}, \bibinfo {author} {\bibfnamefont {Akshunna~S.}\ \bibnamefont {Dogra}},
  \bibinfo {author} {\bibfnamefont {Rikab}\ \bibnamefont {Gambhir}}, \bibinfo
  {author} {\bibfnamefont {Abiy}\ \bibnamefont {Tasissa}}, \ and\ \bibinfo
  {author} {\bibfnamefont {Jesse}\ \bibnamefont {Thaler}},\ }\bibfield  {title}
  {\enquote {\bibinfo {title} {{SHAPER: can you hear the shape of a jet?}}}\
  }\href {\doibase 10.1007/JHEP06(2023)195} {\bibfield  {journal} {\bibinfo
  {journal} {JHEP}\ }\textbf {\bibinfo {volume} {06}},\ \bibinfo {pages} {195}
  (\bibinfo {year} {2023})},\ \Eprint {http://arxiv.org/abs/2302.12266}
  {arXiv:2302.12266 [hep-ph]} \BibitemShut {NoStop}%
\bibitem [{\citenamefont {Komiske}\ \emph
  {et~al.}(2018{\natexlab{a}})\citenamefont {Komiske}, \citenamefont
  {Metodiev},\ and\ \citenamefont {Thaler}}]{Komiske:2017aww}%
  \BibitemOpen
  \bibfield  {author} {\bibinfo {author} {\bibfnamefont {Patrick~T.}\
  \bibnamefont {Komiske}}, \bibinfo {author} {\bibfnamefont {Eric~M.}\
  \bibnamefont {Metodiev}}, \ and\ \bibinfo {author} {\bibfnamefont {Jesse}\
  \bibnamefont {Thaler}},\ }\bibfield  {title} {\enquote {\bibinfo {title}
  {{Energy flow polynomials: A complete linear basis for jet substructure}},}\
  }\href {\doibase 10.1007/JHEP04(2018)013} {\bibfield  {journal} {\bibinfo
  {journal} {JHEP}\ }\textbf {\bibinfo {volume} {04}},\ \bibinfo {pages} {013}
  (\bibinfo {year} {2018}{\natexlab{a}})},\ \Eprint
  {http://arxiv.org/abs/1712.07124} {arXiv:1712.07124 [hep-ph]} \BibitemShut
  {NoStop}%
\bibitem [{\citenamefont {Bahdanau}\ \emph {et~al.}(2016)\citenamefont
  {Bahdanau}, \citenamefont {Cho},\ and\ \citenamefont
  {Bengio}}]{bahdanau2016neural}%
  \BibitemOpen
  \bibfield  {author} {\bibinfo {author} {\bibfnamefont {Dzmitry}\ \bibnamefont
  {Bahdanau}}, \bibinfo {author} {\bibfnamefont {Kyunghyun}\ \bibnamefont
  {Cho}}, \ and\ \bibinfo {author} {\bibfnamefont {Yoshua}\ \bibnamefont
  {Bengio}},\ }\href@noop {} {\enquote {\bibinfo {title} {Neural machine
  translation by jointly learning to align and translate},}\ } (\bibinfo {year}
  {2016}),\ \Eprint {http://arxiv.org/abs/1409.0473} {arXiv:1409.0473 [cs.CL]}
  \BibitemShut {NoStop}%
\bibitem [{\citenamefont {Cheng}\ \emph {et~al.}(2016)\citenamefont {Cheng},
  \citenamefont {Dong},\ and\ \citenamefont {Lapata}}]{cheng2016long}%
  \BibitemOpen
  \bibfield  {author} {\bibinfo {author} {\bibfnamefont {Jianpeng}\
  \bibnamefont {Cheng}}, \bibinfo {author} {\bibfnamefont {Li}~\bibnamefont
  {Dong}}, \ and\ \bibinfo {author} {\bibfnamefont {Mirella}\ \bibnamefont
  {Lapata}},\ }\href@noop {} {\enquote {\bibinfo {title} {Long short-term
  memory-networks for machine reading},}\ } (\bibinfo {year} {2016}),\ \Eprint
  {http://arxiv.org/abs/1601.06733} {arXiv:1601.06733 [cs.CL]} \BibitemShut
  {NoStop}%
\bibitem [{\citenamefont {Vaswani}\ \emph {et~al.}(2023)\citenamefont
  {Vaswani}, \citenamefont {Shazeer}, \citenamefont {Parmar}, \citenamefont
  {Uszkoreit}, \citenamefont {Jones}, \citenamefont {Gomez}, \citenamefont
  {Kaiser},\ and\ \citenamefont {Polosukhin}}]{vaswani2023attention}%
  \BibitemOpen
  \bibfield  {author} {\bibinfo {author} {\bibfnamefont {Ashish}\ \bibnamefont
  {Vaswani}}, \bibinfo {author} {\bibfnamefont {Noam}\ \bibnamefont {Shazeer}},
  \bibinfo {author} {\bibfnamefont {Niki}\ \bibnamefont {Parmar}}, \bibinfo
  {author} {\bibfnamefont {Jakob}\ \bibnamefont {Uszkoreit}}, \bibinfo {author}
  {\bibfnamefont {Llion}\ \bibnamefont {Jones}}, \bibinfo {author}
  {\bibfnamefont {Aidan~N.}\ \bibnamefont {Gomez}}, \bibinfo {author}
  {\bibfnamefont {Lukasz}\ \bibnamefont {Kaiser}}, \ and\ \bibinfo {author}
  {\bibfnamefont {Illia}\ \bibnamefont {Polosukhin}},\ }\href@noop {} {\enquote
  {\bibinfo {title} {Attention is all you need},}\ } (\bibinfo {year} {2023}),\
  \Eprint {http://arxiv.org/abs/1706.03762} {arXiv:1706.03762 [cs.CL]}
  \BibitemShut {NoStop}%
\bibitem [{\citenamefont {Gallicchio}\ and\ \citenamefont
  {Schwartz}(2011)}]{Gallicchio:2011xq}%
  \BibitemOpen
  \bibfield  {author} {\bibinfo {author} {\bibfnamefont {Jason}\ \bibnamefont
  {Gallicchio}}\ and\ \bibinfo {author} {\bibfnamefont {Matthew~D.}\
  \bibnamefont {Schwartz}},\ }\bibfield  {title} {\enquote {\bibinfo {title}
  {{Quark and Gluon Tagging at the LHC}},}\ }\href {\doibase
  10.1103/PhysRevLett.107.172001} {\bibfield  {journal} {\bibinfo  {journal}
  {Phys. Rev. Lett.}\ }\textbf {\bibinfo {volume} {107}},\ \bibinfo {pages}
  {172001} (\bibinfo {year} {2011})},\ \Eprint {http://arxiv.org/abs/1106.3076}
  {arXiv:1106.3076 [hep-ph]} \BibitemShut {NoStop}%
\bibitem [{\citenamefont {Gras}\ \emph {et~al.}(2017)\citenamefont {Gras},
  \citenamefont {H\"oche}, \citenamefont {Kar}, \citenamefont {Larkoski},
  \citenamefont {L\"onnblad}, \citenamefont {Pl\"atzer}, \citenamefont
  {Si\'odmok}, \citenamefont {Skands}, \citenamefont {Soyez},\ and\
  \citenamefont {Thaler}}]{Gras:2017jty}%
  \BibitemOpen
  \bibfield  {author} {\bibinfo {author} {\bibfnamefont {Philippe}\
  \bibnamefont {Gras}}, \bibinfo {author} {\bibfnamefont {Stefan}\ \bibnamefont
  {H\"oche}}, \bibinfo {author} {\bibfnamefont {Deepak}\ \bibnamefont {Kar}},
  \bibinfo {author} {\bibfnamefont {Andrew}\ \bibnamefont {Larkoski}}, \bibinfo
  {author} {\bibfnamefont {Leif}\ \bibnamefont {L\"onnblad}}, \bibinfo {author}
  {\bibfnamefont {Simon}\ \bibnamefont {Pl\"atzer}}, \bibinfo {author}
  {\bibfnamefont {Andrzej}\ \bibnamefont {Si\'odmok}}, \bibinfo {author}
  {\bibfnamefont {Peter}\ \bibnamefont {Skands}}, \bibinfo {author}
  {\bibfnamefont {Gregory}\ \bibnamefont {Soyez}}, \ and\ \bibinfo {author}
  {\bibfnamefont {Jesse}\ \bibnamefont {Thaler}},\ }\bibfield  {title}
  {\enquote {\bibinfo {title} {{Systematics of quark/gluon tagging}},}\ }\href
  {\doibase 10.1007/JHEP07(2017)091} {\bibfield  {journal} {\bibinfo  {journal}
  {JHEP}\ }\textbf {\bibinfo {volume} {07}},\ \bibinfo {pages} {091} (\bibinfo
  {year} {2017})},\ \Eprint {http://arxiv.org/abs/1704.03878} {arXiv:1704.03878
  [hep-ph]} \BibitemShut {NoStop}%
\bibitem [{\citenamefont {Sjostrand}\ \emph {et~al.}(2006)\citenamefont
  {Sjostrand}, \citenamefont {Mrenna},\ and\ \citenamefont
  {Skands}}]{Sjostrand:2006za}%
  \BibitemOpen
  \bibfield  {author} {\bibinfo {author} {\bibfnamefont {Torbjorn}\
  \bibnamefont {Sjostrand}}, \bibinfo {author} {\bibfnamefont {Stephen}\
  \bibnamefont {Mrenna}}, \ and\ \bibinfo {author} {\bibfnamefont {Peter~Z.}\
  \bibnamefont {Skands}},\ }\bibfield  {title} {\enquote {\bibinfo {title}
  {{PYTHIA 6.4 Physics and Manual}},}\ }\href {\doibase
  10.1088/1126-6708/2006/05/026} {\bibfield  {journal} {\bibinfo  {journal}
  {JHEP}\ }\textbf {\bibinfo {volume} {05}},\ \bibinfo {pages} {026} (\bibinfo
  {year} {2006})},\ \Eprint {http://arxiv.org/abs/hep-ph/0603175}
  {arXiv:hep-ph/0603175} \BibitemShut {NoStop}%
\bibitem [{\citenamefont {Sj\"ostrand}\ \emph {et~al.}(2015)\citenamefont
  {Sj\"ostrand}, \citenamefont {Ask}, \citenamefont {Christiansen},
  \citenamefont {Corke}, \citenamefont {Desai}, \citenamefont {Ilten},
  \citenamefont {Mrenna}, \citenamefont {Prestel}, \citenamefont {Rasmussen},\
  and\ \citenamefont {Skands}}]{Sjostrand:2014zea}%
  \BibitemOpen
  \bibfield  {author} {\bibinfo {author} {\bibfnamefont {Torbj\"orn}\
  \bibnamefont {Sj\"ostrand}}, \bibinfo {author} {\bibfnamefont {Stefan}\
  \bibnamefont {Ask}}, \bibinfo {author} {\bibfnamefont {Jesper~R.}\
  \bibnamefont {Christiansen}}, \bibinfo {author} {\bibfnamefont {Richard}\
  \bibnamefont {Corke}}, \bibinfo {author} {\bibfnamefont {Nishita}\
  \bibnamefont {Desai}}, \bibinfo {author} {\bibfnamefont {Philip}\
  \bibnamefont {Ilten}}, \bibinfo {author} {\bibfnamefont {Stephen}\
  \bibnamefont {Mrenna}}, \bibinfo {author} {\bibfnamefont {Stefan}\
  \bibnamefont {Prestel}}, \bibinfo {author} {\bibfnamefont {Christine~O.}\
  \bibnamefont {Rasmussen}}, \ and\ \bibinfo {author} {\bibfnamefont
  {Peter~Z.}\ \bibnamefont {Skands}},\ }\bibfield  {title} {\enquote {\bibinfo
  {title} {{An introduction to PYTHIA 8.2}},}\ }\href {\doibase
  10.1016/j.cpc.2015.01.024} {\bibfield  {journal} {\bibinfo  {journal}
  {Comput. Phys. Commun.}\ }\textbf {\bibinfo {volume} {191}},\ \bibinfo
  {pages} {159--177} (\bibinfo {year} {2015})},\ \Eprint
  {http://arxiv.org/abs/1410.3012} {arXiv:1410.3012 [hep-ph]} \BibitemShut
  {NoStop}%
\bibitem [{\citenamefont {Cacciari}\ \emph {et~al.}(2008)\citenamefont
  {Cacciari}, \citenamefont {Salam},\ and\ \citenamefont
  {Soyez}}]{Cacciari:2008gp}%
  \BibitemOpen
  \bibfield  {author} {\bibinfo {author} {\bibfnamefont {Matteo}\ \bibnamefont
  {Cacciari}}, \bibinfo {author} {\bibfnamefont {Gavin~P.}\ \bibnamefont
  {Salam}}, \ and\ \bibinfo {author} {\bibfnamefont {Gregory}\ \bibnamefont
  {Soyez}},\ }\bibfield  {title} {\enquote {\bibinfo {title} {{The anti-$k_t$
  jet clustering algorithm}},}\ }\href {\doibase 10.1088/1126-6708/2008/04/063}
  {\bibfield  {journal} {\bibinfo  {journal} {JHEP}\ }\textbf {\bibinfo
  {volume} {04}},\ \bibinfo {pages} {063} (\bibinfo {year} {2008})},\ \Eprint
  {http://arxiv.org/abs/0802.1189} {arXiv:0802.1189 [hep-ph]} \BibitemShut
  {NoStop}%
\bibitem [{\citenamefont {Cacciari}\ \emph {et~al.}(2012)\citenamefont
  {Cacciari}, \citenamefont {Salam},\ and\ \citenamefont
  {Soyez}}]{Cacciari:2011ma}%
  \BibitemOpen
  \bibfield  {author} {\bibinfo {author} {\bibfnamefont {Matteo}\ \bibnamefont
  {Cacciari}}, \bibinfo {author} {\bibfnamefont {Gavin~P.}\ \bibnamefont
  {Salam}}, \ and\ \bibinfo {author} {\bibfnamefont {Gregory}\ \bibnamefont
  {Soyez}},\ }\bibfield  {title} {\enquote {\bibinfo {title} {{FastJet User
  Manual}},}\ }\href {\doibase 10.1140/epjc/s10052-012-1896-2} {\bibfield
  {journal} {\bibinfo  {journal} {Eur. Phys. J. C}\ }\textbf {\bibinfo {volume}
  {72}},\ \bibinfo {pages} {1896} (\bibinfo {year} {2012})},\ \Eprint
  {http://arxiv.org/abs/1111.6097} {arXiv:1111.6097 [hep-ph]} \BibitemShut
  {NoStop}%
\bibitem [{\citenamefont {Metodiev}\ and\ \citenamefont
  {Thaler}(2018)}]{Metodiev:2018ftz}%
  \BibitemOpen
  \bibfield  {author} {\bibinfo {author} {\bibfnamefont {Eric~M.}\ \bibnamefont
  {Metodiev}}\ and\ \bibinfo {author} {\bibfnamefont {Jesse}\ \bibnamefont
  {Thaler}},\ }\bibfield  {title} {\enquote {\bibinfo {title} {{Jet Topics:
  Disentangling Quarks and Gluons at Colliders}},}\ }\href {\doibase
  10.1103/PhysRevLett.120.241602} {\bibfield  {journal} {\bibinfo  {journal}
  {Phys. Rev. Lett.}\ }\textbf {\bibinfo {volume} {120}},\ \bibinfo {pages}
  {241602} (\bibinfo {year} {2018})},\ \Eprint
  {http://arxiv.org/abs/1802.00008} {arXiv:1802.00008 [hep-ph]} \BibitemShut
  {NoStop}%
\bibitem [{\citenamefont {Komiske}\ \emph
  {et~al.}(2018{\natexlab{b}})\citenamefont {Komiske}, \citenamefont
  {Metodiev},\ and\ \citenamefont {Thaler}}]{Komiske:2018vkc}%
  \BibitemOpen
  \bibfield  {author} {\bibinfo {author} {\bibfnamefont {Patrick~T.}\
  \bibnamefont {Komiske}}, \bibinfo {author} {\bibfnamefont {Eric~M.}\
  \bibnamefont {Metodiev}}, \ and\ \bibinfo {author} {\bibfnamefont {Jesse}\
  \bibnamefont {Thaler}},\ }\bibfield  {title} {\enquote {\bibinfo {title} {{An
  operational definition of quark and gluon jets}},}\ }\href {\doibase
  10.1007/JHEP11(2018)059} {\bibfield  {journal} {\bibinfo  {journal} {JHEP}\
  }\textbf {\bibinfo {volume} {11}},\ \bibinfo {pages} {059} (\bibinfo {year}
  {2018}{\natexlab{b}})},\ \Eprint {http://arxiv.org/abs/1809.01140}
  {arXiv:1809.01140 [hep-ph]} \BibitemShut {NoStop}%
\bibitem [{\citenamefont {Komiske}\ \emph {et~al.}(2017)\citenamefont
  {Komiske}, \citenamefont {Metodiev},\ and\ \citenamefont
  {Schwartz}}]{Komiske:2016rsd}%
  \BibitemOpen
  \bibfield  {author} {\bibinfo {author} {\bibfnamefont {Patrick~T.}\
  \bibnamefont {Komiske}}, \bibinfo {author} {\bibfnamefont {Eric~M.}\
  \bibnamefont {Metodiev}}, \ and\ \bibinfo {author} {\bibfnamefont
  {Matthew~D.}\ \bibnamefont {Schwartz}},\ }\bibfield  {title} {\enquote
  {\bibinfo {title} {{Deep learning in color: towards automated quark/gluon jet
  discrimination}},}\ }\href {\doibase 10.1007/JHEP01(2017)110} {\bibfield
  {journal} {\bibinfo  {journal} {JHEP}\ }\textbf {\bibinfo {volume} {01}},\
  \bibinfo {pages} {110} (\bibinfo {year} {2017})},\ \Eprint
  {http://arxiv.org/abs/1612.01551} {arXiv:1612.01551 [hep-ph]} \BibitemShut
  {NoStop}%
\bibitem [{ATL(2017)}]{ATL-PHYS-PUB-2017-017}%
  \BibitemOpen
  \href {http://cds.cern.ch/record/2275641} {\emph {\bibinfo {title} {{Quark
  versus Gluon Jet Tagging Using Jet Images with the ATLAS Detector}}}},\
  \bibinfo {type} {Tech. Rep.}\ \bibinfo {number} {ATL-PHYS-PUB-2017-017}\
  (\bibinfo  {institution} {CERN},\ \bibinfo {address} {Geneva},\ \bibinfo
  {year} {2017})\BibitemShut {NoStop}%
\bibitem [{\citenamefont {Cheng}(2017)}]{Cheng:2017rdo}%
  \BibitemOpen
  \bibfield  {author} {\bibinfo {author} {\bibfnamefont {Taoli}\ \bibnamefont
  {Cheng}},\ }\bibfield  {title} {\enquote {\bibinfo {title} {{Recursive Neural
  Networks in Quark/Gluon Tagging}},}\ }\href {\doibase
  10.1007/s41781-018-0007-y} {\  (\bibinfo {year} {2017}),\
  10.1007/s41781-018-0007-y},\ \Eprint {http://arxiv.org/abs/1711.02633}
  {arXiv:1711.02633 [hep-ph]} \BibitemShut {NoStop}%
\bibitem [{\citenamefont {Luo}\ \emph {et~al.}(2019)\citenamefont {Luo},
  \citenamefont {Luo}, \citenamefont {Wang}, \citenamefont {Xu},\ and\
  \citenamefont {Zhu}}]{Luo:2017ncs}%
  \BibitemOpen
  \bibfield  {author} {\bibinfo {author} {\bibfnamefont {Hui}\ \bibnamefont
  {Luo}}, \bibinfo {author} {\bibfnamefont {Ming-Xing}\ \bibnamefont {Luo}},
  \bibinfo {author} {\bibfnamefont {Kai}\ \bibnamefont {Wang}}, \bibinfo
  {author} {\bibfnamefont {Tao}\ \bibnamefont {Xu}}, \ and\ \bibinfo {author}
  {\bibfnamefont {Guohuai}\ \bibnamefont {Zhu}},\ }\bibfield  {title} {\enquote
  {\bibinfo {title} {{Quark jet versus gluon jet: fully-connected neural
  networks with high-level features}},}\ }\href {\doibase
  10.1007/s11433-019-9390-8} {\bibfield  {journal} {\bibinfo  {journal} {Sci.
  China Phys. Mech. Astron.}\ }\textbf {\bibinfo {volume} {62}},\ \bibinfo
  {pages} {991011} (\bibinfo {year} {2019})},\ \Eprint
  {http://arxiv.org/abs/1712.03634} {arXiv:1712.03634 [hep-ph]} \BibitemShut
  {NoStop}%
\bibitem [{\citenamefont {Kasieczka}\ \emph {et~al.}(2019)\citenamefont
  {Kasieczka}, \citenamefont {Kiefer}, \citenamefont {Plehn},\ and\
  \citenamefont {Thompson}}]{Kasieczka:2018lwf}%
  \BibitemOpen
  \bibfield  {author} {\bibinfo {author} {\bibfnamefont {Gregor}\ \bibnamefont
  {Kasieczka}}, \bibinfo {author} {\bibfnamefont {Nicholas}\ \bibnamefont
  {Kiefer}}, \bibinfo {author} {\bibfnamefont {Tilman}\ \bibnamefont {Plehn}},
  \ and\ \bibinfo {author} {\bibfnamefont {Jennifer~M.}\ \bibnamefont
  {Thompson}},\ }\bibfield  {title} {\enquote {\bibinfo {title} {{Quark-Gluon
  Tagging: Machine Learning vs Detector}},}\ }\href {\doibase
  10.21468/SciPostPhys.6.6.069} {\bibfield  {journal} {\bibinfo  {journal}
  {SciPost Phys.}\ }\textbf {\bibinfo {volume} {6}},\ \bibinfo {pages} {069}
  (\bibinfo {year} {2019})},\ \Eprint {http://arxiv.org/abs/1812.09223}
  {arXiv:1812.09223 [hep-ph]} \BibitemShut {NoStop}%
\bibitem [{\citenamefont {Komiske}\ \emph
  {et~al.}(2019{\natexlab{b}})\citenamefont {Komiske}, \citenamefont
  {Metodiev},\ and\ \citenamefont {Thaler}}]{Komiske:2018cqr}%
  \BibitemOpen
  \bibfield  {author} {\bibinfo {author} {\bibfnamefont {Patrick~T.}\
  \bibnamefont {Komiske}}, \bibinfo {author} {\bibfnamefont {Eric~M.}\
  \bibnamefont {Metodiev}}, \ and\ \bibinfo {author} {\bibfnamefont {Jesse}\
  \bibnamefont {Thaler}},\ }\bibfield  {title} {\enquote {\bibinfo {title}
  {{Energy Flow Networks: Deep Sets for Particle Jets}},}\ }\href {\doibase
  10.1007/JHEP01(2019)121} {\bibfield  {journal} {\bibinfo  {journal} {JHEP}\
  }\textbf {\bibinfo {volume} {01}},\ \bibinfo {pages} {121} (\bibinfo {year}
  {2019}{\natexlab{b}})},\ \Eprint {http://arxiv.org/abs/1810.05165}
  {arXiv:1810.05165 [hep-ph]} \BibitemShut {NoStop}%
\bibitem [{\citenamefont {Lee}\ \emph {et~al.}(2019{\natexlab{a}})\citenamefont
  {Lee}, \citenamefont {Park}, \citenamefont {Watson},\ and\ \citenamefont
  {Yang}}]{Lee:2019cad}%
  \BibitemOpen
  \bibfield  {author} {\bibinfo {author} {\bibfnamefont {Jason Sang~Hun}\
  \bibnamefont {Lee}}, \bibinfo {author} {\bibfnamefont {Inkyu}\ \bibnamefont
  {Park}}, \bibinfo {author} {\bibfnamefont {Ian~James}\ \bibnamefont
  {Watson}}, \ and\ \bibinfo {author} {\bibfnamefont {Seungjin}\ \bibnamefont
  {Yang}},\ }\bibfield  {title} {\enquote {\bibinfo {title} {{Quark-Gluon Jet
  Discrimination Using Convolutional Neural Networks}},}\ }\href {\doibase
  10.3938/jkps.74.219} {\bibfield  {journal} {\bibinfo  {journal} {J. Korean
  Phys. Soc.}\ }\textbf {\bibinfo {volume} {74}},\ \bibinfo {pages} {219--223}
  (\bibinfo {year} {2019}{\natexlab{a}})},\ \Eprint
  {http://arxiv.org/abs/2012.02531} {arXiv:2012.02531 [hep-ex]} \BibitemShut
  {NoStop}%
\bibitem [{\citenamefont {Lee}\ \emph {et~al.}(2019{\natexlab{b}})\citenamefont
  {Lee}, \citenamefont {Lee}, \citenamefont {Lee}, \citenamefont {Park},
  \citenamefont {Watson},\ and\ \citenamefont {Yang}}]{Lee:2019ssx}%
  \BibitemOpen
  \bibfield  {author} {\bibinfo {author} {\bibfnamefont {Jason Sang~Hun}\
  \bibnamefont {Lee}}, \bibinfo {author} {\bibfnamefont {Sang~Man}\
  \bibnamefont {Lee}}, \bibinfo {author} {\bibfnamefont {Yunjae}\ \bibnamefont
  {Lee}}, \bibinfo {author} {\bibfnamefont {Inkyu}\ \bibnamefont {Park}},
  \bibinfo {author} {\bibfnamefont {Ian~James}\ \bibnamefont {Watson}}, \ and\
  \bibinfo {author} {\bibfnamefont {Seungjin}\ \bibnamefont {Yang}},\
  }\bibfield  {title} {\enquote {\bibinfo {title} {{Quark Gluon Jet
  Discrimination with Weakly Supervised Learning}},}\ }\href {\doibase
  10.3938/jkps.75.652} {\bibfield  {journal} {\bibinfo  {journal} {J. Korean
  Phys. Soc.}\ }\textbf {\bibinfo {volume} {75}},\ \bibinfo {pages} {652--659}
  (\bibinfo {year} {2019}{\natexlab{b}})},\ \Eprint
  {http://arxiv.org/abs/2012.02540} {arXiv:2012.02540 [hep-ph]} \BibitemShut
  {NoStop}%
\bibitem [{\citenamefont {Moreno}\ \emph {et~al.}(2020)\citenamefont {Moreno},
  \citenamefont {Cerri}, \citenamefont {Duarte}, \citenamefont {Newman},
  \citenamefont {Nguyen}, \citenamefont {Periwal}, \citenamefont {Pierini},
  \citenamefont {Serikova}, \citenamefont {Spiropulu},\ and\ \citenamefont
  {Vlimant}}]{Moreno:2019bmu}%
  \BibitemOpen
  \bibfield  {author} {\bibinfo {author} {\bibfnamefont {Eric~A.}\ \bibnamefont
  {Moreno}}, \bibinfo {author} {\bibfnamefont {Olmo}\ \bibnamefont {Cerri}},
  \bibinfo {author} {\bibfnamefont {Javier~M.}\ \bibnamefont {Duarte}},
  \bibinfo {author} {\bibfnamefont {Harvey~B.}\ \bibnamefont {Newman}},
  \bibinfo {author} {\bibfnamefont {Thong~Q.}\ \bibnamefont {Nguyen}}, \bibinfo
  {author} {\bibfnamefont {Avikar}\ \bibnamefont {Periwal}}, \bibinfo {author}
  {\bibfnamefont {Maurizio}\ \bibnamefont {Pierini}}, \bibinfo {author}
  {\bibfnamefont {Aidana}\ \bibnamefont {Serikova}}, \bibinfo {author}
  {\bibfnamefont {Maria}\ \bibnamefont {Spiropulu}}, \ and\ \bibinfo {author}
  {\bibfnamefont {Jean-Roch}\ \bibnamefont {Vlimant}},\ }\bibfield  {title}
  {\enquote {\bibinfo {title} {{JEDI-net: a jet identification algorithm based
  on interaction networks}},}\ }\href {\doibase 10.1140/epjc/s10052-020-7608-4}
  {\bibfield  {journal} {\bibinfo  {journal} {Eur. Phys. J. C}\ }\textbf
  {\bibinfo {volume} {80}},\ \bibinfo {pages} {58} (\bibinfo {year} {2020})},\
  \Eprint {http://arxiv.org/abs/1908.05318} {arXiv:1908.05318 [hep-ex]}
  \BibitemShut {NoStop}%
\bibitem [{\citenamefont {Qu}\ and\ \citenamefont
  {Gouskos}(2020{\natexlab{b}})}]{Qu:2019gqs}%
  \BibitemOpen
  \bibfield  {author} {\bibinfo {author} {\bibfnamefont {Huilin}\ \bibnamefont
  {Qu}}\ and\ \bibinfo {author} {\bibfnamefont {Loukas}\ \bibnamefont
  {Gouskos}},\ }\bibfield  {title} {\enquote {\bibinfo {title} {{ParticleNet:
  Jet Tagging via Particle Clouds}},}\ }\href {\doibase
  10.1103/PhysRevD.101.056019} {\bibfield  {journal} {\bibinfo  {journal}
  {Phys. Rev. D}\ }\textbf {\bibinfo {volume} {101}},\ \bibinfo {pages}
  {056019} (\bibinfo {year} {2020}{\natexlab{b}})},\ \Eprint
  {http://arxiv.org/abs/1902.08570} {arXiv:1902.08570 [hep-ph]} \BibitemShut
  {NoStop}%
\bibitem [{\citenamefont {Mikuni}\ and\ \citenamefont
  {Canelli}(2020)}]{Mikuni:2020wpr}%
  \BibitemOpen
  \bibfield  {author} {\bibinfo {author} {\bibfnamefont {Vinicius}\
  \bibnamefont {Mikuni}}\ and\ \bibinfo {author} {\bibfnamefont {Florencia}\
  \bibnamefont {Canelli}},\ }\bibfield  {title} {\enquote {\bibinfo {title}
  {{ABCNet: An attention-based method for particle tagging}},}\ }\href
  {\doibase 10.1140/epjp/s13360-020-00497-3} {\bibfield  {journal} {\bibinfo
  {journal} {Eur. Phys. J. Plus}\ }\textbf {\bibinfo {volume} {135}},\ \bibinfo
  {pages} {463} (\bibinfo {year} {2020})},\ \Eprint
  {http://arxiv.org/abs/2001.05311} {arXiv:2001.05311 [physics.data-an]}
  \BibitemShut {NoStop}%
\bibitem [{\citenamefont {He}\ and\ \citenamefont {Wang}(2023)}]{He:2023cfc}%
  \BibitemOpen
  \bibfield  {author} {\bibinfo {author} {\bibfnamefont {Minxuan}\ \bibnamefont
  {He}}\ and\ \bibinfo {author} {\bibfnamefont {Daohan}\ \bibnamefont {Wang}},\
  }\bibfield  {title} {\enquote {\bibinfo {title} {{Quark/Gluon Discrimination
  and Top Tagging with Dual Attention Transformer}},}\ }\href@noop {} {\
  (\bibinfo {year} {2023})},\ \Eprint {http://arxiv.org/abs/2307.04723}
  {arXiv:2307.04723 [hep-ph]} \BibitemShut {NoStop}%
\bibitem [{\citenamefont {Dolan}\ \emph {et~al.}(2023)\citenamefont {Dolan},
  \citenamefont {Gargalionis},\ and\ \citenamefont {Ore}}]{Dolan:2023abg}%
  \BibitemOpen
  \bibfield  {author} {\bibinfo {author} {\bibfnamefont {Matthew~J.}\
  \bibnamefont {Dolan}}, \bibinfo {author} {\bibfnamefont {John}\ \bibnamefont
  {Gargalionis}}, \ and\ \bibinfo {author} {\bibfnamefont {Ayodele}\
  \bibnamefont {Ore}},\ }\bibfield  {title} {\enquote {\bibinfo {title}
  {{Quark-versus-gluon tagging in CMS Open Data with CWoLa and TopicFlow}},}\
  }\href@noop {} {\  (\bibinfo {year} {2023})},\ \Eprint
  {http://arxiv.org/abs/2312.03434} {arXiv:2312.03434 [hep-ph]} \BibitemShut
  {NoStop}%
\bibitem [{\citenamefont {Datta}\ and\ \citenamefont
  {Larkoski}(2018)}]{Datta:2017lxt}%
  \BibitemOpen
  \bibfield  {author} {\bibinfo {author} {\bibfnamefont {Kaustuv}\ \bibnamefont
  {Datta}}\ and\ \bibinfo {author} {\bibfnamefont {Andrew~J.}\ \bibnamefont
  {Larkoski}},\ }\bibfield  {title} {\enquote {\bibinfo {title} {{Novel Jet
  Observables from Machine Learning}},}\ }\href {\doibase
  10.1007/JHEP03(2018)086} {\bibfield  {journal} {\bibinfo  {journal} {JHEP}\
  }\textbf {\bibinfo {volume} {03}},\ \bibinfo {pages} {086} (\bibinfo {year}
  {2018})},\ \Eprint {http://arxiv.org/abs/1710.01305} {arXiv:1710.01305
  [hep-ph]} \BibitemShut {NoStop}%
\bibitem [{\citenamefont {Ellis}\ \emph {et~al.}(1992)\citenamefont {Ellis},
  \citenamefont {Kunszt},\ and\ \citenamefont {Soper}}]{Ellis:1992qq}%
  \BibitemOpen
  \bibfield  {author} {\bibinfo {author} {\bibfnamefont {Stephen~D.}\
  \bibnamefont {Ellis}}, \bibinfo {author} {\bibfnamefont {Zoltan}\
  \bibnamefont {Kunszt}}, \ and\ \bibinfo {author} {\bibfnamefont {Davison~E.}\
  \bibnamefont {Soper}},\ }\bibfield  {title} {\enquote {\bibinfo {title}
  {{Jets at hadron colliders at order $\alpha-s^{3:}$ A Look inside}},}\ }\href
  {\doibase 10.1103/PhysRevLett.69.3615} {\bibfield  {journal} {\bibinfo
  {journal} {Phys. Rev. Lett.}\ }\textbf {\bibinfo {volume} {69}},\ \bibinfo
  {pages} {3615--3618} (\bibinfo {year} {1992})},\ \Eprint
  {http://arxiv.org/abs/hep-ph/9208249} {arXiv:hep-ph/9208249} \BibitemShut
  {NoStop}%
\bibitem [{\citenamefont {Abe}\ \emph {et~al.}(1993)\citenamefont {Abe} \emph
  {et~al.}}]{CDF:1992cus}%
  \BibitemOpen
  \bibfield  {author} {\bibinfo {author} {\bibfnamefont {F.}~\bibnamefont
  {Abe}} \emph {et~al.} (\bibinfo {collaboration} {CDF}),\ }\bibfield  {title}
  {\enquote {\bibinfo {title} {{A Measurement of jet shapes in $p\bar{p}$
  collisions at $\sqrt{s} = 1.8$ TeV}},}\ }\href {\doibase
  10.1103/PhysRevLett.70.713} {\bibfield  {journal} {\bibinfo  {journal} {Phys.
  Rev. Lett.}\ }\textbf {\bibinfo {volume} {70}},\ \bibinfo {pages} {713--717}
  (\bibinfo {year} {1993})}\BibitemShut {NoStop}%
\bibitem [{\citenamefont {Larkoski}\ \emph {et~al.}(2013)\citenamefont
  {Larkoski}, \citenamefont {Salam},\ and\ \citenamefont
  {Thaler}}]{Larkoski:2013eya}%
  \BibitemOpen
  \bibfield  {author} {\bibinfo {author} {\bibfnamefont {Andrew~J.}\
  \bibnamefont {Larkoski}}, \bibinfo {author} {\bibfnamefont {Gavin~P.}\
  \bibnamefont {Salam}}, \ and\ \bibinfo {author} {\bibfnamefont {Jesse}\
  \bibnamefont {Thaler}},\ }\bibfield  {title} {\enquote {\bibinfo {title}
  {{Energy Correlation Functions for Jet Substructure}},}\ }\href {\doibase
  10.1007/JHEP06(2013)108} {\bibfield  {journal} {\bibinfo  {journal} {JHEP}\
  }\textbf {\bibinfo {volume} {06}},\ \bibinfo {pages} {108} (\bibinfo {year}
  {2013})},\ \Eprint {http://arxiv.org/abs/1305.0007} {arXiv:1305.0007
  [hep-ph]} \BibitemShut {NoStop}%
\bibitem [{\citenamefont {Wallace}(1958)}]{10.1214/aoms/1177706528}%
  \BibitemOpen
  \bibfield  {author} {\bibinfo {author} {\bibfnamefont {David~L.}\
  \bibnamefont {Wallace}},\ }\bibfield  {title} {\enquote {\bibinfo {title}
  {{Asymptotic Approximations to Distributions}},}\ }\href {\doibase
  10.1214/aoms/1177706528} {\bibfield  {journal} {\bibinfo  {journal} {The
  Annals of Mathematical Statistics}\ }\textbf {\bibinfo {volume} {29}},\
  \bibinfo {pages} {635 -- 654} (\bibinfo {year} {1958})}\BibitemShut {NoStop}%
\bibitem [{\citenamefont {Maas}(2013)}]{Maas2013RectifierNI}%
  \BibitemOpen
  \bibfield  {author} {\bibinfo {author} {\bibfnamefont {Andrew~L.}\
  \bibnamefont {Maas}},\ }\bibfield  {title} {\enquote {\bibinfo {title}
  {Rectifier nonlinearities improve neural network acoustic models},}\ \
  }(\bibinfo {year} {2013})\BibitemShut {NoStop}%
\bibitem [{\citenamefont {Kluyver}\ \emph {et~al.}(2016)\citenamefont
  {Kluyver}, \citenamefont {Ragan-Kelley}, \citenamefont {P{\'e}rez},
  \citenamefont {Granger}, \citenamefont {Bussonnier}, \citenamefont
  {Frederic}, \citenamefont {Kelley}, \citenamefont {Hamrick}, \citenamefont
  {Grout}, \citenamefont {Corlay}, \citenamefont {Ivanov}, \citenamefont
  {Avila}, \citenamefont {Abdalla},\ and\ \citenamefont
  {Willing}}]{Kluyver:2016aa}%
  \BibitemOpen
  \bibfield  {author} {\bibinfo {author} {\bibfnamefont {Thomas}\ \bibnamefont
  {Kluyver}}, \bibinfo {author} {\bibfnamefont {Benjamin}\ \bibnamefont
  {Ragan-Kelley}}, \bibinfo {author} {\bibfnamefont {Fernando}\ \bibnamefont
  {P{\'e}rez}}, \bibinfo {author} {\bibfnamefont {Brian}\ \bibnamefont
  {Granger}}, \bibinfo {author} {\bibfnamefont {Matthias}\ \bibnamefont
  {Bussonnier}}, \bibinfo {author} {\bibfnamefont {Jonathan}\ \bibnamefont
  {Frederic}}, \bibinfo {author} {\bibfnamefont {Kyle}\ \bibnamefont {Kelley}},
  \bibinfo {author} {\bibfnamefont {Jessica}\ \bibnamefont {Hamrick}}, \bibinfo
  {author} {\bibfnamefont {Jason}\ \bibnamefont {Grout}}, \bibinfo {author}
  {\bibfnamefont {Sylvain}\ \bibnamefont {Corlay}}, \bibinfo {author}
  {\bibfnamefont {Paul}\ \bibnamefont {Ivanov}}, \bibinfo {author}
  {\bibfnamefont {Dami{\'a}n}\ \bibnamefont {Avila}}, \bibinfo {author}
  {\bibfnamefont {Safia}\ \bibnamefont {Abdalla}}, \ and\ \bibinfo {author}
  {\bibfnamefont {Carol}\ \bibnamefont {Willing}},\ }\bibfield  {title}
  {\enquote {\bibinfo {title} {Jupyter notebooks -- a publishing format for
  reproducible computational workflows},}\ }in\ \href@noop {} {\emph {\bibinfo
  {booktitle} {Positioning and Power in Academic Publishing: Players, Agents
  and Agendas}}},\ \bibinfo {editor} {edited by\ \bibinfo {editor}
  {\bibfnamefont {F.}~\bibnamefont {Loizides}}\ and\ \bibinfo {editor}
  {\bibfnamefont {B.}~\bibnamefont {Schmidt}}}\ (\bibinfo {organization} {IOS
  Press},\ \bibinfo {year} {2016})\ pp.\ \bibinfo {pages} {87 --
  90}\BibitemShut {NoStop}%
\bibitem [{\citenamefont {Chollet}(2017)}]{keras}%
  \BibitemOpen
  \bibfield  {author} {\bibinfo {author} {\bibfnamefont {Fraciois}\
  \bibnamefont {Chollet}},\ }\bibfield  {title} {\enquote {\bibinfo {title}
  {Keras},}\ }\href@noop {} {\bibfield  {journal} {\bibinfo  {journal} {GitHub
  repository}\ } (\bibinfo {year} {2017})}\BibitemShut {NoStop}%
\bibitem [{\citenamefont {Abadi}\ \emph {et~al.}(2016)\citenamefont {Abadi},
  \citenamefont {Barham}, \citenamefont {Chen}, \citenamefont {Chen},
  \citenamefont {Davis}, \citenamefont {Dean}, \citenamefont {Devin},
  \citenamefont {Ghemawat}, \citenamefont {Irving}, \citenamefont {Isard} \emph
  {et~al.}}]{tensorflow}%
  \BibitemOpen
  \bibfield  {author} {\bibinfo {author} {\bibfnamefont {Mart{\'\i}n}\
  \bibnamefont {Abadi}}, \bibinfo {author} {\bibfnamefont {Paul}\ \bibnamefont
  {Barham}}, \bibinfo {author} {\bibfnamefont {Jianmin}\ \bibnamefont {Chen}},
  \bibinfo {author} {\bibfnamefont {Zhifeng}\ \bibnamefont {Chen}}, \bibinfo
  {author} {\bibfnamefont {Andy}\ \bibnamefont {Davis}}, \bibinfo {author}
  {\bibfnamefont {Jeffrey}\ \bibnamefont {Dean}}, \bibinfo {author}
  {\bibfnamefont {Matthieu}\ \bibnamefont {Devin}}, \bibinfo {author}
  {\bibfnamefont {Sanjay}\ \bibnamefont {Ghemawat}}, \bibinfo {author}
  {\bibfnamefont {Geoffrey}\ \bibnamefont {Irving}}, \bibinfo {author}
  {\bibfnamefont {Michael}\ \bibnamefont {Isard}},  \emph {et~al.},\ }\bibfield
   {title} {\enquote {\bibinfo {title} {Tensorflow: A system for large-scale
  machine learning.}}\ }in\ \href@noop {} {\emph {\bibinfo {booktitle}
  {OSDI}}},\ Vol.~\bibinfo {volume} {16}\ (\bibinfo {year} {2016})\ pp.\
  \bibinfo {pages} {265--283}\BibitemShut {NoStop}%
\bibitem [{\citenamefont {Lu}(2020)}]{dyingrelu}%
  \BibitemOpen
  \bibfield  {author} {\bibinfo {author} {\bibfnamefont {Lu}~\bibnamefont
  {Lu}},\ }\bibfield  {title} {\enquote {\bibinfo {title} {Dying {ReLU} and
  initialization: Theory and numerical examples},}\ }\href {\doibase
  10.4208/cicp.oa-2020-0165} {\bibfield  {journal} {\bibinfo  {journal}
  {Communications in Computational Physics}\ }\textbf {\bibinfo {volume}
  {28}},\ \bibinfo {pages} {1671--1706} (\bibinfo {year} {2020})}\BibitemShut
  {NoStop}%
\bibitem [{\citenamefont {He}\ \emph {et~al.}(2015)\citenamefont {He},
  \citenamefont {Zhang}, \citenamefont {Ren},\ and\ \citenamefont
  {Sun}}]{he2015delving}%
  \BibitemOpen
  \bibfield  {author} {\bibinfo {author} {\bibfnamefont {Kaiming}\ \bibnamefont
  {He}}, \bibinfo {author} {\bibfnamefont {Xiangyu}\ \bibnamefont {Zhang}},
  \bibinfo {author} {\bibfnamefont {Shaoqing}\ \bibnamefont {Ren}}, \ and\
  \bibinfo {author} {\bibfnamefont {Jian}\ \bibnamefont {Sun}},\ }\href@noop {}
  {\enquote {\bibinfo {title} {Delving deep into rectifiers: Surpassing
  human-level performance on imagenet classification},}\ } (\bibinfo {year}
  {2015}),\ \Eprint {http://arxiv.org/abs/1502.01852} {arXiv:1502.01852
  [cs.CV]} \BibitemShut {NoStop}%
\bibitem [{\citenamefont {Kingma}\ and\ \citenamefont
  {Ba}(2017)}]{kingma2017adam}%
  \BibitemOpen
  \bibfield  {author} {\bibinfo {author} {\bibfnamefont {Diederik~P.}\
  \bibnamefont {Kingma}}\ and\ \bibinfo {author} {\bibfnamefont {Jimmy}\
  \bibnamefont {Ba}},\ }\href@noop {} {\enquote {\bibinfo {title} {Adam: A
  method for stochastic optimization},}\ } (\bibinfo {year} {2017}),\ \Eprint
  {http://arxiv.org/abs/1412.6980} {arXiv:1412.6980 [cs.LG]} \BibitemShut
  {NoStop}%
\bibitem [{\citenamefont {de~Favereau}\ \emph {et~al.}(2014)\citenamefont
  {de~Favereau}, \citenamefont {Delaere}, \citenamefont {Demin}, \citenamefont
  {Giammanco}, \citenamefont {Lema\^\i{}tre}, \citenamefont {Mertens},\ and\
  \citenamefont {Selvaggi}}]{deFavereau:2013fsa}%
  \BibitemOpen
  \bibfield  {author} {\bibinfo {author} {\bibfnamefont {J.}~\bibnamefont
  {de~Favereau}}, \bibinfo {author} {\bibfnamefont {C.}~\bibnamefont
  {Delaere}}, \bibinfo {author} {\bibfnamefont {P.}~\bibnamefont {Demin}},
  \bibinfo {author} {\bibfnamefont {A.}~\bibnamefont {Giammanco}}, \bibinfo
  {author} {\bibfnamefont {V.}~\bibnamefont {Lema\^\i{}tre}}, \bibinfo {author}
  {\bibfnamefont {A.}~\bibnamefont {Mertens}}, \ and\ \bibinfo {author}
  {\bibfnamefont {M.}~\bibnamefont {Selvaggi}} (\bibinfo {collaboration}
  {DELPHES 3}),\ }\bibfield  {title} {\enquote {\bibinfo {title} {{DELPHES 3, A
  modular framework for fast simulation of a generic collider experiment}},}\
  }\href {\doibase 10.1007/JHEP02(2014)057} {\bibfield  {journal} {\bibinfo
  {journal} {JHEP}\ }\textbf {\bibinfo {volume} {02}},\ \bibinfo {pages} {057}
  (\bibinfo {year} {2014})},\ \Eprint {http://arxiv.org/abs/1307.6346}
  {arXiv:1307.6346 [hep-ex]} \BibitemShut {NoStop}%
\end{thebibliography}%

\end{document}